\documentclass[twocolumn]{aa}

\def\HISTORY{
Reference for this paper: Maris et al., 2021, A\&A, 647 (2021) A104
}

\usepackage{url,lineno,microtype,subcaption,tikz,supertabular,longtable,booktabs}
\usepackage[draft]{hyperref}
\usepackage[varg]{txfonts}
\usepackage{wasysym}
\usepackage{rotating}
\usepackage{makecell}
\usepackage{siunitx}

\usepackage{natbib}

\author{
        Michele Maris\inst{\ref{oats}}\thanks{Corresponding author \email{michele.maris@inaf.it}.} 
\and Erik Romelli\inst{\ref{oats}} 
\and Maurizio Tomasi\inst{\ref{unimi}} 
\and Anna Gregorio\inst{\ref{units},\ref{oats}}  
\and Maura Sandri\inst{\ref{oasbo}} 
\and Samuele Galeotta\inst{\ref{oats}}
\and Daniele Tavagnacco\inst{\ref{oats}} 
\and Marco Frailis\inst{\ref{oats}} 
\and Gianmarco Maggio\inst{\ref{oats}} 
\and Andrea Zacchei\inst{\ref{oats}}
}

\institute{
INAF/Trieste Astronomical Observatory, Via G.B.Tiepolo 11 - 34143, Trieste, Italy\label{oats}
\and 
Dipartimento di Fisica ''Aldo Pontremoli'', Universit\`a{} degli Studi di Milano, Via G.Celoria 16, 20133, Milano, Italy\label{unimi}
\and 
Trieste University: Physics Department, Via A. Valerio 2 - 34127, Trieste, Italy\label{units}
\and 
INAF/Bologna Astronomical Observatory, Via Gobetti 93/3 - 40129, Bologna, Italy\label{oasbo}
}

\graphicspath{{images/}}

\newcommand{\Planck}{\textit{Planck}}

\newcommand{\tablenote}[2]{\parbox{\columnwidth}{$^{#1}$\ #2}}
\newcommand{\widetablenote}[2]{\parbox{\textwidth}{$^{#1}$\ #2}}
\newcommand{\tablenoteskip}{\vspace{1em}}

\newcommand{\SKIPTHISLATEX}[1]{   }

\def\water{\mathrm{H}_2\mathrm{O}} \def\SulfidricAcid{\mathrm{H}_2\mathrm{S}} \def\Methane{\mathrm{C}\mathrm{H}_4} \def\Ammonia{\mathrm{N}\mathrm{H}_3} \def\Phosphine{\mathrm{P}\mathrm{H}_3} \def\HidrogenTwo{\mathrm{H}_2}

\def\FcentCrit{\nu_{\mathrm{cent},1}}

                                     \def\GRASP{\textrm{GRASP}}

\def\Planck{\textit{Planck}}   \def\Herschel{\textit{Herschel}}

\def\kboltzman{k_{\mathrm{b}}}  

\def\deg{^{\circ}}

\def\sterad{\mathrm{sterad}}

\def\Krj{\mathrm{K}_{\mathrm{RJ}}}
\def\Kcmb{\mathrm{K}_{\mathrm{cmb}}}
\def\mKcmb{\mathrm{mK}_{\mathrm{cmb}}}
\def\muK{\mu\mathrm{K}}

\def\BandPass{\tau}                          \def\Fcent{\nu_{\mathrm{cent}}} \def\Fcenti{\nu_{\mathrm{cent},i}}  \def\BandWidth{\Delta\nu}       

\def\FcentEff{\nu_{\mathrm{cent},\mathrm{eff}}}

\def\Beam{\gamma_{\nu}}                    

\def\BeamBA{\gamma^{(\mathrm{ba})}}                                      \def\BeamBAdip{\gamma_{\mathrm{dip}}^{(\mathrm{ba})}}                   \def\BeamBAp{\gamma_{\mathrm{p}}^{(\mathrm{ba})}}                   \def\BeamGRASP{\gamma_{\mathrm{GRASP},\nu}}

\def\Win{\Delta I_{\mathrm{in}}}                     \def\WinBck{\Delta I_{\mathrm{in},\mathrm{bck}}} \def\WinP{\Delta I_{\mathrm{in},\mathrm{p}}}     \def\WinBlk{\Delta I_{\mathrm{in},\mathrm{block}}} \def\DeltaWdip{\Delta I_{\mathrm{dip}}}          \def\DeltaWinZero{\Delta I_{\mathrm{0}}}           

\def\SED{S}                               \def\SEDba{S^{(\mathrm{ba})}}                           \def\SEDbabck{S_{\mathrm{bck}}^{(\mathrm{ba})}}

\def\DipoleWMAP{A_{\mathrm{WMAP}}}        \def\DipolePlanck{A_{\mathrm{Planck}}}    \def\wmapTOplanck{G_{\mathrm{WMAP},\mathrm{Planck}}}

\def\SpinAxis{\mathbf{\hat{S}}}                      \def\PlanckLocation{\mathbf{R}_{\mathrm{S}}}                                                   

                                                       \def\PlanetDirectiont{\mathbf{\hat{\Delta}}_t}  \def\PlanetVector{\mathbf{\Delta}}                       \def\PlanetDistance{\Delta}                                               \def\PlanetBoresightDot{\dot{\beta}}                    \def\PlanetBoresight{\beta}                    \def\OmegaPlanet{\Omega_{\mathrm{p}}}          \def\OmegaPlanetEquatorial{\Omega_{\mathrm{p}}^{(\mathrm{eq})}}          \def\AngDiam{\Theta_{\mathrm{p}}}          \def\Rsun{R_{\mathrm{h}}}                      

\def\PlanetDistanceFiducial{\tilde{\Delta}}                           \def\OmegaPlanetFiducial{\widetilde{\Omega}_{\mathrm{p}}}          

\def\PlanetRadiusEquatorial{R_{\mathrm{eq}}} \def\PlanetRadiusPolar{R_{\mathrm{pol}}}           

\def\geometricalCorrectionFactor{f_{\mathrm{geom}}} \def\aspectCorrectionFactor{f_{\mathrm{asp}}} \def\effAsp{\aspectCorrectionFactor}

\def\Fsync{F_{\mathrm{sync}}}                                          

\def\SyncOneBA{F_{\mathrm{sync},1}^{\mathrm{(ba)}}}

\def\PlanetAspectAngle{D_{\mathrm{P}}}

\def\PlanetLocation{\mathbf{R}_{\textit{pl}}}

        \def\boresight{\beta_{\mathbf{fh}}}                 

\def\OmegaBeam{\Omega_{\mathrm{beam}}}     \def\OmegaBeamNu{\Omega_{\mathrm{beam},\nu}}     \def\OmegaBeamba{\Omega_{\mathrm{beam}}^{(\mathrm{ba})}} \def\OmegaBeambadip{\Omega_{\mathrm{beam},\mathrm{dip}}^{(\mathrm{ba})}} \def\OmegaBeambacmb{{\Omega}_{\mathrm{beam},\mathrm{cmb}}^{(\mathrm{ba})}}  \def\OmegaBeambap{\Omega_{\mathrm{beam},\mathrm{p}}^{(\mathrm{ba})}} 

\def\OmegaBeamFiducial{\tilde{\Omega}_{\mathrm{beam}}}

\def\effSL{f_{\mathrm{SL}}}                                    \def\effAper{f_{\mathrm{aper}}}                                

\def\effTbBa{f_{\mathrm{TbBa}}}                

\def\effEtaBeam{f_{\mathrm{\eta}}}                \def\effEtaBeami{f_{\mathrm{\eta},i}}                

\def\effEtaBeamCorrection{x_{\mathrm{\eta}}}                                

\def\ROI{R_{\mathrm{ROI}}}
\def\ROIplanet{R_{\mathrm{ROI-I}}}
\def\ROIavoid{R_{\mathrm{ROI-II}}}

\def\Bnu{B_{\nu}}                          \def\invBnu{B_{\nu}^{-1}}                  \def\BnuBA{B_{\nu}^{\mathrm{(ba)}}}                                        \def\BnuBAi{B_{\nu,i}^{\mathrm{(ba)}}}

\def\DeltaTdip{\Delta T_{\mathrm{dip}}}

\def\DeltaTantP{\Delta T_{\mathrm{ant},\mathrm{p}}}
\def\DeltaTantPstar{\Delta T_{\mathrm{ant},\mathrm{p}}^*}
\def\DeltaTantPstarFiducial{\widetilde{\Delta T}_{\mathrm{ant},\mathrm{p}}}

\def\Tcmb{T_{\mathrm{cmb}}}
\def\Tant{T_{\mathrm{ant}}}

\def\Tb{T_{\mathrm{b}}}

\def\Tantt{T_{\mathrm{ant},t}}

\def\TbRJWMAP{T_{\mathrm{b},\mathrm{rj}}^{(\mathrm{wmap})}} 

 \def\OmegaPlanetFiducial{\tilde{\Omega}_{\mathrm{p}}} \def\OmegaPlanetFiducialSync{\tilde{\Omega}_{\mathrm{p},\mathrm{sync}}} 

\def\TbRT{T_{\mathrm{b}}^{(\mathrm{RT})}}

\def\BrBArt{B^{\mathrm{(ba)}}_{\mathrm{RT}}}         

\def\BrThSync{B_{\mathrm{RT+sync}}}         

\def\TbRJ{T_{\mathrm{b},\mathrm{rj}}} \def\TbRJt{T_{\mathrm{b},\mathrm{rj},t}} \def\TbRJrnd{\delta_{\mathrm{rnd}} T_{\mathrm{b},\mathrm{rj}}} 

 \def\TbMono{T_{\mathrm{b},\mathrm{c}}} \def\TbBA{T_{\mathrm{b}}^{(\mathrm{ba})}}  

\def\DeltaTant{\Delta T_{\mathrm{ant}}}

\def\DeltaTantt{\Delta T_{\mathrm{ant},t}}

\def\DeltaTantBlk{\Delta T_{\mathrm{ant},\mathrm{block}}}

\def\BeamOrientation{\mathbf{\hat{\Theta}}} 

\def\Pointing{\mathbf{\hat{P}}}

\def\Pointingt{\mathbf{\hat{P}}_t}

\def\PointingPrime{\mathbf{\hat{P}}'}

\def\versorex{\mathbf{\hat{e}}_{\mathrm{\mathbf{x}}}} \def\versorey{\mathbf{\hat{e}}_{\mathrm{\mathbf{y}}}}  \def\versorezbrf{\mathbf{\hat{e}}_{\mathrm{\mathbf{z}}}^{\mathrm{brf}}} 

\def\SourceStructure{u}

\def\UeclBeam{\mathbf{U}_{\mathrm{ecl},\mathrm{beam}}}

\def\UeclBeamInv{\mathbf{U}_{\mathrm{beam},\mathrm{ecl}}}
\def\UeclBeamInvt{\mathbf{U}_{\mathrm{beam},\mathrm{ecl},t}}

\def\gammabat{g_{t}^{(\mathrm{ba})}}
\def\gammabapt{g_{\mathrm{p},t}^{(\mathrm{ba})}}

\def\gammabacmbt{g_{\mathrm{cmb},t}^{(\mathrm{ba})}}

\def\fwhm{\mathrm{FWHM}} 

 \def\tsmp{\delta t_{\mathrm{samp}}}

\def\BrBA{B^{\mathrm{(ba)}}}                  \def\BrBAplanet{B_{\mathrm{p}}}         \def\BrBAplaneti{B_{\mathrm{p},i}}

\def\BcmbBA{B_{\nu,\mathrm{cmb}}^{\mathrm{ba}}}         \def\Brj{B_{\nu,\mathrm{rj}}}                      \def\BrjOne{B_{\nu,\mathrm{rj},1}}           \def\BrjBAOne{B_{\nu,\mathrm{rj},1}^{\mathrm{ba}}}           \def\BrjOneRad{B_{\nu,\mathrm{rj},1,r}}           

                                 \def\dBdTcmbba{\left(\frac{dB_{\nu}}{dT}\right)_{\mathrm{cmb}}^{(\mathrm{ba})}} 

\def\dBdTcmb{\left(\frac{dB_{\nu}}{dT}\right)_{\mathrm{cmb}}}

\def\rmsBackground{\mathrm{RMS}_\mathrm{background}} \def\BackgroundScaling{\alpha_{\mathrm{bck}}} \def\BackgroundZero{z_{\mathrm{bck}}} 

\def\Nsmear{N_{\mathrm{smear}}}

\def\BrjOneAver{\tilde{B}_{\nu,\mathrm{rj},1}}             \def\Tdisk{T_{\mathrm{d}}}           \def\TbdiskMono{T_{\mathrm{d},\mathrm{c}}}           \def\TbdiskBA{T_{\mathrm{d}}^{\mathrm{(ba)}}}           \def\Tring{T_{\mathrm{r}}}                                                       \def\Omegauc{\Omega_{\mathrm{uc}}}            \def\Omegacr{\Omega_{\mathrm{c},r}}           \def\Omegaunr{\Omega_{\mathrm{uh},r}}         \def\wDisk{w_{\mathrm{D}}}           \def\wRing{w_{\mathrm{R}}}           \def\wDiskt{w_{\mathrm{D},t}}           \def\wRingt{w_{\mathrm{R},t}}

\begin{document}

\title{Revised planet brightness temperatures using the Planck/LFI 2018 data release\footnote{\HISTORY}}

\abstract
{}
{
We present new estimates of the brightness temperatures of 
Jupiter, Saturn, Uranus, and Neptune based on the measurements carried in 2009--2013 by \Planck/LFI at 30, 44, and 70\,GHz and released to the public in 2018.
This work extends the results presented
in the 2013 and 2015 \Planck/LFI Calibration Papers, based on the data acquired in 2009--2011. 
}
{
\Planck{} observed each planet up to eight times during the nominal mission. 
We processed time-ordered data from the 22 LFI radiometers to derive planet antenna temperatures for each planet and transit. We accounted for the beam shape, radiometer bandpasses, and several systematic effects. We compared our results with the results from the ninth year of WMAP, \Planck/HFI observations, and existing data and models for planetary microwave emissivity.
}
{
For Jupiter, we obtain 
$\Tb = 144.9$, 159.8, 170.5 $K$ ($\pm0.2\;$\,K at $1\sigma$, with temperatures expressed using the Rayleigh-Jeans scale) at 30, 44 and 70 GHz, 
respectively, or equivalently a band averaged Planck temperature $\TbBA=144.7$, 160.3, 171.2\,K
in good agreement with WMAP and existing models. 
A slight excess at 30\,GHz with respect to models is interpreted as an effect of synchrotron emission. 
Our measures for Saturn agree with the results from WMAP for rings 
$\Tb = 9.2 \pm 1.4, 12.6 \pm 2.3, 16.2 \pm 0.8$\,K, 
 while for the disc we obtain 
$\Tb = 140.0 \pm 1.4, 147.2 \pm 1.2, 150.2 \pm 0.4$\,K,
or equivalently a $\TbBA=139.7$, 147.8, 151.0\,K.
Our measures for Uranus 
($\Tb = 152 \pm 6, 145 \pm 3, 132.0 \pm 2$\,K,
or $\TbBA=152$, 145, 133\,K) 
and Neptune 
($\Tb = 154 \pm 11, 148 \pm 9, 128 \pm 3$\,K,
or $\TbBA=154$, 149, 128\,K) 
agree closely with WMAP and previous 
data in literature.
}
{}


\titlerunning{Revised Planck/LFI 2018 planet brightness temperatures}
\authorrunning{Maris, Romelli, Tomasi, et al.} 

\date{
March 16, 2021 
\\
{\tiny
\\
DOI: 10.1051/0004-6361/202037788;\\
arxiv: 2012.04504;\\
Journal: A\&A, 2021, 647, A104\\
\\
\\
Versioning:\\
\begin{tabular}{lll}
V2 & March 16, 2021 & as published on the journal \\
V1 & December 9, 2021 & as accepted for the pubblication on A\&A Dec 3, 2020
\end{tabular}{l}
}
}

\maketitle

\section{Introduction}

The \Planck{} mission was led by the European Space Agency (ESA) and measured the intensity and polarization of the microwave radiation from the sky in a wide frequency range (30--850\,GHz). The primary scientific purpose of the mission was to fully characterize the spatial anisotropies of the flux of the cosmic microwave background (CMB) over the full sky sphere and to measure the polarization anisotropies of the CMB itself. Secondary science done with \Planck{} data has provided important results in several domains of astrophysics such as the characterization of Galactic cold clumps and detection of Sunyaev-Zeldovich sources. The \Planck{} spacecraft orbited around the $L_2$ Lagrangian point of the Sun-Earth system and measured the full sky sphere once every six months. The spacecraft hosted two instruments: the High Frequency Instrument (HFI) was an array of bolometers working in the 100--850\,GHz range, while the Low Frequency Instrument (LFI) was an array of High Electron Mobility Transistors (HEMT)-based polarimeters working in the 30--70\,GHz range. 
Because of the design of the $100\;\mathrm{mK}$ cooling system
used to cool down its bolometers, HFI was able to perform its measurements until January 2012. On the other hand, LFI was operated without significant interruptions for four years, completing eight surveys of the sky.

In this work, we present new estimates for the flux densities of 
Jupiter, Saturn, Uranus, and Neptune in the frequency range 30--70\,GHz, obtained using the LFI on board the \Planck{} spacecraft. This work follows \citet{planck.intermediate.52.planet.flux.densities}, which presented estimates for the same planets using HFI data at higher frequencies (100--850\,GHz). The Planck observations were carried out over the period from August 2009 to September 2013. Each planet was observed seven or eight times and each observation lasted a few days. We used the data included in the latest \Planck{} data release \citep{planck.2018.results.I}, which implements the most recent and accurate calibration and systematics removal algorithms, as described in the \Planck{} Explanatory Supplement\footnote{\url{https//wiki.cosmos.esa.int/planck-legacy-archive/index.php/Main_Page}}.

There are several reasons why planetary measurements for a mission like \Planck{} are important.
The first one is that planets like Jupiter and Saturn are bright sources when observed at the frequencies used by CMB experiments: the signal-to-noise ratio (S/N) for measurements of the flux of Jupiter using LFI can be greater than 300. Thus, the measurement of their flux can be used as a way to calibrate the instrument or to assess the quality and stability of the calibration. Moreover, it can be used to compare the calibration among different experiments.
The second is that planets are nearly point sources when observed with the beams used in a typical CMB experiment: the largest apparent radius of a planet is always less than one arcminute, 
thus smaller than the typical resolution of CMB surveys.
This fact, combined with the remarkable brightness of planets like Jupiter and Saturn, permits us to calibrate the response of the optical system.
The third is that we can put constraints on radiative transfer modelling of gaseous planets like Jupiter and Saturn, which are useful to better understand their structure.

We did not use planets to calibrate the LFI detectors in any of the \Planck{} data releases \citep{planck.2013.05.LFI.calibration,planck.2015.05.LFI.calibration}. The Doppler effect caused by the motion of the spacecraft with respect to the rest frame of the CMB produces a dipolar signature in the CMB itself that is better suited for the calibration of LFI and HFI. If compared with Jupiter and other point-like bright sources, the dipole is always visible and its spectrum is identical to the CMB anisotropies. As a consequence, the scanning strategy adopted by \Planck{} was not optimized to observe planets.
The observation of any planet occurred when \Planck{} beams were sufficiently close to the planet itself.
This happened roughly twice per year for each of the planets considered in this work, that is, Jupiter, Saturn, Uranus, and Neptune.
In this paper, we do not present results about Mars. Owing to its larger proper motion and time variability, the analysis of its observations requires a more complex approach, which we postpone to a future work.

In \citet{planck.2013.04.LFI.beams} and \citet{planck.2015.04.LFI.beams}, we used observations of Jupiter to characterize the beam response of each LFI detector. For the kind of beams used in experiments like Planck, beam responses are characterized by a nearly Gaussian peak centred along the beam axis, whose full width half maximum (FWHM) characterizes the angular resolution of the instrument. Far from the beam axis, the beam response is significantly smaller (roughly 0.1--0.4\,\%), but its characterization is still important because it can lead to non-negligible systematics \citep{planck.2013.03.lfi.systematics,planck.2015.03.lfi.systematics}. Therefore, \citet{planck.2013.04.LFI.beams} and \citet{planck.2015.04.LFI.beams} used numerical simulations to estimate the beam response over the $4\pi$ sphere and used the Jupiter measurement to validate the simulations within a few degrees from the beam axis in the regions called the ``main beam'' and ``intermediate beam'' (as explained in Sect.~\ref{sec:selection:of:samples}).

The structure of this paper is the following: In Sect.~\ref{sec:definitions} we present a general review of the terms and conventions used in the field, the geometry of observations, and a description of the way LFI radiometers measure the signal from the sky. In section~\ref{sec:dataAnalysis}, we explain how we derived estimates of planet antenna temperatures from the timelines acquired by the LFI radiometers. In particular, section~\ref{sec:brightnessTemperatures} contains a description of the method we used to convert antenna temperatures into brightness temperatures, which are physically more significant.  Section~\ref{sec:results} uses the estimates derived in Sect.~\ref{sec:brightnessTemperatures} to compare our estimated spectral energy distributions (SEDs) with those produced by the 
Wilkinson Microwave Anisotropy Probe (WMAP) team.
Finally, Sect.~\ref{sec:conclusions} sums up the results of this work.
Appendix~\ref{sec:technicalInformation} contains detailed information about our data analysis pipeline.

\section{Methodology and models used in the analysis}
\label{sec:definitions}

In this section, we define the frame of reference and conventions that we use in the following sections to describe the observing conditions and the planet signal. When possible, we adhere to the conventions used in \citet{planck.intermediate.52.planet.flux.densities}.
Our approach to the analysis of planetary signals is the following: We model how the  SED of a planet produces a signal that is measured as an antenna temperature, and from this result we provide a chi-squared formula to derive the best estimate of the SED using the observations. When we have an estimate of the SED, it is then possible to derive an estimate of the brightness of the planet.

\subsection{\Planck/LFI focal plane, scanning strategy, and observing conditions}
\label{sec:ScStrategyAndObservingConditions}

The timing and geometry of planets transits depend on the focal plane geometry, scanning strategy, and orbit of \Planck, these are fully described in 
\citet{planck.2013.01.results} and \citet{planck.2013.04.LFI.beams}.
We recall that during nominal operations, \Planck\ scanned the sky spinning at a nearly constant rate of about one~rotation per minute around its spin axis $\SpinAxis$. The vector $\SpinAxis$ was kept stable for some time, equivalent to 30--60 rotations, and then de-pointed by a small amount.  This provides a fundamental timescale for the analysis of the \Planck{} observations. This ``pointing period'' is composed of a short period with unstable spin axis and unreliable attitude reconstruction followed by a long stable period when attitude information can be derived reliably. 

The focal plane of \Planck/LFI contained 22 beams, which belonged to 11 horns. Each beam was sensitive to one of the two orthogonal linear polarizations of each horn and fed a dedicated radiometric chain. The two polarizations are denoted in many ways in papers by the \Planck\ Collaboration,  for example, S/M, 1/0, and X/Y. For instance, 27-1, 27X and 27S are the same polarized beam in horn 27\footnote{This can be summarized by the so-called six (S-1-X) rule.}.
Beams in the focal plane where aimed at fixed positions with respect to $\SpinAxis$ and the spacecraft structure, so that each beam scanned the sky in circles with radii defined by their boresight angle $\boresight$, which is the angle between the effective spin axis $\SpinAxis$ of the spacecraft and the 
pointing direction $\Pointing$ of the beam.
 
Horns on the focal plane where paired according to the scan direction. The pairs in order of increasing boresight angles are listed as
    LFI18/23, LFI19/22, LFI20/21 (70\,GHz); 
LFI25/26 (44\,GHz);
LFI24 (44\,GHz), and LFI27/28 (30\,GHz).
        We note that LFI24 (44\,GHz) was alone and was nearly aligned with the LFI27/28 pair. 
Paired horns saw a source in the sky nearly at the same time. However, owing to different boresight angles, the same source transited through different pairs at different times. 
The direction of the orbital motion of the \Planck{} spacecraft splits a scan circle into a 
``leading'' and a ``trailing'' side, the former being the side towards which \Planck{} was moving. Transits are classified accordingly.
For planets, in leading transits the angle between the planet and the spin axis increased in time, so the planet was observed at first by LFI18/23 and at last by  LFI27/28 plus LFI24.
The opposite occurred in the trailing case.
However, the geometry of the transits was such that a pair with a larger boresight angle 
observed the planet when it was nearer to the spacecraft than a pair with a smaller boresight angle,
irrespective of the fact that the transit was leading or trailing.  Therefore, LFI27/28 and LFI24 always saw a planet with a smaller solid angle than LFI18/23.

\begin{figure*}
        \centering
        \begin{tabular}{c@{\hspace{2cm}}c}
        \includegraphics[width=0.33\textwidth]{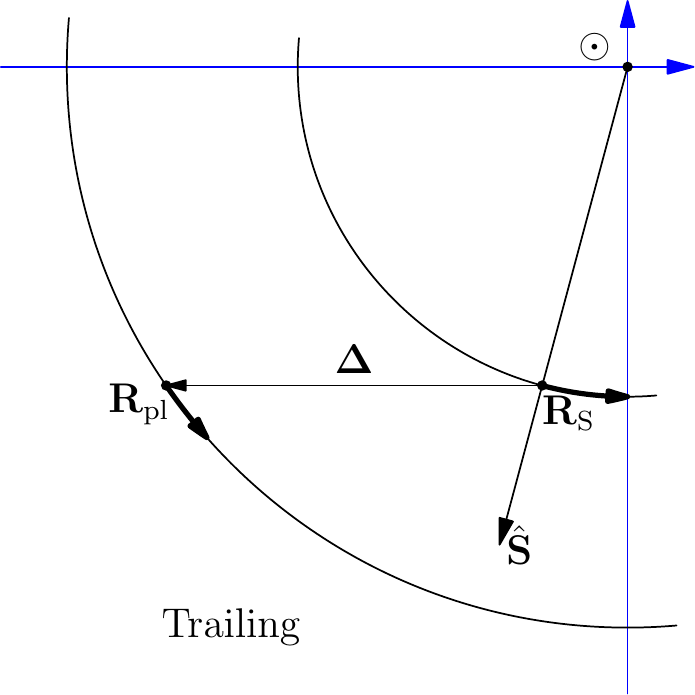} 
        &
        \includegraphics[width=0.33\textwidth]{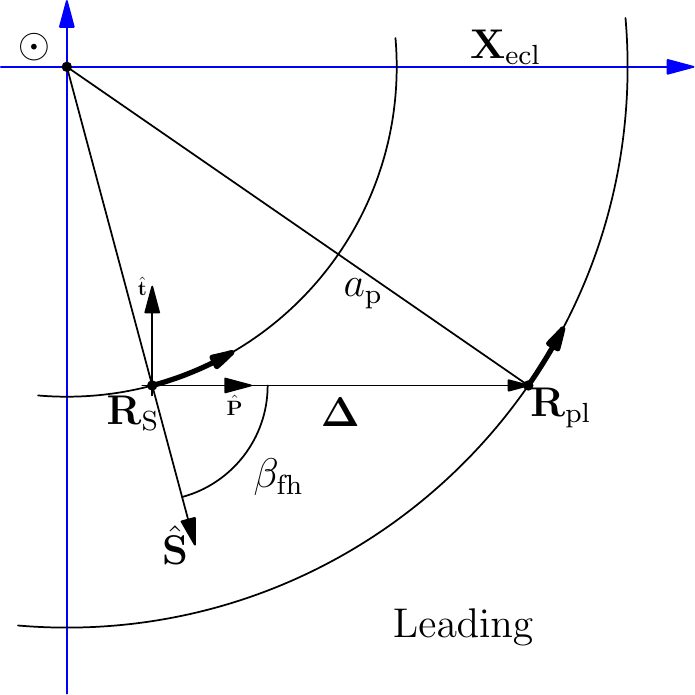} 
        \end{tabular}
\caption{\label{fig:observing:geometry} Geometrical configuration of a planet observation by \Planck{}. Left and right frames refer to trailing and leading observations respectively. The position of the spacecraft is denoted by $\PlanckLocation$, the position of the planet by $\PlanetLocation$. The Sun is indicated with the symbol $\odot$.
        Both the spacecraft and the planet revolve counter-clockwise around the Sun in circular and coplanar orbits. 
        For a detailed discussion of the symbols, see the text.
}
\end{figure*}

The apparent motion of a planet in the reference frame of a beam was complex. The \Planck{} team implemented a number of predictors and used these at different stages of mission planning \citep{Maris:Burigana:2009}. The principle behind these predictors can be derived from 
Fig.~\ref{fig:observing:geometry}, which
shows the most important parameters that describe a transit within a beam:
(1) the beam boresight angle $\boresight$,
(2) the location of the spacecraft at the epoch of observation within the Solar System
$\PlanckLocation$, and (3) the corresponding planet location $\PlanetLocation$.
The figure defines the spacecraft-planet vector
\begin{equation}\label{eq:planet:vector}
\PlanetVector = \PlanetLocation-\PlanckLocation,
\end{equation}
and the instantaneous  planet boresight angle $\PlanetBoresight$ 
\begin{equation}\label{eq:planet:boresight}
\cos \PlanetBoresight = \SpinAxis \cdot \PlanetLocation.
\end{equation}
Using these quantities, the condition for a transit is written as 
\begin{equation}\label{eq:planet:boresight:conditions}
 |\PlanetBoresight - \boresight| \le \fwhm.
\end{equation}

\begin{figure*}
        \centering
        \includegraphics[width=\textwidth]{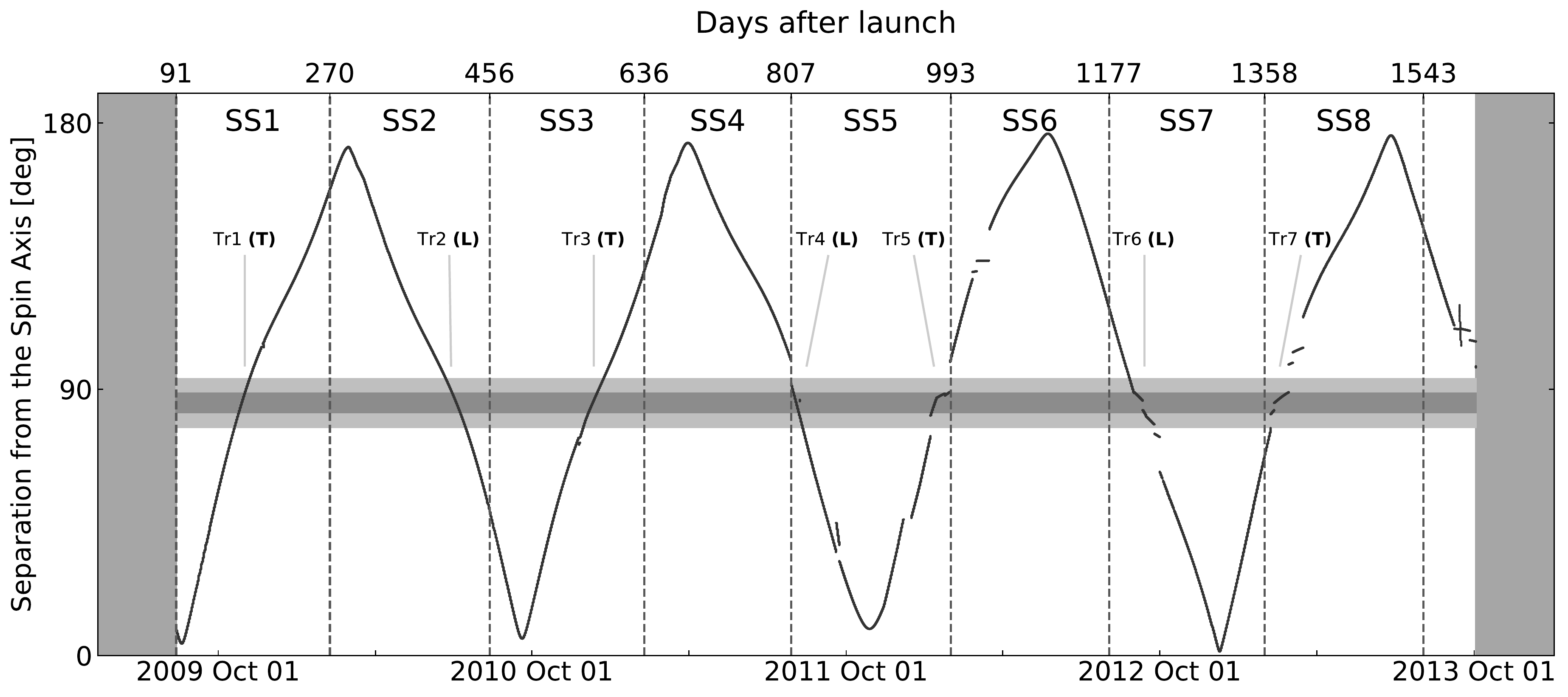}
        \caption{\label{fig:epoch:observation} Time dependence of the angle between Jupiter's direction and the spin axis of the \Planck{} spacecraft. The darker horizontal bar indicates the angular region of the 11 LFI beam axes, while the lighter bar is enlarged by $\pm 5^\circ$. Saturn, Uranus, and Neptune show a similar pattern. 
The labels SS1\ldots{}SS8 denote \Planck{} sky surveys, as defined in 
        \protect\citet{planck.2015.05.LFI.calibration}, from which the figure is taken.
        Tr1\ldots{}Tr7 denote the transits; letters T and L indicate whether it was a trailing or leading transit, according to Fig.~\ref{fig:observing:geometry}.
}
\end{figure*}

Figure~\ref{fig:epoch:observation} is adapted from \citet{planck.2015.05.LFI.calibration} and depicts $\PlanetBoresight$ (continuous line) as a function of observational epoch. Jumps and interruptions in the line denote changes in the scanning strategy.
The grey band in the figure represents the range of $\boresight$ angles for the
whole set of the \Planck/LFI feed-horns.
It is important to note that LFI27/28 (30\,GHz) and LFI24 (44\,GHz) have the smallest $\boresight$, LFI25/26 (44\,GHz) have the largest $\boresight$, and LFI18--23 (70\,GHz) have $\boresight$ within these extremes; \Planck/HFI beams fall in the latter category too.
Sometimes transits are indicated either with $(\mathrm{L})$ or $(\mathrm{T})$, 
whether the planet encounters the scan
circle in its leading or trailing sides, defined with respect to the direction of the \Planck\ orbital motion.
In a $(\mathrm{L})$ transit, 
the planet enters the scan circle from outside, that is, 
 $\PlanetBoresightDot<0$, while in a $(\mathrm{T})$ transit 
 the planet exits the scan circle from inside. 
The labels SS1\ldots{}SS8 are used to indicate the eight \Planck{} sky surveys. In general, planet transits are labelled sequentially as Tr1\ldots{}Tr8, but there is no one-to-one correspondence between transits and surveys. For example, no Jupiter transits occurred in SS4, but two transits occurred in SS5 (Tr4 and Tr5). In Fig.~\ref{fig:epoch:observation}, as in the rest of the paper, we follow the convention of marking epochs in \Planck{} Julian days (PJD), which is the number of Julian days after the launch; therefore,
\begin{equation}\label{eq:pjd}
 \mathrm{PJD} =  \mathrm{JD} - 2454964.5 \;.
\end{equation}

In Sect.~\ref{sec:results}, we tabulate the geometrical quantities described in this section for each planet and transit: see 
Tables~\ref{tab:jupiter:observing:conditions} (Jupiter), 
\ref{tab:saturn:observing:conditions} (Saturn), 
\ref{tab:uranus:observing:conditions} (Uranus),
and
\ref{tab:neptune:observing:conditions} (Neptune). The meaning of the columns is the following:
1) ``Tr'' lists the transit;
2) ``Epoch'' is the calendar date of the middle of the transit;
3) ``PJD\_Start'' 
refers to the epoch when the planet enters in one of the main beams for the first time, and  ``PJD\_End'' refers to the last time the planet is seen, PJD is defined in Eq.~(\ref{eq:pjd});
4) ``Nsmp'' is the number of samples in the timeline that were acquired while the planet was within a main beam;
5) ``EcLon'' and ``EcLat'' are the ecliptic coordinates of the planet as seen from \Planck{};
6) ``GlxLat'' is the Galactic latitude of the planet as seen from \Planck{};
7) $|\PlanetLocation|$ is the Sun-planet distance;
8) $\Delta$ is the \Planck{}-planet distance;
9) $\AngDiam$ is the apparent angular diameter of the planet;
10) $\PlanetAspectAngle$ is the aspect angle of the planet as observed by \Planck{} ($0\deg$/$90\deg$ means that the planet is seen along the equator/poles), but this quantity also represents the sub-\Planck{} latitude observed from the planet at the epoch when the radiation observed by \Planck{} left the planet (see Sect.~\ref{sec:geometric:correction}).
All the time-dependent quantities are evaluated in the middle of the transit period, which corresponds approximately to the epoch in which the planet transits at the centre of the focal plane. These are computed using the 
{\tt Horizons} web service\footnote{\url{https://ssd.jpl.nasa.gov/?ephemerides}}.

\subsection{Modelling of planet signals}

The power collected by a horn pointing towards some direction
$\Pointing$ close to a planet is the sum of four components:
\begin{equation}
\label{eq:poweCollectedByHorn}
\Win = \WinP+\WinBck-\WinBlk + \DeltaWinZero,
\end{equation}
where $\WinP$ is the power delivered by the planet,
$\WinBck$ the power from the background minus
$\WinBlk$ the radiation coming from the background 
but blocked by the planet, and 
$\DeltaWinZero$ the noise from the instrument.

The signal from a generic source with spatial brightness distribution $\SourceStructure(\Pointing)$ (flux over solid angle) and SED $\SED(\nu)$ is written as
\begin{equation}\label{eq:win}
\Win = \int_0^\infty\mathrm{d}\nu\int_{4\pi}\mathrm{d}^3\PointingPrime\, \BandPass(\nu)\,\SED(\nu)\,
\Beam\Bigl(\UeclBeamInv(\Pointing,\BeamOrientation)\cdot\PointingPrime\Bigr)
\,\SourceStructure(\PointingPrime),
\end{equation}
where $\BandPass(\nu)$ is the instrumental bandpass;
$\Beam(\mathbf{\hat{x}})$ is the pattern of beam response at frequency $\nu$ for a pointing direction $\mathbf(\hat{x}))$ in the beam reference frame; and
$\UeclBeamInv(\Pointing,\BeamOrientation)$ is the matrix describing the transformation
from the ecliptic reference frame to the beam reference 
frame\footnote{In this section and in the following we denote with $\mathbf{U}_{x,y}$ the transformation $y \rightarrow x$, from reference frame $y$ to reference frame $x$.}, accounting for the beam pointing direction 
$\Pointing$ and orientation $\BeamOrientation$ at the time of observation\footnote{
Usually, convolution is denoted as 
$\int \Beam(\PointingPrime-\Pointing)
\SourceStructure(\PointingPrime)\,\mathrm{d}^3\PointingPrime$. However, this notation fails to underline that
the beam is convolved over the $4\pi$ sphere and does not explicitly include $\BeamOrientation$. 
}.
We assume that $\BandPass(\nu)\le1$, with total bandwidth
$\BandWidth=\int\BandPass(\nu)\,\mathrm{d}\nu$ and central frequency $\Fcent=\int\BandPass(\nu)\nu\,\mathrm{d}\nu / \BandWidth$.
In the following, the dependence on $\Pointing$ and $\BeamOrientation$ is omitted.
If $\versorezbrf$ is the versor of the Z-axis of the beam reference frame, aligned with the beam optical axis
$\Pointing=\UeclBeam\versorezbrf$, then
  $\Beam(\versorezbrf)$ is the peak value of the beam. The quantity 
\begin{equation}
\OmegaBeamNu = \frac{\int \mathrm{d}^3\Pointing\,\Beam(\Pointing)}{\Beam(\versorezbrf)}
\end{equation}
is the beam solid angle at frequency $\nu$. 
If beam normalization is assumed to have 
$\int \mathrm{d}^3\Pointing\,\Beam(\Pointing)=1$
then $\Beam(\versorezbrf)=1/\OmegaBeamNu$.
In this paper, we follow the usual convention to map the main beam over a Cartesian $(u, v)$ system
drawn on a plane normal to $\versorezbrf$ in the beam reference frame, so that pointing $\Pointing$ corresponds to the following $(u, v)$ coordinates:
\begin{equation}\label{eq:pointing2uv}
        \left\{
                \begin{array}{ccl}
                        u &= \versorex \cdot \UeclBeamInv(\Pointing,\BeamOrientation) \Pointing; \\
                        v &= \versorey \cdot \UeclBeamInv(\Pointing,\BeamOrientation) \Pointing.
                \end{array}
        \right.
\end{equation}

We indicate band-integrated quantities using the apex $\mathbf{\cdot}^{\mathrm{(ba)}}$, such as $\OmegaBeamba$, $\SEDba$, and so on. Therefore, for a generic source it holds that
\begin{align}
\SEDba &= \frac{1}{\BandWidth}\int \tau(\nu) \SED(\nu)\,\mathrm{d}\nu,
 \\
\BeamBA(\Pointing) &= \frac{1}{\BandWidth\SEDba}\int \tau(\nu) \SED(\nu) \Beam(\Pointing) \,\mathrm{d}\nu,
 \\
\OmegaBeamba &=\frac{1}{\BeamBA(\versorezbrf)}.
\end{align}

\subsection{Estimation of planet signals}
\label{sec:estimationPlanetSignals}

We now tackle the problem of connecting the quantities in Eq.~\eqref{eq:poweCollectedByHorn} to the SEDs of the planets, background, and blocking radiation. For this purpose, we now detail the model behind each of the  terms in that equation. Using the conventions presented in the previous paragraphs, the integrated power for planet, background and blocking terms are written as
\begin{align}
\WinP &= \frac{\OmegaPlanet}{\OmegaBeambap} \BnuBA(\TbBA) \BandWidth \, \gammabapt,
 \\
\WinBck &= \SEDbabck \BandWidth, 
 \\
\WinBlk &= \frac{\OmegaPlanet}{\OmegaBeambacmb} \BnuBA(\Tcmb) \BandWidth \, \gammabacmbt,
\end{align}
where $\BnuBA(\nu)$ is the band averaged black-body brightness; it is 
assumed that the planet is an extended source with solid angle $\OmegaPlanet = \int \mathrm{d}^3\PointingPrime \SourceStructure(\PointingPrime) \ll \OmegaBeam$ and that most of the blocked radiation is the CMB with SED
$\Bnu(\Tcmb,\nu)$, so that 
\begin{equation}
\BnuBA(\Tcmb) = \frac{1}{\BandWidth} \int_0^\infty \mathrm{d}\nu\,\BandPass(\nu)\,\Bnu(\Tcmb,\nu).
\end{equation}
In the equations above, we used the following definition:
\begin{equation}
\gammabat =  \BeamBA(\UeclBeamInvt\cdot\PlanetDirectiont),
\end{equation}
which denotes the band-averaged beam response for a planet located within the main beam at epoch $t$. This stems from the fact that $\UeclBeamInvt \PlanetDirectiont$ is the position of the planet with respect to the beam reference frame, where
$\PlanetDirectiont$ is the direction in which the planet is seen at time $t$ in the 
ecliptical reference frame centred on the spacecraft.
The difference between $\gammabapt$ and $\gammabacmbt$ is in the SED used to compute the band-averaged integral. Usually, $\gammabat$ and $\OmegaBeamba$ are averaged accounting for the background SED, but in the following sections we do not account for this detail. 

\subsection{Converting signals to antenna temperatures}
\label{sec:signalsToTant}

We now provide the equations we used to connect SEDs to antenna temperatures, which are the quantities that are actually measured by the instrument.
Calibration of radiometers maps the measured input power $\Win$ onto a scale of antenna temperature variations based on the cosmological dipole, whose antenna temperature $\DeltaTdip$ depends on the pointing direction $\Pointing$ 
\citep{planck.2013.05.LFI.calibration,planck.2015.05.LFI.calibration}.
If we assume that the gain is linear, applying Eq.~(\ref{eq:win}) to the cosmological dipole
\begin{equation}
\DeltaWdip(\Pointing) = 
\dBdTcmbba
\DeltaTdip(\Pointing)
\BandWidth,
\end{equation}
where $\DeltaTdip$ is the temperature fluctuation of the cosmological dipole, convolved 
with the appropriate band-averaged beam pattern 
\begin{align}
\dBdTcmbba &= \frac{1}{\BandWidth}\int \mathrm{d}\nu \, \tau(\nu) \dBdTcmb(\nu) ,
 \\
 \BeamBAdip(\Pointing) &= \frac{1}{\dBdTcmbba\BandWidth}\int \mathrm{d}\nu\, \tau(\nu) \dBdTcmb(\nu) \Beam(\Pointing)
\\
 \OmegaBeambadip &= \frac{1}{\BeamBAdip(\versorezbrf)}.
\end{align}
Therefore, the planet signal is mapped onto an equivalent variation of thermodynamic temperature through $\DeltaTantP/\DeltaTdip=\WinP/\DeltaWdip$.
Assuming that the planet is aligned with the centre of the beam, the variation of antenna temperature caused by the presence of the planet is given by
\begin{equation}\label{eq:DeltaTantPstar}
\DeltaTantPstar = 
        \frac{\OmegaPlanet 
        \BnuBA(\TbBA)
        }{\OmegaBeambap\dBdTcmbba }.
\end{equation}
During a transit, the planet motion within the beam causes a time modulation of the antenna temperature 
$\Delta\Tantt\propto\gammabat \DeltaTantP$.
Therefore, the planet antenna temperature $\DeltaTantPstar$ for each transit and radiometer can be estimated through the minimization of the quantity
\begin{equation}\label{eq:chisq:naive}
\chi^2 = \sum_t \frac{1}{\sigma_t^2}
 \left( 
   \DeltaTantPstar \, \gammabapt  
+ b^{\mathrm{m}}_t - \DeltaTantt
 \right)^2,
\end{equation}
where $\sigma_t$is  the confusion noise for the sample at time $t$, 
$b^{\mathrm{m}}_t$ the background model discussed in Sect.~\ref{sec:background:modelling},
and $\gammabat$ the beam model described in Sect.~\ref{sect:bandpass:and:beam:pattern} and Sect.~\ref{sec:smearing}.
A rigorous treatment would also include a term to account for the blocked radiation 
\begin{equation}\label{eq:DeltaTantBlk}
\DeltaTantBlk= 
        \frac{\OmegaPlanet 
        \BnuBA(\Tcmb)
        }{\OmegaBeambacmb\dBdTcmbba }
        ,
\end{equation}
by the addition of a term $-\DeltaTantBlk \, \gammabacmbt$ in Eq.~\eqref{eq:chisq:naive}, as shown in Sect.~\ref{sec:blocking}. 
This would lead to an estimate for $\DeltaTantPstar$ that is already corrected for the blocking factor. However, since blocking is a minor effect, it is customary to correct it later. We chose to follow this approach, and therefore in this work $\DeltaTantPstar$ does not include correction for blocking. This convention introduces a small systematic effect, since $\gammabapt \ne \gammabacmbt$.   Table~\ref{tab:photo:parameters} summarizes all the radiometer-dependent quantities that are relevant for photometric analysis, which we presented in this section, together with other parameters that are discussed later.

\begin{table*}\centering
\caption{\label{tab:photo:parameters}
Photometric parameters for \Planck/LFI radiometers and band averaged beams. }
{\footnotesize
\begin{tabular}{lrrrrrrrrrr}
\hline\hline
Radiometer$^a$&$\Fcent^b$&$\Delta\nu$&$\BcmbBA$&$\dBdTcmbba$&$\BrjOne$&$\BrjBAOne$&${\OmegaBeamba}^c$&$\effAper$&$\effEtaBeam$&${\SyncOneBA}^d$\\
&[GHz]&[GHz]&[MJy/sr]&[MJy/sr/K]&[MJy/sr/K]&[MJy/sr/K]&[$\times 10^5$ sr]&[$\times 10^3$]&[$\times 10^3$]&\\
\midrule
     70-18M &   71.738 &       7.945 &    214.15 &       139.05 &   158.114 &     159.220 &          1.673 &      8.236 &         3.386 &        0.693 \\
     70-18S &   70.096 &       9.775 &    208.01 &       133.64 &   150.959 &     152.245 &          1.703 &      5.624 &         2.779 &        0.699 \\
     70-19M &   67.513 &       8.865 &    198.49 &       124.95 &   140.041 &     140.790 &          1.625 &      8.023 &         3.035 &        0.709 \\
     70-19S &   69.695 &       7.316 &    206.69 &       132.15 &   149.237 &     150.048 &          1.610 &      9.004 &         4.013 &        0.700 \\
     70-20M &   69.174 &       8.194 &    204.73 &       130.43 &   147.013 &     147.837 &          1.549 &      9.527 &         3.209 &        0.703 \\
     70-20S &   69.585 &       8.611 &    206.25 &       131.82 &   148.767 &     149.668 &          1.553 &      9.090 &         3.559 &        0.701 \\
     70-21M &   70.412 &       8.879 &    209.29 &       134.60 &   152.325 &     153.337 &          1.537 &      9.538 &         3.163 &        0.698 \\
     70-21S &   69.696 &      11.674 &    206.63 &       132.20 &   149.244 &     150.201 &          1.559 &      8.317 &         3.221 &        0.701 \\
     70-22M &   71.483 &       9.500 &    213.30 &       138.14 &   156.994 &     157.908 &          1.586 &      6.779 &         2.423 &        0.693 \\
     70-22S &   72.788 &       8.732 &    218.07 &       142.56 &   162.777 &     163.794 &          1.605 &      6.417 &         2.831 &        0.689 \\
     70-23M &   70.764 &       6.717 &    210.74 &       135.63 &   153.852 &     154.481 &          1.679 &      6.271 &         2.623 &        0.696 \\
     70-23S &   71.322 &       6.874 &    212.77 &       137.55 &   156.288 &     157.049 &          1.693 &      5.554 &         2.786 &        0.694 \\
     44-24M &   44.451 &       3.098 &    109.13 &        57.84 &    60.708 &      60.907 &          5.080 &      2.108 &         0.841 &        0.838 \\
     44-24S &   44.060 &       3.068 &    107.65 &        56.91 &    59.643 &      59.876 &          4.961 &      2.202 &         0.977 &        0.841 \\
     44-25M &   43.995 &       3.051 &    107.40 &        56.72 &    59.469 &      59.665 &          8.250 &      1.362 &         1.671 &        0.841 \\
     44-25S &   44.184 &       3.146 &    108.11 &        57.17 &    59.979 &      60.161 &          8.723 &      1.566 &         1.481 &        0.840 \\
     44-26M &   43.949 &       2.529 &    107.24 &        56.65 &    59.344 &      59.599 &          8.276 &      1.295 &         1.610 &        0.842 \\
     44-26S &   44.074 &       2.582 &    107.68 &        56.89 &    59.682 &      59.845 &          8.699 &      1.646 &         1.568 &        0.841 \\
     30-27M &   28.345 &       2.594 &     52.00 &        24.29 &    24.685 &      24.809 &         10.011 &      8.795 &         2.381 &        1.004 \\
     30-27S &   28.536 &       2.970 &     52.67 &        24.66 &    25.018 &      25.200 &         10.074 &      7.794 &         2.276 &        1.002 \\
     30-28M &   28.790 &       2.465 &     53.44 &        25.06 &    25.466 &      25.616 &         10.050 &      9.545 &         2.522 &        0.998 \\
     30-28S &   28.155 &       3.184 &     51.47 &        24.03 &    24.355 &      24.541 &         10.068 &      7.476 &         2.268 &        1.007 \\
\bottomrule
\end{tabular}
\tablenoteskip\\
\widetablenote{a}{Radiometers are identified by their frequency channel, either 30, 44 or 70~GHz; the feedhorn number, between 18--28; and the polarization arm, either S or M.}
\widetablenote{b}{Central frequency.}
\widetablenote{c}{The radiometric quantities $\OmegaBeamba$, $\effAper$, and $\effEtaBeam$ refer to a $u^2$ SED.}
\widetablenote{d}{Band average of the synchrotron spectral dependence $\nu^{-0.4}$ (see Eq.~\protect\ref{eq:thermal:plus:syncrotron}) for the 30 GHz and 40\,GHz channels.}
}
\end{table*}
  
\section{Data analysis}
\label{sec:dataAnalysis}

In this section, we describe the data analysis procedures used to implement the equations presented in Sect.~\ref{sec:definitions}. The results of our analysis are discussed in Sect.~\ref{sec:results}.
Since it is not possible to list the full set of measurements per planet, transit, and radiometer in this paper, we present only summary plots showing data at various data reduction steps. 
The technical details of our data analysis pipeline are explained in Appendix~\ref{sec:technicalInformation}.

\subsection{Characteristics of the input data}
\label{sec:input:data}

In our analysis, we used the Planck 2018 data release, whose timelines were calibrated using the procedure described in \citet{planck.2018.lfi.processing}. We do not detail the procedure used to produce these data, it is sufficient to recall that in the Planck/LFI 2018 data processing pipeline
i.) the timelines are cleaned of the dipole signal; ii.) the Galactic pick-up through beam sidelobes has been removed; iii.) ADC non-linearities are corrected, iv.) the pointing is corrected for a number of systematics\footnote{https://wiki.cosmos.esa.int/planckpla2015/index.php/Detector\_pointing}.
 Each sample in the LFI timelines consists of the following fields: i.) the UTC time of acquisition; ii.) the antenna temperature $\Tant$, calibrated in $\Kcmb$; iii.) the apparent pointing direction $\Pointingt$ (direction of the beam axis) in the J2000 reference frame; iv.) the beam orientation in the sky; v.) the quality flags; vi.) the absolute address of the sample within the global mission timeline.
The pointing directions and beam orientations can be used to compute the $\UeclBeamInvt$ matrix for the sample. 

To produce sky maps from timelines, the \Planck/LFI pipeline needs to reduce the level of noise in the timelines. Planck/LFI timelines suffer from the presence of correlated noise, whose spectral shape can be approximated by the function
\begin{equation}
\label{eq:oneoverf}
P(f) = \left[1 + \left(\frac{f_k}f\right)^\alpha\right] \frac{\sigma^2}{f_s},
\end{equation}
where $f$ is the frequency, $f_s$ is the sampling frequency of the detector, $\sigma$ is the level of white noise in the data, and $f_k$ is the so-called {knee frequency} of the $1/f$ noise; in the case of the Planck/LFI receivers, $f_k \approx 20\div 60\,\text{mHz}$ \citep{mennella2010}. The presence of $1/f$ noise invalidates many assumptions used in common data analysis tasks, and several works have dealt with the problem of removing it from time streams. One of the most simple yet effective solutions is the {destriping algorithm}, which is able to determine the time dependence of $1/f$ noise through an approximation of the noise time stream with a number of simple basis functions \citep{maino2002,Keihaenen2004}. Each basis function is constrained by the requirement that each pass on the same pixel should yield the same measurement if the noise part in Eq.~\eqref{eq:oneoverf} were negligible. In its simplest incarnation, a destriper uses constant-valued basis functions: in this case, each function is called a {baseline}, and its duration in time must be smaller than $1/f_k$ in order for the destriper to be effective.

Madam \citep{Keihaenen2004}, the map-maker implemented in the Planck/LFI pipeline, uses a destriping technique to produce frequency maps that are cleaned from correlated noise and a set of baselines that approximate the correlated noise in the timeline. However, we were not able to use this information to clean the timelines in our analysis. One of the fundamental assumptions of the destriping algorithm is that the signal measured on the sky must be constant in time. Therefore, the LFI pipeline masks all those samples acquired while a moving object was within the main beam, and these samples are not considered in the application of the destriping algorithm. We must add that the destriping technique is able to find a reliable solution if there are enough crossings of the same point in the sky among different scan circles. We attempted to use destriping on each planet transit within the main beam of each radiometer: as one transit lasts only a few hours, planets can be considered as fixed point sources. However, the quality of the solution was poor because the number of rings was not sufficient to fully constrain the solution. A comparison of the estimates for $\DeltaTantPstar$ obtained with and without the application of destriping show differences within the random errors due to white noise. For this reason, we decided not to use destriping in our pipeline.

\subsection{Overview of the analysis procedure}
\label{sec:synoptic}

\begin{figure*}
        \centering
        \includegraphics[width=\textwidth]{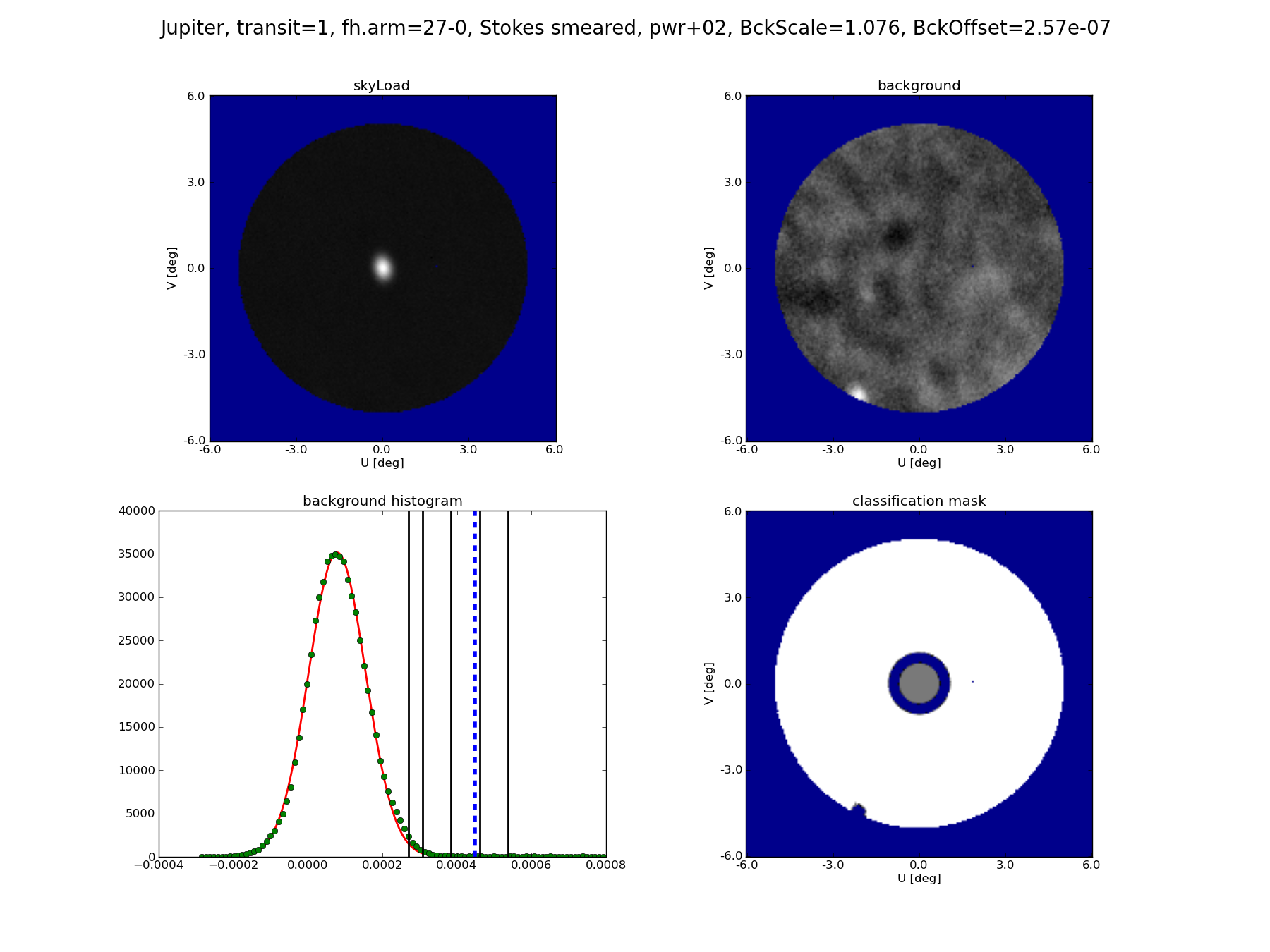} 
        \caption{ \label{fig:Tant:classification}
                Example of a map in the $(u,v)$ reference frame for Jupiter.
This image shows the first transit as seen by radiometer 27-0 (30\,GHz).  
\textbf{Top left}: Map of $\Tant$ in $\Kcmb$ ranging from $-4\times10^{-4} \, \Kcmb$ to $0.4 \Kcmb$. 
                \textbf{Top right}: Map of the background model, expressed as $\Tant$ in $\Kcmb$ ranging from $-4\times10^{-4} \, \Kcmb$ to $1\times10^{-3} \Kcmb$.
                \textbf{Bottom left}: Histogram of $\Tant$ in $\Kcmb$ for the background. The green points indicate the samples in the histogram, the red line indicates the best-fit Gaussian distribution, and the threshold for the classification mask is shown by the dashed blue line.
                \textbf{Bottom right}: Classification mask. The grey region shows the planet ROI, the white annulus is the background ROI, and the blue regions denote unused samples.
        }
\end{figure*}

To estimate the antenna temperature $\DeltaTantPstar$ for the sources considered in this work, we minimized the value of $\chi^2$ shown in Eq.~\eqref{eq:chisq:naive}. We only considered those samples that were acquired when the point source fell within a circular region of interest (ROI) centred on the main axis of the beam (details are provided in Sect.~\ref{sec:selection:of:samples}), whose radius is always $5\deg$, regardless of the radiometer, transit, or planet. An example of the ROI is shown in  Fig.~\ref{fig:Tant:classification}.
As in \citet{planck.2013.05.LFI.calibration} and \citet{planck.2015.05.LFI.calibration},
the background was estimated by splitting the ROI in two concentric circles: the ``planet ROI'' and the ``background ROI'' (see Fig.~\ref{fig:Tant:classification} and Sect.~\ref{sec:selection:of:samples}). However, unlike 
\citet{planck.2013.05.LFI.calibration} and \citet{planck.2015.05.LFI.calibration}, we did not consider the background as a constant but we allowed for 
 spatial variations of the background, as described in Sect.~\ref{sec:background:modelling}. 
This permits  us to remove weak background sources and to mask bright sources, as we show in Fig.~\ref{fig:Tant:classification}. 
We modelled the beam $\gammabat$ using a band-averaged map of the main beam, described in Sec.~\ref{sect:bandpass:and:beam:pattern}. We accounted for the apparent motion of the planet and the background within the beam during the acquisition of a sample using the so-called {smearing} algorithm, which is described in Sect.~\ref{sec:smearing}.

\begin{figure*}
        \centering
        \includegraphics[width=\textwidth]{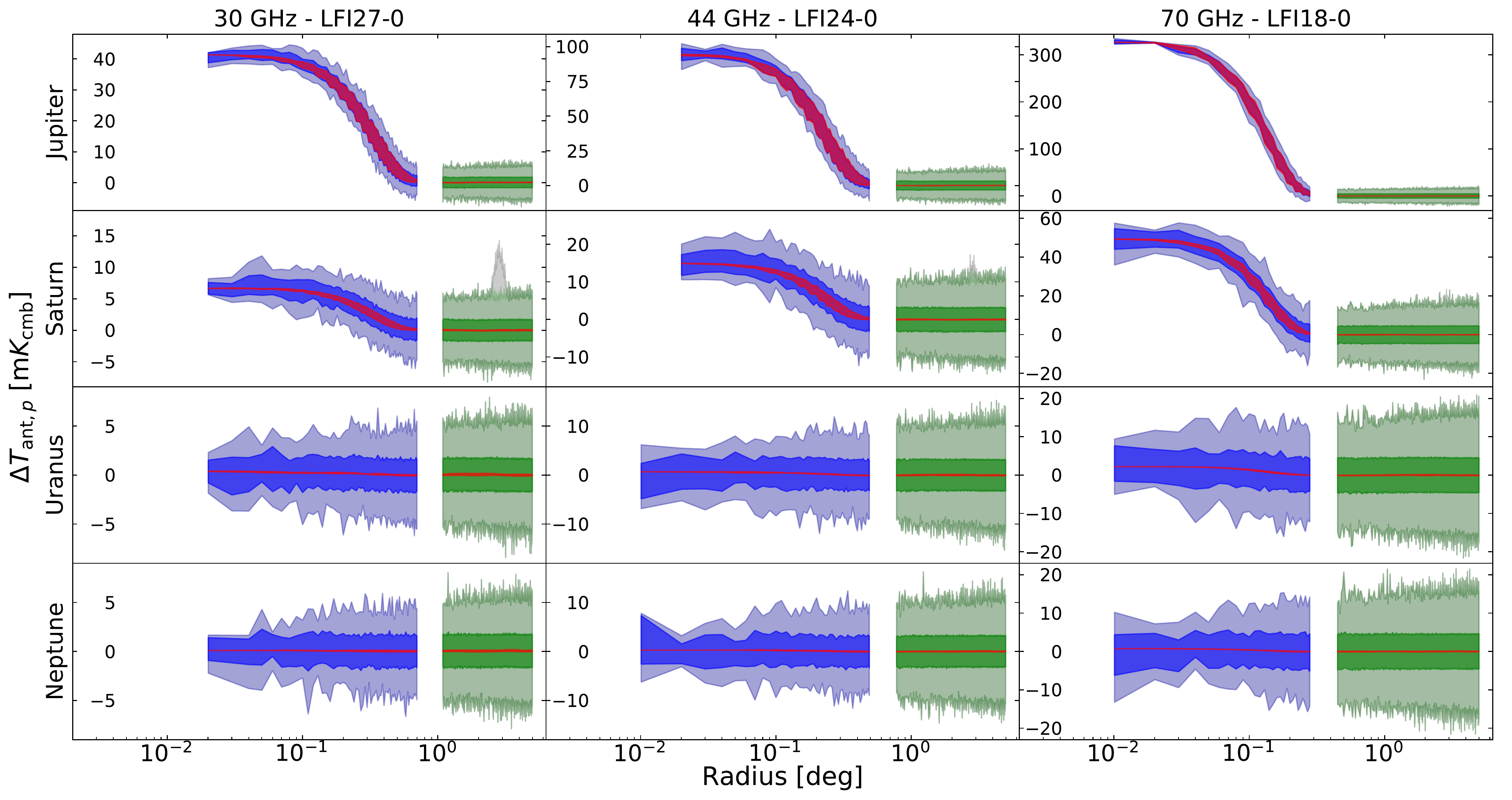}
        \caption{ \label{fig:Tant:regression}
                Antenna temperature estimates $\Tant$ for Jupiter, 
                Saturn, Uranus, and Neptune (top to bottom) and for three radiometers representative
                of the 30\,GHz (left), 44\,GHz (centre), and 70\,GHz (right) channels, as a function of the angular distance from the beam centre.
                The blue bands show the distribution of samples in the planet ROI (dark blue: $1\sigma$ region; light blue: peak-to-peak variation). The green bands have the same interpretation, but indicate the background ROI. The grey bands show the data before having been $\sigma$-clipped; for the case of Saturn observed by LFI27-0 a point source is present that was removed before the analysis (not present in the green line). The separation between the blue and green lines indicates the presence of the avoidance ROI, not included in our fits. The red line shows the best-fit model, and its width is the root mean square (RMS) of the model due to the ellipticity of the beam.
        }
\end{figure*}

Figure~\ref{fig:Tant:regression} shows the regression of $\DeltaTantPstar$ for Jupiter, 
Saturn, Uranus, and Neptune for the first transit and for the three radiometers LFI27-0, LFI24-0, and LFI18-0, which are representative of the 30\,GHz, 44\,GHz, and 70\,GHz frequency channels, respectively.
Samples are plotted as a function of the radial distance between the planet and the beam centre.
The blue and green points represent samples in the planet and background ROIs, while the grey points represent samples not used in the fit; the best-fit model is represented by red points. The dispersion of red points as a function of radial distance is mainly caused by the ellipticity of the beam. This did not occur for WMAP, as the WMAP team used a symmetrized beam \citep{WMAP:PLANETS:2011,WMAP:PLANETS:2013}.
We note that there is an apparent increase in dispersion for large radius. This is not due to an actual increase in the variance of the samples, but to the fact that at larger distances the population of samples increases in size, thus widening the spanning of the plotted points.
The LFI data for Jupiter and Saturn show a S/N that is high enough to be seen in raw data. The same does not hold for Uranus and Neptune.

\begin{figure*}
        \centering
        \includegraphics[width=\textwidth]{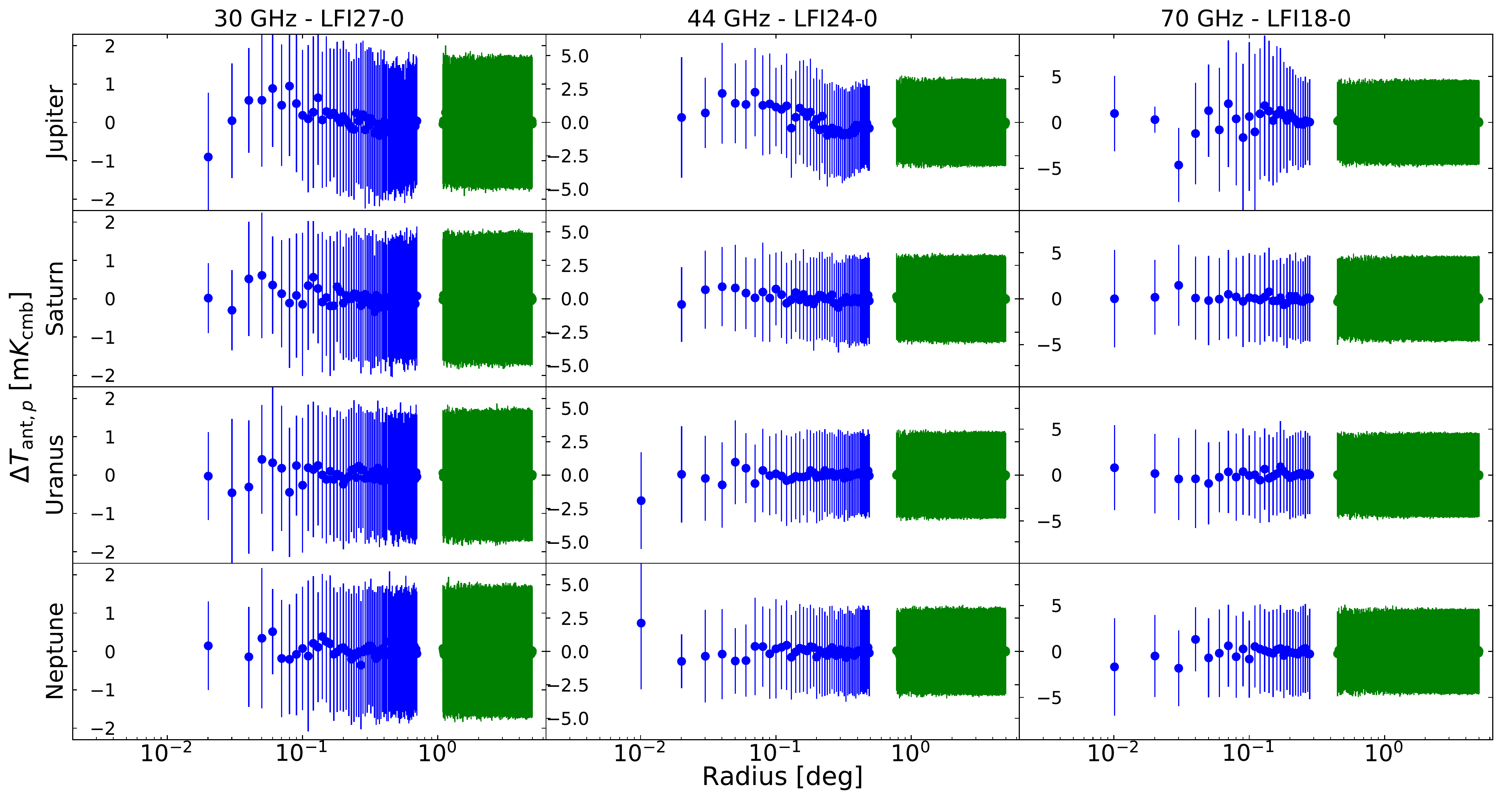}
        \caption{ \label{fig:synoptic_tant_residual_radius_averages_multitransit_plot}
                Radial pattern of residuals averaged over the whole set of transits.
}
\end{figure*}

Figure~\ref{fig:synoptic_tant_residual_radius_averages_multitransit_plot} shows 
the distribution of the residuals of the fit, radially averaged in constant-width bins; the bars denote the RMS of the residuals in each bin.
In most cases, the radial pattern of the residuals is nearly flat, apart from Jupiter 24 and 27, which show
a systematic error with a peak-to-peak amplitude $\lesssim 10^{-3}\,\Kcmb$ (to be compared with a temperature of $\approx 0.3\,\Kcmb$).
We chose to neglect this residual, as at this stage it is not easy to understand whether this effect is due to uncertainties in the beam model or bandpass or other perturbations. Moreover, the definition of a new beam model for \Planck/LFI is outside the purpose of this paper.

\begin{figure*}
        \includegraphics[width=\textwidth]{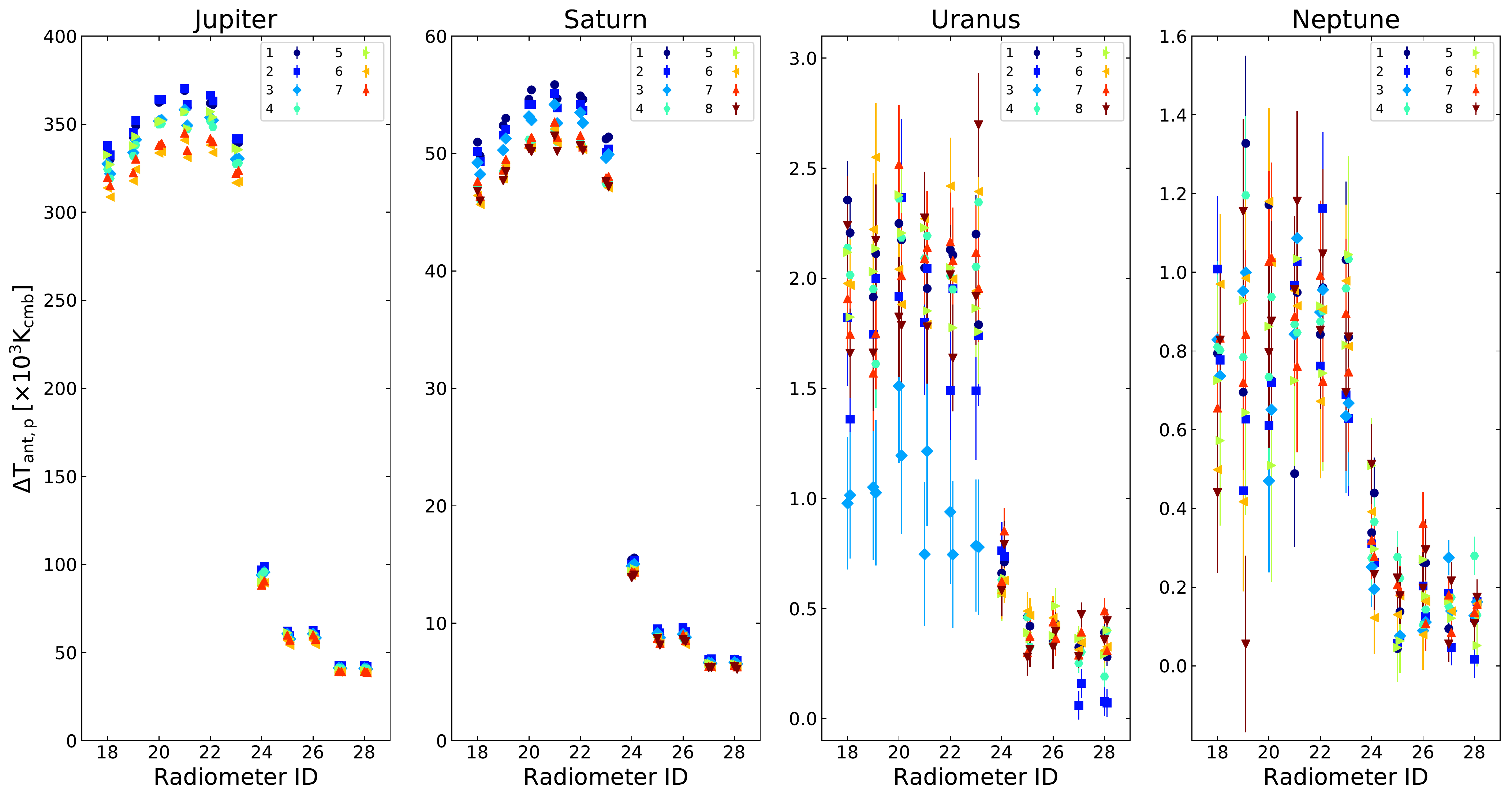}
        \caption{ \label{fig:synoptic_of_transits_tant_stokes_background}
                Values of $\DeltaTantPstar$ for transits of each planet and radiometer.
                The X-axis is the radiometer index in form $hh.p$ with $hh=18$, 19, 20, 21, 22, 23 for the 70\,GHz channel,
                24, 25, 26 for the 44\,GHz channel, and 27, 28 for the 30~GHZ channels.
                The $p$ index accounts for polarization with $0$ for M (Y) polarization and 1 for S (X) polarization.
The error bars account for noise. 
                For Jupiter and Saturn,
                 they 
                 are smaller than the size of the symbols.
}
\end{figure*}

\begin{table}
\centering
\footnotesize
\caption{\label{tab:tant:errors:distribution}
Error bars for $\DeltaTantP$.
}
\begin{tabular}{lr@{ -- }lr@{ -- }lr@{ -- }l}
\hline\hline
Planet& \multicolumn{2}{c}{30\,GHz$^a$}& \multicolumn{2}{c}{44\,GHz$^a$}& \multicolumn{2}{c}{70\,GHz$^a$}\\
\hline
Jupiter& 15& 59& 37& 120& 75& 280\\
Saturn& 26& 62& 52& 120& 150& 300\\
Uranus& 40& 68& 63& 130& 160& 360\\
Neptune& 42& 61& 62& 120& 170& 300\\
\hline
\end{tabular}
\tablenoteskip\\
\tablenote{a}{Values in $\mu\Kcmb$.}
\end{table}
 
Figure~\ref{fig:synoptic_of_transits_tant_stokes_background} provides a summary of our measures for $\DeltaTantPstar$ for the whole set of planets, transits, and radiometers.
For Jupiter and Saturn 
the dispersion of $\DeltaTantPstar$ is only partially affected by random noise,
which introduces a RMS scatter in $\DeltaTantPstar$ of at most a few $10^{-4}\,\Kcmb$.
When converted into a relative error per planet, transit, and radiometer, the order of magnitude for Jupiter is $10^{-3}$, for Saturn $10^{-2}$, 
about $5\times10^{-1}$ for Uranus, 
and up to $1$ for Neptune. The range of errors for our estimates are provided in Table~\ref{tab:tant:errors:distribution}.

Because of the small S/N, in some cases the signal for Uranus and Neptune is consistent with zero. This occurs when the confusion noise from the instrument and the background are larger than the signal induced by the planet. Whenever this happened, we removed the affected data from our analysis.

\subsection{Estimation of a fiducial antenna temperature}
\label{sec:FiducialTantReduction}

Figure~\ref{fig:synoptic_of_transits_tant_stokes_background} shows some variability among transits and radiometers for the same planet, 
with a clear pattern in the variation of $\DeltaTantPstar$ within the same frequency channel and transit.  
As an example, $\DeltaTantPstar$ for LFI20/21 is larger than for LFI19/22, which in turn is larger than for LFI18/23.
The first reason for these discrepancies is the difference in the value of $\OmegaBeam$ among various radiometers because this value is largest for the radiometers located far from the centre of the focal plane, and produces the bent pattern of the 70~GHz channel or the jump between horn 24 and horn 25 and 26. 
Secondly, we must consider changes in the circumstances of the observation among different radiometers and transits,
which leads to differences in the \Planck--planet distance 
$\left|\PlanetVector\right|$ (Eq.~\ref{eq:planet:vector}), and so in $\OmegaPlanet$, producing the relative shift of the measurements between one transit and the other.
We considered the change in $\left|\PlanetVector\right|$ among different transits and the change occurring while observing the same transit from different horns (refer to Sect.~\ref{sec:ScStrategyAndObservingConditions}). 
Since planets are not spherical and their polar axis are tilted on their orbital planes, 
varying observing conditions led to different apparent aspect ratios of the shape of the planets.
In addition we have to take care of systematics of the beam model as its numerical efficiency and the beam aperture. We can reduce the antenna temperature to standardized conditions, using the following formula:
\begin{equation}\label{eq:deltaTantPstarFiducial}
\DeltaTantPstarFiducial = 
\frac{\OmegaBeamba / \OmegaBeamFiducial}{\OmegaPlanet / \OmegaPlanetFiducial}
\frac{
        (1+\effAper)
(1+\effEtaBeamCorrection)
}{1+
\effAsp
}
\DeltaTantP^*,
\end{equation}
where tilted quantities indicate fiducial values. 
 %
 %
For each channel we take as a fiducial value $\OmegaBeamFiducial$, the median of the $\OmegaBeamba$ for that channel from Table~\ref{tab:photo:parameters}.
The actual values we used are $1.006\times 10^{-4}\,\mathrm{sterad}$ (30\,GHz), $8.263\times 10^{-5}\,\mathrm{sterad}$ (44\,GHz), and $1.607\times 10^{-5}\,\mathrm{sterad}$ (70\,GHz). 
Since the planet solid angle $\OmegaPlanet$ depends on the observer-to-planet distance, $\left|\PlanetVector\right|$ (Eq.~\ref{eq:planet:vector}), the reduction to a fiducial solid angle is equivalent to reduction to a fiducial distance.
In Table~\ref{tab:fiducial:geometry}, we list the values we used for planet radii, distances to the observer, and solid angles of the planets.
Several conventions and approximations are used in the literature to measure distances and solid angles. As an example, distances to Jupiter can range from $4.04$\,AU to 
$5.2$\,AU. To ease comparisons, we use the same scale of distances and solid angles as WMAP 
\citep{WMAP:PLANETS:2011,WMAP:PLANETS:2013}.
The quantity $\effAsp$ in Eq.~(\ref{eq:deltaTantPstarFiducial}) is the aspect correction factor described in Sect.~\ref{sec:geometric:correction}; this accounts for the fact that the aspect ratio of the planet seen by \Planck{} changes in time.
The parameter $\effAper$ is the aperture correction described in Sect.~\ref{sec:aperture:correction}, and it corrects for the loss of signal in the background ROI.
The quantity $\effEtaBeamCorrection$ is a correction factor for the lack of numerical efficiency of the beam.
As detailed in Sect.~\ref{sec:beam:efficiency},
the limited accuracy in the numerical computation of the beam 
induces a systematic in the measured fluxes at the level of $\sim 10^{-3}$.
The precise value of $\effEtaBeamCorrection$ cannot be determined precisely, 
but it is in the range $\pm\effEtaBeam$ given in 
Table~\ref{tab:photo:parameters}.
For this reason, we did not apply the correction, thus assuming $\effEtaBeamCorrection = 0$, and we included this in the overall uncertainty. 
We provide more details in Sect.~\ref{sec:beam:efficiency} and Sect.~\ref{appendix:averaged:brightness}.
In the \citet{planck.2013.05.LFI.calibration} and \citet{planck.2015.05.LFI.calibration},
a correction factor $\effSL$ was introduced to account for sidelobes.
In this work, this correction is no longer needed because the \GRASP{} beam model already includes the effect of side lobes; Sect.~\ref{sec:sidelobes} provides more details.

\begin{table}\centering
\footnotesize
\caption{\label{tab:fiducial:geometry}
Fiducial geometric parameters.
}
\begin{tabular}{lrrrr}
\hline\hline
Planet&$\PlanetRadiusEquatorial^a$&$\PlanetRadiusPolar^b$&$\PlanetDistanceFiducial^c$&$\OmegaPlanetFiducial^d$\\
 &[km]&[km]&[AU]&[$\mathrm{sterad}$]\\
\midrule
Jupiter& 71492& 66854& 5.2& $2.481\times10^{-8}$\\
Saturn& 60268& 54364& 9.5& $5.096\times10^{-9}$\\
Uranus& 25559& 24973& 19.0& $2.482\times10^{-10}$\\
Neptune& 24764& 24341& 29.0& $1.006\times10^{-10}$\\
\bottomrule
\end{tabular}
\tablenoteskip\\
\tablenote{a}{Equatorial radius of the planet.}
\tablenote{b}{Polar radius of the planet}
\tablenote{c}{Fiducial distance of the planet.}
\tablenote{d}{Solid angle subtended by the planet.}
\end{table}

\begin{figure*}
        \includegraphics[width=\textwidth]{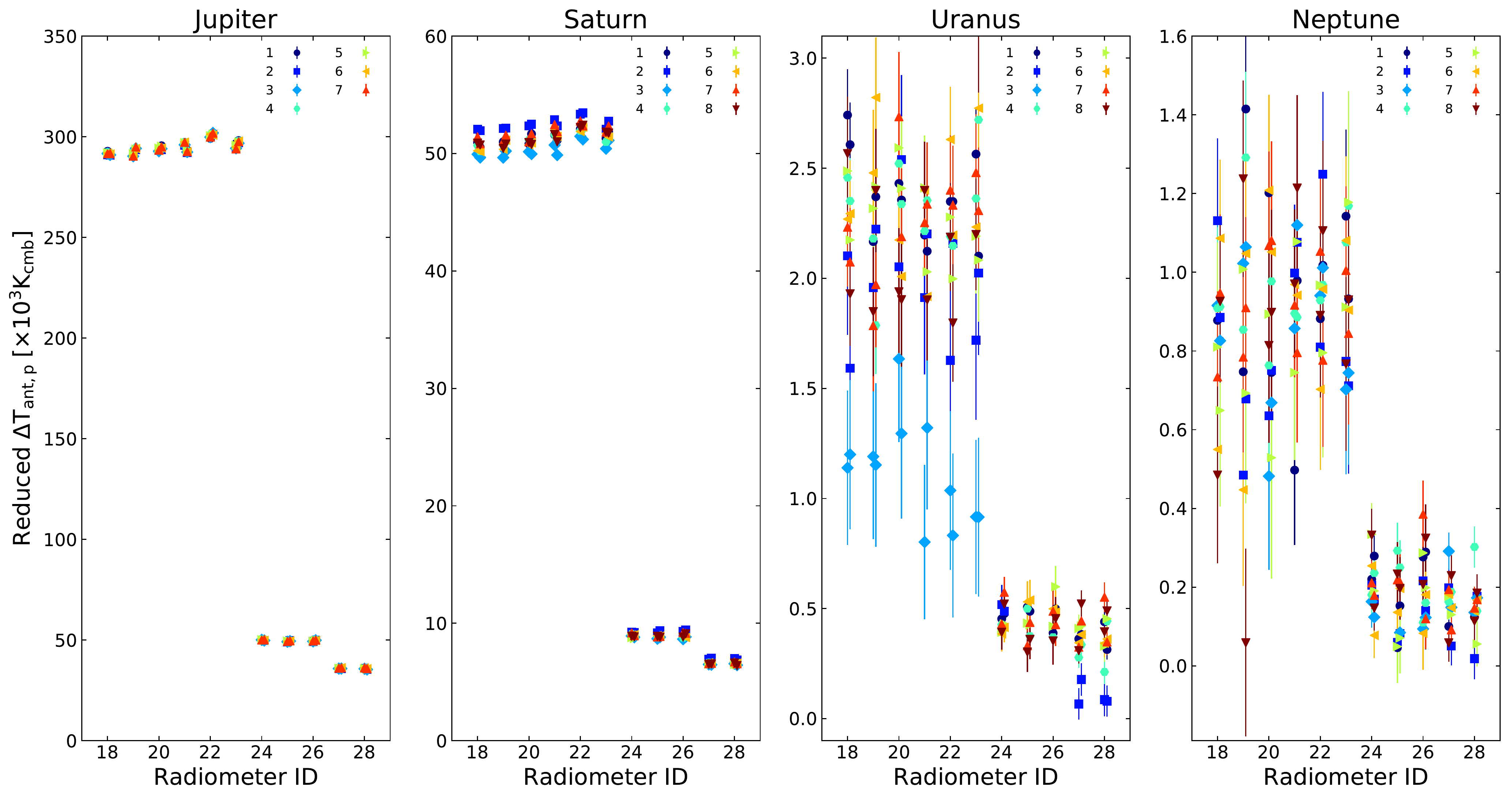}
        \caption{ \label{fig:synoptic_of_tant_after_reduction_to_same_omegabeam}
                Values of $\DeltaTantP$ after reduction to fiducial observing conditions and standardized $\OmegaBeam$ and $\OmegaPlanet$. 
        }
\end{figure*}

Figure~\ref{fig:synoptic_of_tant_after_reduction_to_same_omegabeam} shows the derived distribution of the values $\DeltaTantPstarFiducial$ (Eq.\ref{eq:deltaTantPstarFiducial}). The dispersion within the same frequency channel is significantly reduced for the 70\,GHz and nearly flattens, and all the 44\,GHz radiometers are now consistent.  Geometric corrections do not affect the dispersion in 30\,GHz channels significantly.

\subsection{Reduction of antenna temperatures to brightness temperatures}
\label{sec:brightnessTemperatures}

The result of our estimate is expected to be the brightness of the planet, expressed as a brightness temperature. The brightness for each radiometer and transit can be derived from $\DeltaTantPstarFiducial$ with the formula
\begin{equation}
\label{eq:TantToTbr}
\BrBAplanet = 
\frac{\OmegaBeamFiducial}{\OmegaPlanetFiducial}
\dBdTcmbba
\DeltaTantPstarFiducial
+ \BnuBA(\Tcmb),
\end{equation}
where $\BnuBA(\Tcmb)$ is the correction for the blocked radiation; see also Sect.~\ref{sec:blocking} and Table~\ref{tab:photo:parameters}. We note that the factor $\OmegaBeamFiducial / \OmegaPlanetFiducial$ removes the corresponding correction for standardized observing conditions.

We now turn to the problem of properly defining what we mean with ``brightness temperature'' $\Tb$, as several definitions are available in the literature.
One widely used convention is to define a Rayleigh-Jeans (RJ) brightness temperature as \begin{equation}\label{eq:tb:rj}
\TbRJ = \frac{\BrBAplanet}{\BrjOne},
\end{equation}
where $\BrjOne=2 \kboltzman \Fcent^2/c^2 $ is the RJ brightness at 1\,K estimated at frequency $\Fcent$ (see also Table~\ref{tab:photo:parameters}). This is the convention followed by WMAP \citep{WMAP:PLANETS:2011,WMAP:PLANETS:2013}.
On the other hand, when data are used to model planetary atmospheres, it is better 
to define $\Tb$ through the inversion of a Planckian  curve
\citep{dePater:Dunn:2003,Gibson:dePater:2005,dePater:etal:2016a,Karim:etal:2018,dePater:etal:2019} as follows:
\begin{equation}\label{eq:tb:mono}
\Bnu(\TbMono,\Fcent) = \BrBAplanet,
\end{equation}
where ``c'' denotes one of the frequency channels 30, 44, or 70\,GHz.
In some cases, the following band-averaged formula can be used to define $\TbBA$:
\begin{equation}\label{eq:tb:ba}
\BnuBA\left(\TbBA\right) = \BrBAplanet,
\end{equation}
where $\BnuBA(\Tb)$ is the band-averaged SED of a Planckian black body. Its inversion is described in Sect.~\ref{sec:band:averaged:bnu}.
Conversion among the different conventions is not difficult, but a detailed model of the instrument bandpass must be taken in account. 
To simplify the comparison between our results and those from WMAP, and to produce numbers useful for atmospheric modelling, we provide the three quantities $\TbRJ$, $\TbMono$, and $\TbBA$ when needed\footnote{In the abstract we followed the WMAP convention and we quoted $\TbRJ$ as $\Tb$.}.

\begin{figure*}
        \includegraphics[width=\textwidth]{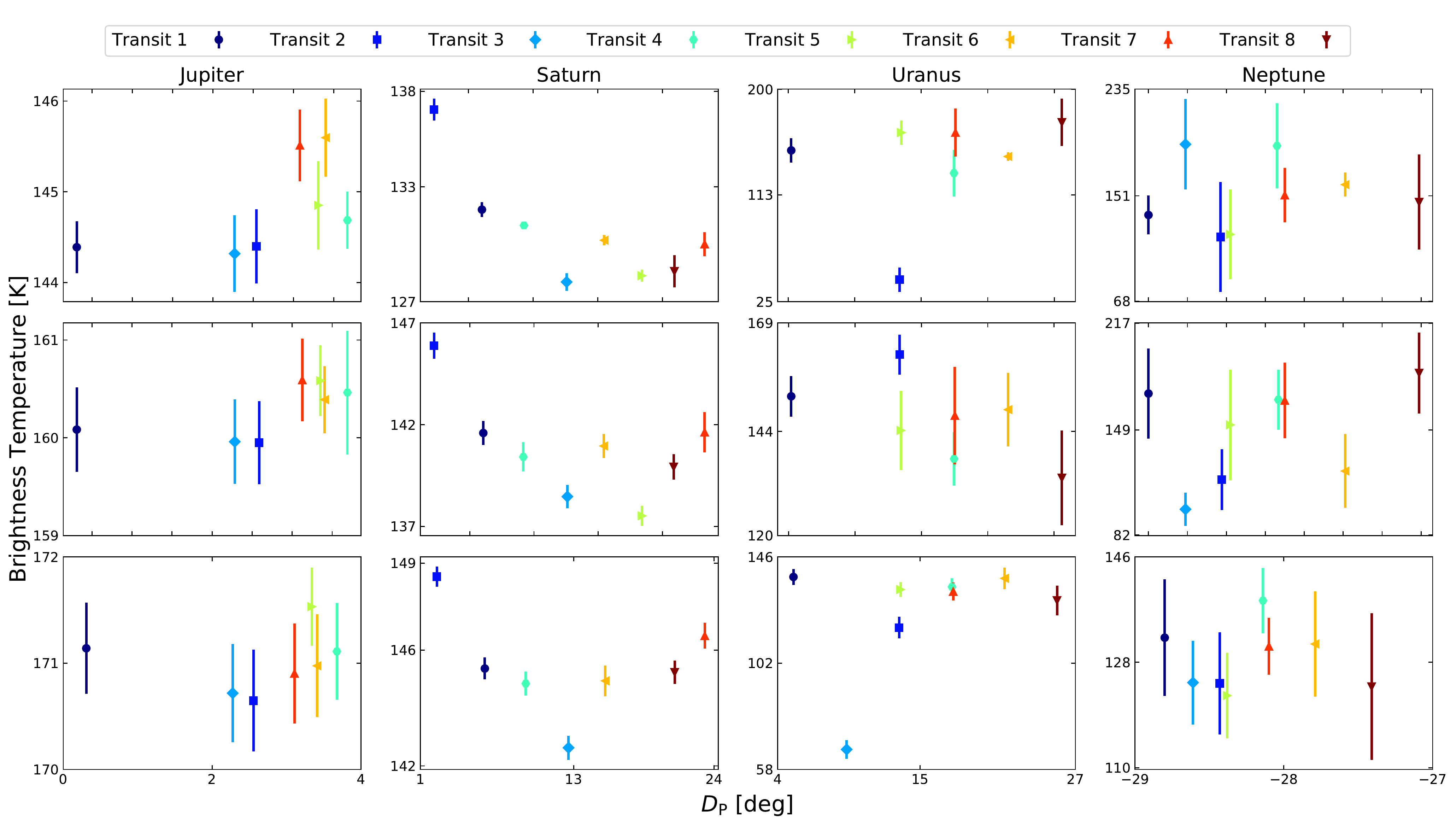}
        \caption{ \label{fig:synoptic_of_tb_ba_transit_channel_averages}
                Values of channel-averaged $\TbBA$ per transit as a function of $\PlanetAspectAngle$,
                for 30\,GHz (top), 44\,GHz (middle), and 70\,GHz (bottom) channels. The high variability in the estimates for Saturn is mainly due to the presence of the rings, which were not  removed in this plot.
        }
\end{figure*}

Figure~\ref{fig:synoptic_of_tb_ba_transit_channel_averages}
is a summary of the channel-averaged $\TbBA$ for each single transit and planet as a function of the quantity $\PlanetAspectAngle$, the sub-\Planck{} latitude at the epoch of the observation as seen from the planet; it represents the planet aspect angle as seen from \Planck. Since we already include the effect of band-averaging in Eq.~\eqref{eq:TantToTbr}, we do not need any colour-correction factor.

\begin{table}
\centering
\footnotesize
\caption{\label{tab:most:used:symbols}
List of symbols used in Sect.~\protect\ref{sec:results}.
}
\begin{tabular}{cp{5cm}}
\hline\hline
$\BandPass(\nu)$, $\BandWidth$, $\Fcent$, $\FcentEff$& Band pass, bandwidth, central frequency and effective central frequency.\\
$\Delta$, $\PlanetAspectAngle$& \Planck-planet range and planet aspect angle.\\
$\OmegaPlanet$, $\OmegaPlanetFiducial$& Planet solid angle and its reference value.\\
$\effAsp$, $\effAper$, $\effEtaBeam$& Corrections for planet flattening, beam aperture and beam numerical efficiency.\\
$\DeltaTant$& Variation of antenna temperature.\\
  $\Tb$& Brightness temperature.\\
  $\TbRJ$& Brightness temperature in the Rayleigh-Jeans scale.\\
  $\TbMono$& Monochromatic brightness temperature.\\
  $\TbBA$&Band-averaged brightness temperature.\\
$\BnuBA$ &  Model band-averaged brightness.\\
$\BrBAplanet$ &  Measured brightness.\\
\bottomrule
\end{tabular}
\end{table}

\section{Results}\label{sec:results}
In comparing our results with those from WMAP, we must take in account the different value of the dipole amplitude used by \Planck{} and WMAP, as this leads to a mismatch in the absolute calibration level: the \Planck{} team used the value $\DipolePlanck = 3364 \pm 2\,\muK$ \citep{planck.2013.05.LFI.calibration,planck.2015.05.LFI.calibration,planck.2018.lfi.processing},
 while the WMAP  team used $\DipoleWMAP = 3355 \pm 8\,\muK$
\citep{wmap.maps.2009}. Therefore, we scaled the WMAP estimates of $\TbRJ$ by the factor 1.002831. 
Moreover, WMAP reported $\TbRJ$ rather than $\TbMono$ or $\TbBA$. When needed,
we used the WMAP bandpasses to derive $\TbMono$ or $\TbBA$ from $\TbRJ$, according to the procedure outlined in Sect.~\ref{eq:appendix:wmap:tbrj:to:tbba}.
Each of the quantities $\TbRJ$, $\TbMono$, $\TbBA$ includes the correction for blocking radiation, as explained in Sect.~\ref{sec:blocking}.
The definition of the main symbols is provided in 
Table~\ref{tab:most:used:symbols}.

\subsection{Jupiter}\label{sec:Jupiter}

\begin{table*}
  \centering
  \footnotesize
  \caption{\label{tab:jupiter:observing:conditions}
    Observing conditions of Jupiter per transit.
}
\begin{tabular}{rlrrrrrrrrrr}
    \hline\hline
    Transit& Epoch& PJD\_Start& PJD\_End& Nsmp& EcLon& EcLat& GlxLat& $\Rsun$& $\Delta$& $\AngDiam$& $\PlanetAspectAngle$\\
                    &  &  &  &  & [deg]& [deg]& [deg]& [AU]& [AU]& [arcsec]& [deg]\\
    \midrule
    1&  2009-10-28&     164.66&   171.47&  8421& 317.4&  $-1.0$&  $-40.3$&    5.02&     4.73&      41.65&                 0.31 \\
    2&  2010-07-05&     413.31&   422.26& 11040&   2.7&  $-1.3$&  $-61.4$&    4.97&     4.71&      41.85&                 2.56 \\
    3&  2010-12-11&     571.72&   581.43& 12104& 354.2&  $-1.4$&  $-61.0$&    4.95&     4.79&      41.18&                 2.28 \\
    4&  2011-08-04&     810.48&   816.28&  6839&  39.1&  $-1.3$&  $-43.2$&    4.95&     4.82&      40.93&                 3.68 \\
    5&  2012-01-18&     971.09&   988.22& 37035&  31.0&  $-1.1$&  $-48.9$&    4.98&     4.81&      41.02&                 3.34 \\
    6&  2012-09-07&    1208.18&  1218.63& 22852&  75.1&  $-0.8$&  $-13.3$&    5.03&     4.93&      40.03&                 3.41 \\
    7&  2013-02-17&    1367.85&  1384.41& 30724&  66.7&  $-0.5$&  $-20.3$&    5.08&     4.88&      40.43&                 3.11 \\
    \bottomrule
  \end{tabular}
\end{table*}

Table~\ref{tab:jupiter:observing:conditions} lists the seven transits of Jupiter that have been observed by LFI; the last three transits were not considered in the analysis presented by \citet{planck.intermediate.52.planet.flux.densities}.
Because of a combination of factors, fewer samples have been acquired in transits 1 and 4.
All the transits occur near the Equator, with $0.3\deg < \PlanetAspectAngle < 3.4\deg$ (see Sect.~\ref{sec:signalsToTant} for the definition of $\PlanetAspectAngle$), so that $\effAsp<3\times10^{-4}$.
The Galactic latitude is always negative, with transit from 1 to 5 between
$-62^\circ$ and $-40^\circ$, transit 6 at $-13^\circ$, and transit 7 approximately at $-20^\circ$.
The last two transits are sufficiently close to the Galactic plane to suffer 
larger background contamination; this is particularly true at 30\,GHz, where Jupiter is weaker and the Galactic background is larger.
Figure~\ref{fig:synoptic_of_tb_ba_transit_channel_averages}
shows no evident correlations between brightness temperatures and 
$\PlanetAspectAngle$. 
However, transits 6 and 7 at 30\,GHz depart significantly from the average.
For this reason,
we limited our analysis to the first five transits.
In total there are 
        110 measurements (+44 in transits 6 and 7), of which 
        20 (+8) at 30\,GHz,
        30 (+12) at 44\,GHz, and
        60 (+14) at 70\,GHz.

Table~\ref{tab:jupiter:channel:average} 
reports our values for 
$\BrBAplanet$, $\TbRJ$, $\TbMono$, and $\TbBA$. We computed these as the weighted
averages of the measurements for each frequency channel across the corresponding set of radiometers, still considering five transits.
Adding transits 6 and 7 has a minor impact on the 70\,GHz
channel: 
        $\TbRJ   = 170.40\pm0.16$\,K,
        $\TbMono = 172.08\pm0.16$\,K,
        $\TbBA   = 171.07\pm0.17$\,K. This is a $0.1$\,K reduction in temperature, and a marginal improvement on the error bars.
Since we consider band-averaged quantities, we used
the weighted average of the individual $\Fcent$ or $\FcentEff$ of each radiometer as the reference frequency.
We did not include the effect of the beam numerical efficiency $\effEtaBeam$ (Sect.~\ref{sec:beam:efficiency}) in Table~\ref{tab:jupiter:channel:average},
so we added an uncertainty of $0.3\%$; the calibration uncertainty introduces an additional $0.1\%$ to the error budget.

To derive the averaged values in Table~\ref{tab:jupiter:channel:average}, we had to consider some subtleties in the analysis; these are described in Sect.~\ref{appendix:averaged:brightness}.
Of course, averaging $\BrBAplanet$ and $\TbRJ$ is not the same as averaging $\TbMono$ and $\TbBA$, as these are not additive quantities.
A more rigorous approach requires us to determine the values of $\TbMono$ and $\TbBA$  that fit the observed $\BrBAplanet$; this can be done through the minimization of 
the function of merit in Eq.~\eqref{eq:averaging:tbba},
Sect.~\ref{appendix:averaged:brightness}.
We verified that a simple average agrees with the result of a minimization
within the second decimal figure, given the observing conditions of \Planck/LFI. However, the numbers we report in Table~\ref{tab:jupiter:channel:average}
were derived using the rigorous approach.

Estimating uncertainties is more subtle, as several effects are to be considered. Firstly, there is a large variability in the error bars for $\TbRJ$, 
which are denoted as $\TbRJrnd$: in fact, $\TbRJrnd$ 
varies from 0.06\,K to 0.26\,K ($1\sigma$), 
These variations can look puzzling, but the transit-to-transit variability in
$\TbRJrnd$ is highly correlated with the number of samples $N_{\mathrm{P}}$ in the planet ROI:
the correlation coefficient between $1/\sqrt{N_{\mathrm{P}}}$ and $\TbRJrnd$ is $\geq 0.96$.
If we assume that the average of $\TbRJrnd$ across a channel
is representative of the uncertainties in the data, we should expect overall errors to be 
$\sim 0.12/\sqrt{28}\,\mathrm{K}$, $\sim 0.16/\sqrt{42}\,\mathrm{K}$, and $\sim0.08/\sqrt{84}\,\mathrm{K}$, for the 30, 44, and 70\,GHz channels, respectively. However, this is not what we see in 
Table~\ref{tab:jupiter:channel:average}, as the errors reported are of the order of $0.2$~K, which are comparable to the worst $\TbRJrnd$ on a single measure.

Another indication of some possible systematic error in our data is the scatter of $\TbRJ$ among transits, which exceeds what would be expected from a normal distribution with variance
$\TbRJrnd^2$. The standard deviations for $\TbRJ$ are
$0.800$\,K at 30\,GHz, 
$1.072$\,K at 44\,GHz,
and
$1.439$\,K at 70\,GHz,
while peak-to-peak 
variations are
$3.38$\,K at 30\,GHz, 
$4.13$\,K at 44\,GHz,
and
$6.06$\,K at 70\,GHz.
Moreover, the distribution of the residuals is not Gaussian.

Fig.~\ref{fig:synoptic_of_tant_after_reduction_to_same_omegabeam}
shows that the estimates for $\DeltaTant$ at 70\,GHz are distributed around the mean, but they are not completely compatible with random fluctuations.
A closer inspection reveals that most of the effect comes from 
data collected by the radiometers associated with horns 18 and 22. 
The averaged $\TbRJ$ from horn 18 deviates by $-2.5$~K from the average  for 70\,GHz, while for horn 22 the deviation is $+2.$\,K; for others, the difference is less than 0.5\,K, which is compatible with the hypothesis of random noise fluctuations. 
However, removing these samples does not change 
the results in the table significantly; as an example, we obtained $\TbBA=171.02$\,K instead of $\TbBA=170.17$\,~K (but the $1\sigma$ error decreases from $0.19$\,K to $0.11$\,K).

Part of the observed variability across radiometers is intrinsic to the source, given the relatively wide bandwidth  of our frequency channels, especially at 70\,GHz \citep{planck.2015.05.LFI.calibration}. 
This means that introducing some correction to flatten 
this effect would introduce another kind of distortion in the data.
However when computing uncertainties on channel averaged quantities, the adequacy of 
usual error propagation formula must be assessed. 
To validate our estimates for uncertainties given by least-square fits, we used a bootstrap technique and a Markov chain Monte Carlo (MCMC). In case of significant discrepancies, we picked the largest error estimate.

\begin{table*}
  \centering
  \footnotesize
  \caption{\label{tab:jupiter:channel:average}
    Channel-averaged results$^a$ for Jupiter, excluding transits 6 and 7.
  }
  \begin{tabular}{cccccccc}
    \hline\hline
    ch& ${\Fcent}$& ${\FcentEff}$& ${\BrBAplanet}^b$& ${\TbRJ}^b$& ${\TbMono}^b$& ${\TbBA}^b$\\
      & [GHz]& [GHz]& [MJy/sr]& [K]& [K]& [K]\\
    \midrule
    30 &    28.40 &       28.43 &  $3598.2 \pm        16.4$ &  $144.93 \pm        0.17$ &    $145.62 \pm        0.17$ &  $144.69 \pm      0.19$ \\
    44 &    44.10 &       44.16 &  $9570.0 \pm        23.0$ &  $159.76 \pm        0.19$ &    $160.82 \pm        0.19$ &  $160.27 \pm      0.19$ \\
    70 &    70.40 &       70.36 & $25866.0 \pm       127.8$ &  $170.50 \pm        0.18$ &    $172.18 \pm        0.18$ &  $171.17 \pm      0.19$ \\
    \bottomrule
  \end{tabular}
  \tablenoteskip

  \widetablenote{a}{The effect of $\effEtaBeam$ is not included.}
  \widetablenote{b}{The value includes blocked radiation.}
\end{table*}

Values of $\TbRJ$, $\TbMono$, and $\TbBA$ in
Table~\ref{tab:jupiter:channel:average} are very similar, with
differences smaller than 2\,K ($\sim 1\,\%$). This happens because
the brightness temperature of Jupiter is greater than $140$\,K: since
the radiometers of \Planck \ measure frequencies below $100$\,GHz the
difference between Planck's law and the RJ approximation is not large.
However, the difference exists and explains the fact that
$\TbRJ > \TbBA$ at 30\,GHz and the opposite at 44\,GHz and 70\,GHz.
In fact, below 30\,GHz Planck’s law is sufficiently approximated
by the RJ law with brightness scaling as $\nu^2$; in this case, the
band averaged brightness is larger than the RJ brightness computed at
the central frequency. Consequently, $\TbRJ>\TbBA$ is needed to
explain the same brightness. At higher frequency, the two laws diverge
more significantly, and the band-averaged brightness is always lower
than the RJ brightness at central frequency; therefore, $\TbRJ<\TbBA$
is needed to explain the same brightness. The critical frequency where
this swap occurs is mainly determined by the bandwidth: for 30\,GHz
and 44\,GHz radiometers, the critical frequency is in the range
29--37\,GHz, while for 70\,GHz radiometers is 53--60\,GHz. The central
frequencies for the 30\,GHz channel are just below the critical
frequencies, while the opposite happens for 44\,GHz and 70\,GHz
radiometers, thus explaining the observed difference. We provide a
more quantitative discussion in
Appendix~\ref{appendix:TbBA:and:TbRJ:relations}.

\begin{figure*}
        \centering
        \includegraphics[width=0.85\textwidth]{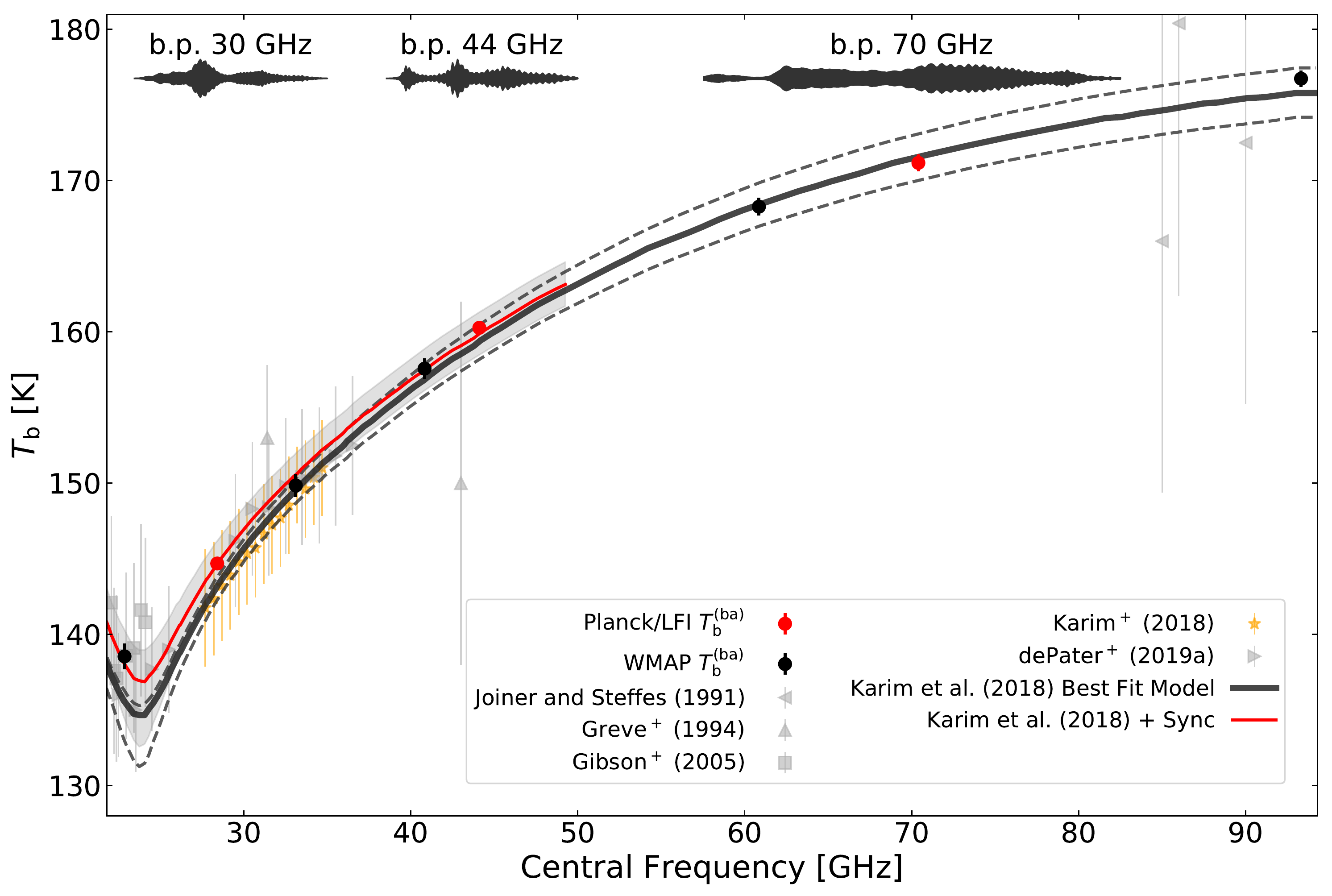}
        \caption{ \label{fig:jupiter:vz:karim} 
                Comparison of Jupiter measurements made in this paper (red circles)
                with WMAP (black points, taken from \protect\citep{WMAP:PLANETS:2011,WMAP:PLANETS:2013} and converted to $\TbBA$) and with 
                measurements from literature 
                and the RT model by \protect\citet{Karim:etal:2018}. 
The model brightness temperature 
                $\TbRT$ is plotted as a dark grey line. 
                The top and bottom dashed lines represent the upper and lower limits of model uncertainties as provided in the same paper.
The RT model plus the synchrotron emission correction 
                is plotted as a red line; the grey
                band represents the lower and upper limits.
For references of ground-based measurements, see the text.
The brightness temperatures reported by \protect\citet{Gibson:dePater:2005} are recalibrated measures originally published by \protect\citet{Klein:Gulkis:1978}.
The violin plots at the top of the figure represent the relative
sensitivity to the frequencies inside each bandpass. }
\end{figure*}

In Fig.~\ref{fig:jupiter:vz:karim} we plot the estimates for $\TbBA$ reported in Table~\ref{tab:jupiter:channel:average} and compare these with a selection of results and models available in the literature. Points are plotted at $\Fcent$ for each frequency channel. The violin plots at the top give insight into how $\BandPass(\nu)$ changes within each  frequency channel.
The quoted error bars are comparable with the size of the symbols, even including 
the effect of the $\pm0.3\%$ $\effEtaBeam$ correction uncertainty. 
The black points in the figure represent the WMAP measurements, taken from 
\citet{WMAP:PLANETS:2011} and \citet{WMAP:PLANETS:2013}, and converted to $\TbBA$ as detailed at the begin of this section and in Sect.~\ref{eq:appendix:wmap:tbrj:to:tbba}. 
 The WMAP results compare well with ours. The plot includes measurements taken from\footnote{
        Results from 
        \protect\citet{dePater:etal:1982} and 
        \protect\citet{Goldin:etal:1997} are not included, as they are outside the frequency range of interest.
} 
\citet{Joiner:Steffes:1991}, \citet{Greve:etal:1994}, \citet{Gibson:dePater:2005}, \citet{Karim:etal:2018} and \citet{dePater:etal:2019a}.
The results from \citet{Gibson:dePater:2005} are measures provided by \citet{Klein:Gulkis:1978} (Tab.II in the paper) and reprocessed. According to the conventions used in this paper they are similar to $\TbMono$.

Apart from WMAP, Fig.~\ref{fig:jupiter:vz:karim} compares our estimates with other results found in the literature. The few measurements above 40\,GHz are consistent with our estimates; the error bars are however large, and the consistency is therefore of little significance. Below 40\,GHz, the situation is much better. In particular, the CARMA measurements in \citet{Karim:etal:2018} cover the 27.7--34.7\,GHz fairly well. Our estimate at 30\,GHz is consistent with CARMA, but we see an excess in $\TbBA$ of nearly 2\,K. As we explain below, this excess
is likely due to the presence of a synchrotron contribution to the Jupiter signal
that has been removed in the CARMA data \citep{Karim:etal:2018}.

The sparse frequency coverage of measurements in the literature
makes it difficult to quantitatively compare our measurements with those of other authors
without adopting an interpolation scheme.
But microwave emission of Jupiter cannot be reduced 
to a simple polynomial expression at \Planck/LFI frequencies.
The emissivity observed outside the atmosphere is the result of 
the radiative transfer of microwave emission produced by different layers 
within the atmosphere that radiate towards the observer and 
are extinguished by the traversed layers \citep[see][]{dePater:Massie:1985,Gibson:dePater:2005}.
At \Planck/LFI frequencies, extinction is dominated by the NH$_3$ absorption.
For this reason, it is interesting to compare our data with representative models in the literature, but before discussing the comparison with models, we note that this paper is devoted to the presentation of Planck/LFI data and not to a detailed discussion of models for planetary thermal microwave emission. 

We take as a reference the Jupiter radiative-transfer (RT) model described in
\citet{Karim:etal:2018}.
The model estimates the full-disc thermal emissivity $\TbRT(\nu)$ of Jupiter for wavelengths 
between 0.3\,cm and 4\,cm and compares well with 
a number of observations, including CARMA and WMAP seven-year measurements. 
In the plot, $\TbRT(\nu)$ is represented by a thick black line. 
The dashed lines are two further models provided by \citet{Karim:etal:2018}, which 
represents an upper and a lower limit for the predicted $\TbRT(\nu)$.
Our estimates and the model agree very well at 
70\,GHz, but we overshoot the model at lower frequencies; in particular, at 30\,GHz the overshoot is almost 2\,K.
This happens because at frequencies below 40\,GHz the measurement is affected by a small synchrotron emission due to solar 
high-energy electrons trapped in radiation belts (analogous to Earth's Van Allen belts)
within a few Jupiter radii from the planet \citep{Klein:Gulkis:1978}.

For Jupiter, the synchrotron emission is mainly concentrated around the equatorial plane, with two emission lobes clearly seen in Very Large Array maps \citep{dePater:1981,dePater:Dunn:2003,Kloosterman:etal:2005},
and this emission is polarized at the level of 20--25\,\%
\citep{dePater:Dunn:2003}.
Gradual changes over time in the total intensity of the emission have been reported by \citet{Klein:etal:2001}, \citet{Dunn:etal:2003} and \citet{Kloosterman:etal:2005}
at 2.3\,GHz and 1.4\,GHz.
These are mainly connected to secular changes in the density of relativistic electrons in the Jupiter magnetosphere \citep{Dunn:etal:2003,Kloosterman:etal:2005}, thereby leading to changes in the synchrotron total intensity but not in its spatial distribution, 
and to a minor extent to changes in viewing geometry.
Abrupt changes in both the intensity and spatial distribution were recorded as a consequence of impacts of minor bodies with Jupiter, as in the case of comet Shoemaker-Levy 9 in 1994 \citep{dePater:etal:1995,Klein:etal:2001} and of an unidentified object in July 2009 \citep{Santos-Costa:etal:2011}, a few months before the first scan of Jupiter from \Planck.
While WMAP did not attempt any removal of this contribution \citep{WMAP:PLANETS:2011}, it is expected to amount to about $1\,\%$ of the thermal emission of the disc at 28.5\,GHz \citep{Karim:etal:2018}. Therefore, this effect is comparable or larger than our error bars.

To include the amount of contamination from synchrotron emission, we follow the formalism in \citet{dePater:Dunn:2003}, \citet{WMAP:PLANETS:2011} and \citet{Karim:etal:2018}. According to this formalism, the synchrotron emission seen by an observer at Earth has a $\nu^{-0.4}$ spectral dependence,
and at the reference frequency of 28.5\,GHz the expected synchrotron flux is 
$\Fsync = 1.5 \pm 0.5$~Jy  \citep{dePater:Dunn:2003,Karim:etal:2018},
assuming Jupiter as seen at
$\Delta=4.04\;\mathrm{AU}$ corresponding to an $\OmegaPlanetFiducialSync=4.11075\times10^{-8}\mathrm{sr}$. 
The total 
brightness is the sum of the thermal and synchrotron components as follows:

\begin{equation}\label{eq:thermal:plus:syncrotron}
\BrThSync(\nu) = \Bnu\left(\TbRT(\nu)\right) + \frac{\Fsync }{\OmegaPlanetFiducialSync} 
\left( 
\frac{\nu}{28.5 \, \mathrm{GHz}}
 \right)^{-0.4},
\end{equation}
where $\TbRT(\nu)$ is derived from the RT 
model.

The addition of the 1.5\,Jy synchrotron emission explains the 30\,GHz overshoot.
To better constrain our data, we left $\Fsync$ as a free parameter and fitted it against
the 30\,GHz and 44\,GHz data taken separately and then together.
We fitted band averaged brightness from models against the individual $\BrBA$ 
for each transit and radiometer.
This is obtained by replacing
$\BrThSync(\nu)$ and the $\nu^{-0.4}$ dependence with the corresponding band-averaged
quantities in Eq.~\eqref{eq:thermal:plus:syncrotron} as follows:
\begin{eqnarray}\label{eq:syncrotron:band:averaged}
\BrBArt & = & \frac{1}{\BandWidth}\int_0^{+\infty} d\nu \; \tau(\nu) \; \Bnu\left(\TbRT(\nu)\right),  \\
\SyncOneBA & = & \frac{1}{\BandWidth}\int_0^{+\infty} d\nu \; \tau(\nu) \; \left( 
                                        \frac{\nu}{28.5 \, \mathrm{GHz}}
                        \right)^{-0.4},
\end{eqnarray}
where $\SyncOneBA$ is tabulated for each radiometer in Table~\ref{tab:photo:parameters}. 

To analyse the effect of the uncertainty on the beam numerical efficiency correction
$\effEtaBeam$, 
we scaled 
$\BrBA$ by $(1\pm\effEtaBeam)$ obtaining an upper and a lower limit for 
$\Fsync$.
Similarly we accounted for the uncertainty in the $\TbRT(\nu)$ model by 
replacing the best-fit model in 
Fig.~\ref{fig:jupiter:vz:karim}
with the upper or the lower limits models represented by the dashed lines.
The best-fit $\Fsync$ and its uncertainties were derived with the fitting methods already discussed above;
we used a bootstrapping algorithm to validate these uncertainties.

Results are shown in the bottom part of 
Table~\ref{jupiter:sync:fit} for the 30\,GHz and 40\,GHz alone
and then taken together.
The top part of the table lists weighted averages of 
$\SyncOneBA/\OmegaPlanetFiducialSync$ taken across the data sets, and of
$\BrBArt$ computed for the final model and its lower and upper limits.
At 30\,GHz the best fit is for $\Fsync=1.50 \pm 0.15\,\mathrm{Jy}$,
to be compared with the expected $\Fsync=1.5 \pm 0.5\,\mathrm{Jy}$.
The uncertainty introduced by the unknown numerical beam efficiency
increases the width of the confidence region to 
$1.15\,\mathrm{Jy} < \Fsync < 1.84\,\mathrm{Jy}$. If we use the best-fit model of \citet{Karim:etal:2018} with the lower or the upper limit, we get $\Fsync = 2.83\,\mathrm{Jy}$ and $\Fsync = 0.47\,\mathrm{Jy,}$ respectively.
The 44\,GHz suggests an higher value, $\Fsync = 2.53\,\mathrm{Jy}$, but the uncertainty is larger; moreover, the upper model would not require any synchrotron component.
Combining 30\,GHz and 44\,GHz gives nearly identical results to the 30\,GHz alone. 
The thermal model plus $\Fsync=1.5\;\mathrm{Jy}$ computed from 
Eq.~\eqref{eq:thermal:plus:syncrotron}
is represented by the red line in 
Fig.~\ref{fig:jupiter:vz:karim}.
The grey band represents the difference between
upper and lower limit models.
The effect of the uncertainty in the 
$\effEtaBeam$ correction
is comparable to the width of the red dots, and it is not displayed.
The inclusion of transits 6 and 7 affects mainly the 30\,GHz; in this case, the best fit leads to $\Fsync=1.75 \pm 0.12 $.

\begin{table*}
  \centering
  \footnotesize
  \caption{\label{jupiter:sync:fit}
    Derivation of $\Fsync$ from the overshooting of 30\,GHz and 44\,GHz.}
  \begin{tabular}{cclcccc}
    \hline\hline
    &   &   & \multicolumn{3}{c}{Data Set} & \\
    &  &        & 30\,GHz              & 44\,GHz     & 30 \& 44\,GHz & \\
    \multicolumn{7}{l}{\bf Model$^a$} \\
    & $\BrBArt$          & model w.a.$^b$    &  3561.7       & 9518.2    & --  &   "    \\
    & $\BrBArt$          & lower model w.a.  &  3539.0       & 9447.1    & --  &   "    \\
    & $\BrBArt$          & upper model w.a.  &  3577.1       & 9584.8    & --  &   "    \\
    & & & & & & \\
    & $\SyncOneBA/\OmegaPlanetFiducialSync $ 
        & w.a. & 24.40         & 20.44     & -- & " \\
    & & & & & & \\
    \multicolumn{7}{l}{\bf Fit$^c$} \\
    &$\Fsync$   & best fit & 1.50                & 2.53             & 1.58            & Jy\\
    &"  & random error     & 0.15                & 0.49             & 0.14  & "\\
    & & & & & & \\
    &"& fit lower $\effEtaBeam$ 
            & 1.15                & 1.97             & 1.22            & " \\ 
    &"& fit upper $\effEtaBeam$ 
            & 1.84                & 3.09             & 1.94            & " \\ 
    & & & & & & \\
    &"&  fit lower model 
            & 2.43                & 6.01             & 0.46            & " \\ 
    &"& fit upper model 
            & 0.87                & $-0.73$          & 2.99            & " \\ 
    \hline
  \end{tabular}
  \tablenoteskip\\
  \widetablenote{a}{Model for $\TbRT(\nu)$. The ``lower model'' and ``upper model'' labels indicate the lower and upper limits of the model.}
  \widetablenote{b}{``W.a.'' denotes a weighted average over the dataset.}
  \widetablenote{c}{The uncertainty in the best fit is divided in three components: (1) random error, (2) upper/lower limits for the effect of the uncertainty on the beam numerical efficiency correction $\effEtaBeam$, and (3) the effect of taking the upper or the lower limit for the model.}
\end{table*}

The presence of some synchrotron could potentially introduce a source of variability 
for the Jupiter disc averaged brightness. However, apart from the case of transits 6 and 7 at 30\,GHz,
we found no other significant correlation with time or with the geometry of observation in our transit-averaged data.
Therefore, we may conclude that during our observations Jupiter behaved as a stable microwave source within $\sim10^{-3}$ over three years, in agreement with \citet{WMAP:PLANETS:2011}.

Before going further, we want to note that \citet{planck.2015.05.LFI.calibration}
reports slightly different results for Jupiter.  
This is caused by a number of small differences 
in data processing; the most important is the evaluation of $\effAper$, as explained in Sect.~\ref{sec:aperture:correction}.
In addition, \citet{planck.2015.05.LFI.calibration} compared $\TbMono$ (Eq.~\ref{eq:tb:mono}) with $\TbRJWMAP$ (including blocking radiation), 
which have relative differences 
of $-5\times10^{-4} \div -3\times10^{-3}$,
equivalent to $-0.07 \div -0.44$\,K at 30\,GHz,
$2\times10^{-3} \div 3\times10^{-3}$ 
equivalent to $0.3$\,K at 44\,GHz,
and $6\times10^{-6} \div 5\times10^{-3}$ 
equivalent to $0.1\div 0.8$\,K at 70\,GHz. 
 
\subsection{Saturn}\label{sec:Saturn}

\begin{table*}\centering
  \footnotesize
  \caption{\label{tab:saturn:observing:conditions}
    Observing conditions of Saturn per transit.
}
  \begin{tabular}{rlrrrrrrrrrr}
    \hline\hline
    transit& Epoch& PJD\_Start& PJD\_End& Nsmp& EcLon& EcLat& GlxLat& $\Rsun$& $\Delta$& $\AngDiam$& $\PlanetAspectAngle$\\
           &  &  &  &  & [deg]& [deg]& [deg]& [AU]& [AU]& [arcsec]& [deg]\\
    \midrule
    1&  2010-01-05&     232.80&   240.74&  9738& 184.5&   2.3&   62.2&    9.48&     9.28&      17.91&                 6.04 \\
    2&  2010-06-16&     393.89&   403.46& 11864& 177.9&   2.4&   62.5&    9.53&     9.45&      17.59&                 2.27 \\
    3&  2011-01-19&     612.40&   619.32&  8440& 197.1&   2.5&   58.3&    9.59&     9.36&      17.75&                12.60 \\
    4&  2011-07-03&     776.55&   785.29& 10743& 190.6&   2.5&   60.9&    9.64&     9.62&      17.28&                 9.23 \\
    5&  2012-01-29&     990.40&   990.61& 31613& 209.3&   2.5&   51.1&    9.70&     9.50&      17.50&                18.41 \\
    6&  2012-07-13&    1154.23&  1159.87&  6807& 202.8&   2.5&   55.1&    9.75&     9.68&      17.16&                15.47 \\
    7&  2013-02-02&    1358.62&  1363.85&  6118& 221.1&   2.5&   42.4&    9.80&     9.72&      17.10&                23.28 \\
    8&  2013-07-23&    1529.49&  1535.02&  6608& 214.8&   2.4&   47.1&    9.84&     9.74&      17.07&                20.91 \\
    \bottomrule
  \end{tabular}
\end{table*}

Saturn was observed in eight transits, all of which occurred with $\PlanetAspectAngle > 0\deg$;
Saturn did not cross the Galactic plane in any case.
The observing circumstances for Saturn are listed in Table~\ref{tab:saturn:observing:conditions};
we note the higher sampling density in transits 2, 4, and 5. Because of changes in the scanning strategy of the \Planck{} spacecraft, only horns 24, 27, and 28 observed Saturn during transit 5.
Transits from 1 to 4 happened simultaneously with \Planck/HFI \citep{planck.intermediate.52.planet.flux.densities},
while transits from 5 to 8 were observed by \Planck/LFI alone.
Transits 1 and 2 occurred near the last two WMAP seasons
\citep{WMAP:PLANETS:2013}.
In total, there are 326 measurements: 96 made by 70\,GHz channels, 50 by 44\,GHz channels, and 36 by 30\,GHz channels. 

\begin{table*}
  \centering
  \footnotesize
  \caption{\label{tab:saturn:tbrj}
    Table of channel-averaged $\TbRJ$ and $\BrBAplanet$ for Saturn, for each transit.}
  \begin{tabular}{lrrrrrrrr}
    \hline\hline
    & Tr.~1 & Tr.~2 & Tr.~3 & Tr.~4 & Tr.~5 & Tr.~6 & Tr.~7 & Tr.~8\\
    \hline
    \multicolumn{9}{c}{\bf{Channel 30}}\\
    \hline 
    $\wDisk$           & 0.9305    & 0.9693     &  0.8841   & 0.9041    & 0.8686    & 0.8728    & 0.8842    & 0.8722\\
    $\wRing$           & 0.3409    & 0.1313     &  0.7164   & 0.5261    & 1.0631    & 0.8875    & 1.3762    & 1.2209\\
    $\TbRJ$ [$\Krj$]   & $132.17$  & $137.30$   & $128.88$  & $131.56$  & $129.96$  & $131.44$  & $132.46$  & $130.60$\\
    error [$\Krj$]     & $0.32$    & $0.56$     & $0.41$    & $0.16$    & $0.30$    & $0.24$    & $0.62$    & $0.82$\\
$\BrBAplanet$ [MJy/sr] & $3280.64$ & $3407.65$  & $3198.91$ & $3265.32$ & $3225.74$ & $3262.69$ & $3287.49$ & $3240.65$\\
error [MJy/sr]         & $37.79$   & $43.81$    & $39.13$   & $27.98$   & $33.23$   & $25.73$   & $44.79$   & $50.35$\\
    \hline
    &   &   &   &   &   &   &   &  \\
    \multicolumn{9}{c}{\bf{Channel 44(24)}}\\
    \hline 
    $\wDisk$           & 0.9305     & 0.9693    & 0.8841    & 0.9040    & 0.8686    & 0.8728    & 0.8842    & 0.8722\\
    $\wRing$           & 0.3404     & 0.1313    & 0.7164    & 0.5266    & 1.0631    & 0.8875    & 1.3762    & 1.2209\\
    $\TbRJ$ [$\Krj$]   & $142.00$   & $144.46$  & $139.53$  & $142.14$  & $138.57$  & $142.33$  & $144.02$  & $141.05$\\
    error [$\Krj$]     & $0.86$     & $0.10$    & $0.72$    & $0.41$    & $0.46$    & $0.32$    & $1.78$    & $0.36$\\
$\BrBAplanet$ [MJy/sr] & $8526.66$  & $8675.34$ & $8378.37$ & $8535.68$ & $8321.22$ & $8547.59$ & $8647.98$ & $8470.06$\\
        error [MJy/sr] & $125.87$   & $81.92$   & $116.58$  & $99.19$   & $100.20$  & $55.53$   & $182.74$  & $95.68$\\
    \hline 
    &   &   &   &   &   &   &   &  \\
    \multicolumn{9}{c}{\bf{Channel 44(25--26)}}\\
    \hline 
    $\wDisk$           & 0.9303     & 0.9715    & 0.8841    & 0.9055    &     ---   & 0.8731    & 0.8845    & 0.8719\\
    $\wRing$           & 0.3421     & 0.1211    & 0.7170    & 0.5152    &     ---   & 0.8816    & 1.3783    & 1.2158\\
    $\TbRJ$ [$\Krj$]   & $140.63$   & $146.20$  & $137.97$  & $138.95$  &     ---   & $140.96$  & $143.27$  & $141.72$\\
    error [$\Krj$]     & $0.77$     & $0.85$    & $0.68$    & $0.28$    &     ---   & $0.94$    & $1.46$    & $1.02$\\
$\BrBAplanet$ [MJy/sr] & $8381.62$  & $8714.18$ & $8223.17$ & $8281.82$ &     ---   & $8401.98$ & $8539.56$ & $8446.73$\\
error [MJy/sr]         & $47.68$    & $63.73$   & $57.67$   & $33.61$   &     ---   & $47.08$   & $80.84$   & $61.53$\\
    \hline 
    &   &   &   &   &   &   &   &  \\
    \multicolumn{9}{c}{\bf{Channel 70}}\\
    \hline 
    $\wDisk$          & 0.9304      & 0.9698    & 0.8841    & 0.9044    &     ---   & 0.8729    & 0.8843    & 0.8721\\
    $\wRing$          & 0.3411      & 0.1288    & 0.7170    & 0.5237    &     ---   & 0.8863    & 1.3767    & 1.2201\\
    $\TbRJ$ [$\Krj$]  & $144.90$    & $147.93$  & $142.73$  & $144.61$  &     ---   & $145.45$  & $148.41$  & $146.65$\\
    error [$\Krj$]    & $0.36$      & $0.34$    & $0.40$    & $0.39$    &     ---   & $0.52$    & $0.43$    & $0.39$\\
$\BrBAplanet$ [MJy/sr]& $22019.76$  &$22474.54$ &$21689.96$ & $21981.11$&     ---   &$22100.77$ &$22558.32$ & $22287.28$\\
error [MJy/sr]        & $258.39$    & $258.11$  & $265.70$  & $253.51$  &     ---   & $269.20$  & $272.30$  & $280.01$\\
    \hline 
  \end{tabular}
\end{table*}

Table~\ref{tab:saturn:tbrj}
lists the weighted average of $\TbRJ$ 
and $\BrBAplanet$
for each transit and channel.
Errors in the averaged $\TbRJ$ 
and $\BrBAplanet$ 
are derived using usual error propagation and are cross-checked both with bootstrap and Monte Carlo simulations.
The 44\,GHz channel is divided in two sub-channels: 44(24) refers to horn 24, and 44(25--26) refers to the average of horns 25 and 26. This split accounts for the fact that the 
transits in horn 24 and in the pair 25--26
occurs about five to nine days apart.
The correction for blocking 
in both 
$\TbRJ$ 
and $\BrBAplanet$
is already introduced. 
The correction for the beam numerical efficiency $\effEtaBeam$ 
is not, this adds an uncertainty 
in $\TbRJ$ or $\BrBAplanet$
of
$\pm0.30$\,K (or $\pm7.45$\,MJy/sr) for the 30\,GHz channel,
$\pm0.13$\,K (or $\pm7.81$\,MJy/sr) for the 44(24)\,GHz sub-channel,
$\pm0.22$\,K (or $\pm12.12$\,MJy/sr) for the 44(25-26)\,GHz sub-channel,
$\pm0.44$\,K (or $\pm66.89$\,MJy/sr) for the 70\,GHz channel independent from the transit down to the second decimal 
figure.
The difference in magnitude for the effect in the 44(24)\,GHz and 44(25-26)\,GHz is connected to 
the location of the feed horns in the focal plane. Horn 24 was between the 30\,GHz, 
while horns 25 and 26 where on the opposite site of the focal plane with respect to horn 24.

The aspect-angle correction we applied to other planets is unreliable in the case of Saturn, because of the presence of the rings.
They emit microwave radiation and 
partially
extinguish the microwave emission radiated from the regions of Saturn's disc along lines of sight intersecting both the ring and the disc. On the other hand, they scatter but do not block background radiation \citep{WMAP:PLANETS:2011}.

Following the approach in 
\citet{WMAP:PLANETS:2011}, \citet{WMAP:PLANETS:2013} and \citet{planck.intermediate.52.planet.flux.densities}, we used an empirical model to separate the disc and the ring contribution as follows:
\begin{equation}\label{eq:saturn:empirical:model}
\TbRJ 
=
\frac{
   \Omegauc+ \sum_{r=1}^7 \Omegacr \exp(-\tau_r|\csc B| ) 
}{
\OmegaPlanetEquatorial
}
\Tdisk
+ 
\frac{
\sum_{r=1}^7 \Omegaunr
}{
\OmegaPlanetEquatorial
}
\Tring,
\end{equation}
where 
$\TbRJ$ are the RJ brightness temperatures quoted in 
Table~\ref{tab:saturn:tbrj} for each frequency channel and transit,
$\Tdisk$ and $\Tring$ are RJ temperatures for the disc and the rings (free parameters of the model),
$\OmegaPlanetEquatorial$ is the equatorial solid angle of the disc,
$\Omegacr$ is the solid angle of the fraction of the disc that is hidden by the rings,
$\Omegauc$ is the solid angle of the unimpeded disc, $\Omegaunr$ is the solid angle of the part of ring $r$ that is not obscured by the disc, and
$B=\PlanetAspectAngle$ is the ring opening angle. All the quantities are calculated  at the epoch of 
the given transit.
Rings are numbered starting from the outermost (ring A is $r = 1$) to the innermost (inner C is $r = 7$).
The radii of the rings and their optical depths $\tau_r$ are fixed parameters of the model and are taken from 
Table~10 of \citet{WMAP:PLANETS:2011},
which follows \citet{Dunn:etal:2002}.
The possibility of considering all the $\tau_r$ as free parameters was discussed in 
\citet{WMAP:PLANETS:2011}, \citet{WMAP:PLANETS:2013}, and \citet{planck.intermediate.52.planet.flux.densities},
without conclusive results; because of our error bars, we decided to keep them as fixed parameters.

For each set of observations, we derived $\Tdisk$ and $\Tring$ through the minimization of the quantity
\begin{equation}\label{eq:saturn:empirical:model:minimization}
\chi^2 = \sum_t \frac{\left( \wDiskt\Tdisk  + \wRingt\Tring  - \TbRJt \right)^2}{\sigma_t^2},
\end{equation}
where $t$ runs over the list of transits, and 
$\wDiskt$ and $\wRingt$ are abbreviations for the coefficients in front of $\Tdisk$ and $\Tring$ in  
Eq.~\eqref{eq:saturn:empirical:model}. In general, $\wDiskt+\wRingt \ne 1$.
The weights $\wDiskt$ and $\wRingt$ are weighted averages of coefficients derived for each radiometer in a given channel and are tabulated in Table~\ref{tab:saturn:tbrj}.

\begin{figure*}
        \centering
        \includegraphics[width=\textwidth]{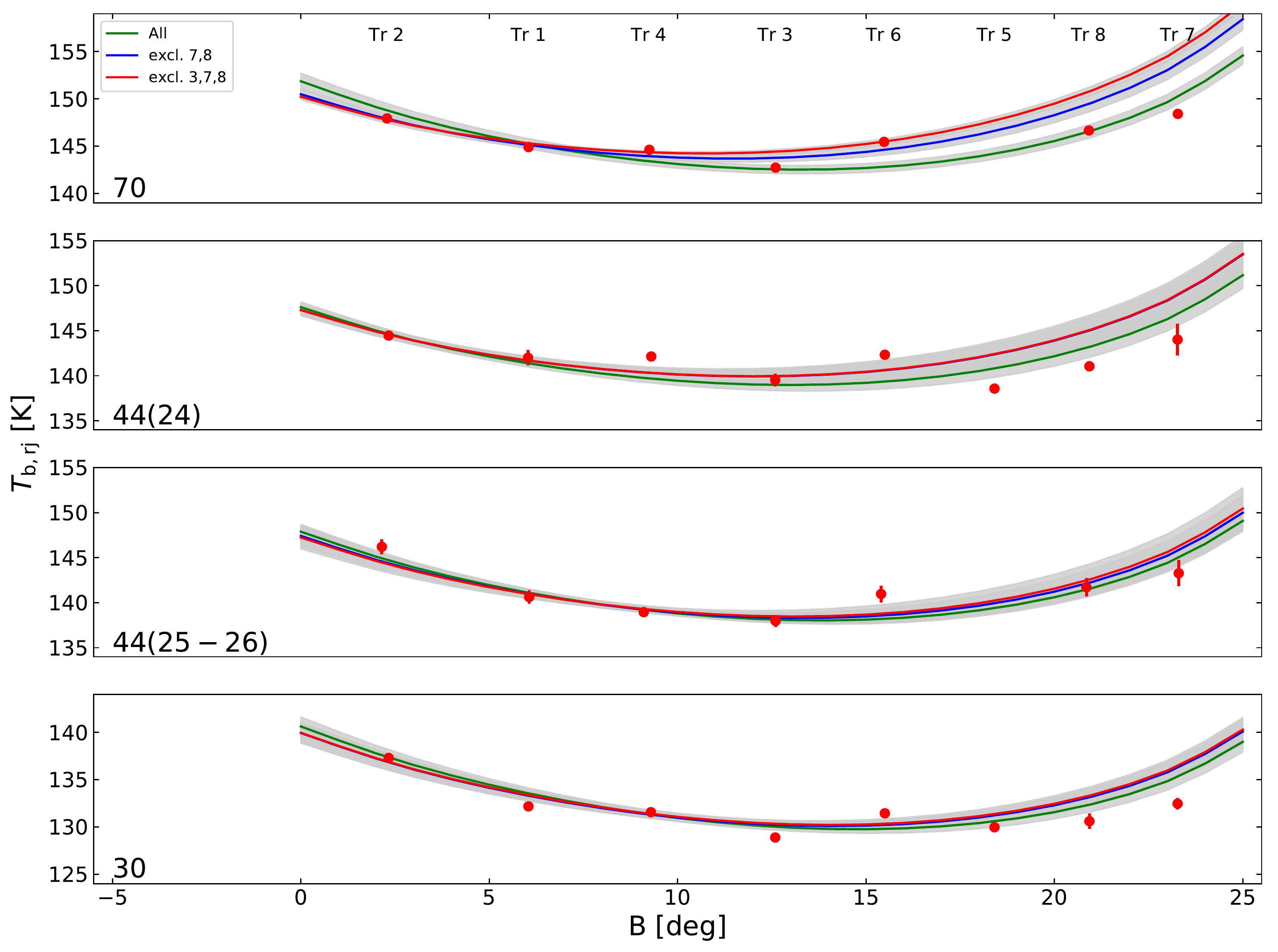}
        \caption{ \label{fig:saturn_effect_of_rings}
                Saturn 
$\TbRJ$ 
                for 70\,GHz frequency channel (first frame from top), 44\,GHz frequency channel (second and third frames), and
                30\,GHz frequency channel (fourth frame) as a function of $\PlanetAspectAngle$.
Labels Tr1, Tr2, $\dots$, Tr8 denotes the transit from which each observation originates.
Continuous curves refers to the best-fit models of $\Tb$ with various selections of
                data: all transits, transits 7 and 8 excluded, and transits 3, 7, and 8 excluded.
The grey bands are $1\sigma$ uncertainties in the models.
        }
\end{figure*}

Figure~\ref{fig:saturn_effect_of_rings} shows $\TbRJ$ for each transit and frequency channel as a function of the planet aspect angle $\PlanetAspectAngle$. 
Continuous curves show the best-fit model of Saturn brightness temperature, obtained with different data cuts: (1) all the transits, (2) all but 7 and 8, and (3) all but 3, 7, and 8. The reason why we considered transit 3 as peculiar is the occurrence of a massive Saturnian storm during the transit \citep{Janssen:etal:2013}.The exclusion of transits 7 and 8 is motivated by the fact that an analysis performed including those transits produces a significantly lower $\Tring$ than expected from WMAP and literature measurements. 
This anomaly is more important at 70\,GHz, but it can be traced in the other channels as well.
Therefore, at 70\,GHz the expectation from WMAP is $\Tring\approx16\,\Krj$, while \Planck/LFI data lead to
$\Tring=11.6\pm1.0\,\Krj$ (all transits),
$\Tring=13.9\pm1.0\,\Krj$ (no 7 and 8), and
$\Tring=16.2\pm0.7\,\Krj$ (no 3, 7, and 8).
Moreover, the reduced $\chi^2$ for the three cases shows a clear progression: 
$\chi^2_{\nu}=12.1$, 4.1, 0.97.
Inspection of Fig~\ref{fig:saturn_effect_of_rings} suggests that 
the reason for this anomaly resides in the fact that $\TbRJ$ for transits 7 and 8 are too low
when compared to the other transits. 
We have no explanation for this result because there were no background sources bright enough to disturb our measurements during those transits, and there were no obvious anomalies in the timelines. 
We note that the massive Saturnian storm was still visible in 2015
\citep{dePater:etal:aspc:2018}; however, without other independent observations to compare,
we decided to tag transits 7 and 8 as anomalous.
In the remaining discussion, transits 3, 7, and 8 are not used in the fit.

\begin{table*}
  \centering
  \footnotesize
  \caption{\label{tab:saturn:td:tr:by:channel}
    Channel-averaged $\Tring$, $\Tdisk$, $\TbdiskMono$ and $\TbdiskBA$ for Saturn from transits 1, 2, 4, 5, and 6.}
  \begin{tabular}{llcccc}
    \hline \hline 
    Channel & $\Fcent$ & $\Tring$ & $\Tdisk$ & $\TbdiskMono$ & $\TbdiskBA$ \\
            & $[\mathrm{GHz}]$ & $[\Krj]$ & $[\Krj]$ & $[\mathrm{K}]$ & $[\mathrm{K}]$ \\
    \hline 
    30 & $28.43$  & $9.21  \pm 1.39$  & $139.95  \pm  1.07$  & $140.64  \pm  1.07$  & $139.74  \pm  1.06$  \\
    44(24) & $44.23$  & $13.60  \pm 1.58$  & $147.29  \pm  0.62$  & $148.35  \pm  0.62$  & $147.82  \pm  0.62$  \\
    44(25--26) & $44.06$  & $11.59  \pm 2.27$  & $147.24  \pm  1.27$  & $148.30  \pm  1.27$  & $147.81  \pm  1.27$  \\
    70 & $70.46$  & $16.18  \pm 0.74$  & $150.22  \pm  0.37$  & $151.95  \pm  0.26$  & $151.02  \pm  0.26$  \\
    \hline \hline 
  \end{tabular}
\end{table*}

Table~\ref{tab:saturn:td:tr:by:channel}
gives the list of fitted $\Tring$, $\Tdisk$
for each channel excluding transits 3, 7, and 8.
As for the other planets, 
disc RJ brightness $\Tdisk$ 
are also converted to 
$\TbdiskMono$ which are equivalent to a $\TbMono$ for the other planets,
and to $\TbdiskBA$ which are equivalent to 
a $\TbBA$.
The former 
are derived from $\Tdisk$ using  
$\TbdiskMono=\invBnu(\Fcent,\BrjOneAver\Tdisk)$, 
where
$\BrjOneAver=\sum_{r,t} w_{t,r} \BrjOneRad/\sum_{t,r} w_{t,r}$ 
and $w_{t,r}$ are the weights per transit and radiometer $(t,r)$
used to derive the $\TbRJ$ in 
Table~\ref{tab:saturn:tbrj} and
the latter are obtained through a numerical inversion of 
$\Tdisk= \int d\nu\;\tilde{F}(\nu)\;\Bnu(\nu,\TbdiskBA)$,
where
$\tilde{F}(\nu) = \sum_{t,r} w_{t,r} \BandPass_r(\nu)/\BrjOneRad/\sum_{t,r} w_{t,r}$.
As our starting point is 
Table~\ref{tab:saturn:tbrj}, $\Tdisk$ is already corrected for blocking, we verified
that adding the blocking correction to $\TbRJ$ has a minor effect on $\Tring$ 
compared to the errorbars.
As usual we estimated errors using error propagation, bootstrap, and Monte Carlo simulations. We find a good 
agreement between the three methods,
but where differences were relevant, we quoted the largest one.
The effect of $\effEtaBeam$ is equivalent to add a systematic uncertainty of 
$\pm0.35$\,K for the 30\,GHz channel,
$\pm0.13$\,K for the 44(24)\,GHz sub-channel,
$\pm0.26$\,K for the 44(25-26)\,GHz sub-channel,
and
$\pm0.46$\,K for the 70\,GHz channel at $\Tdisk$, $\TbdiskMono$ and $\TbdiskBA$.
For $\Tring$ the uncertainty is 
$\pm3.5\times10^{-3}$\,K for the 30\,GHz channel,
$\pm8.6\times10^{-2}$\,K for the 44(24)\,GHz sub-channel,
$\pm2.0\times10^{-2}$\,K for the 44(25-26)\,GHz sub-channel,
$\pm4.6\times10^{-2}$\,K for the 70\,GHz channel.

\begin{figure}
        \centering
        \includegraphics[width=\columnwidth]{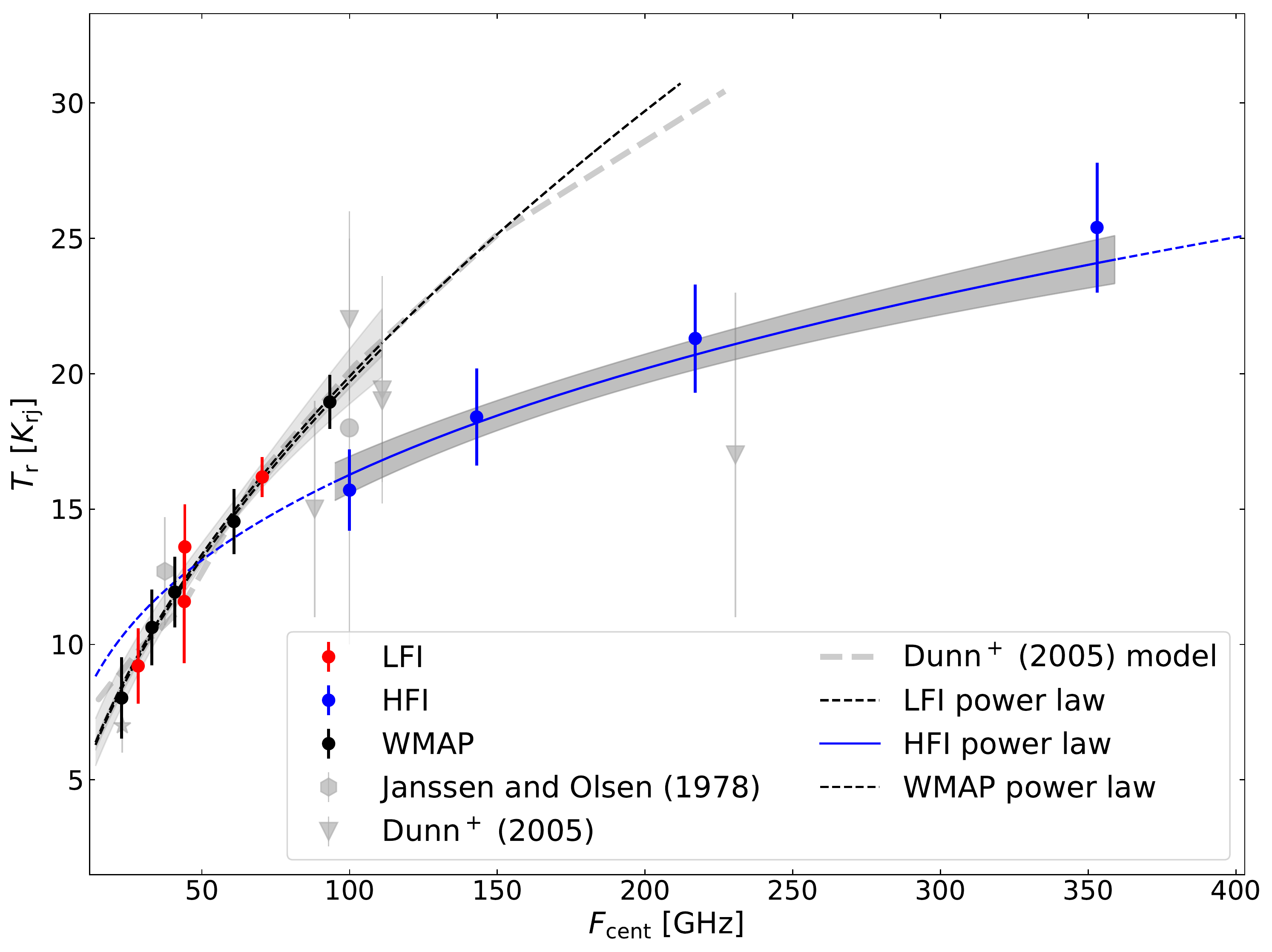}\\
        \caption{ \label{fig:saturn_tring_vz_models}
                Spectral energy distribution of Saturn rings from \Planck/LFI (red), \Planck/HFI (blue), and WMAP (black).
                Data from previous literature outside WMAP and \Planck{} are shown in light grey
                (references in the text).
        }
\end{figure}

Figure~\ref{fig:saturn_tring_vz_models} compares $\Tring$ in 
Table~\ref{tab:saturn:td:tr:by:channel}
with results from 
\citet{WMAP:PLANETS:2011}, \citet{WMAP:PLANETS:2013} and \citet{planck.intermediate.52.planet.flux.densities}.
Data from literature are
presented as grey marks \citep{Janssen:Olsen:1978,Schloerb:etal:1979a,Schloerb:etal:1979b,Epstein:etal:1980,Dunn:etal:2005}.
Moreover, we compared our results with the model of \citet{Dunn:etal:2005}.
Our estimates for Planck/LFI compare well with the other available data.
In particular, both \Planck/LFI and WMAP measurements fit well the
result of \citet{Janssen:Olsen:1978} near 40\,GHz and with 
\citet{Dunn:etal:2005} at 100\,GHz and 110\,GHz.
Also both for the case of Planck/LFI and WMAP 
$\Tring$ matches a power law of the form $\Tring=A \Fcent^\alpha$,
consistent with the model of \citet{Dunn:etal:2005}.
For WMAP 
$A=(1.35 \pm 0.12)\,\Krj$ and $\alpha=0.58\pm0.02$,
while for
\Planck/LFI
$A=(1.36 \pm 0.55)\,\Krj$ and $\alpha=0.58\pm0.10$
.
It is interesting to note that \Planck/HFI data exhibits a less steep power law with $A=(3.88 \pm 0.92)\,\Krj$ and $\alpha=0.311\pm0.044$, thereby predicting a lower $\Tring$ at 
70\,GHz and a higher $\Tring$ at 30\,GHz; the agreement at 44\,GHz is much more significant. 
When taken together, Planck/LFI, WMAP, and \Planck/HFI seem to suggest a change in slope around 
100\,GHz, but the only measure in significant disagreement with \Planck/HFI seems to be that from WMAP 90\,GHz. The remaining data from literature are not sufficiently accurate
to make a decision. This could be an interesting point for a future observing campaign in the 50--150\,GHz frequency range.

\begin{figure*}
        \centering
        \includegraphics[width=0.85\textwidth]{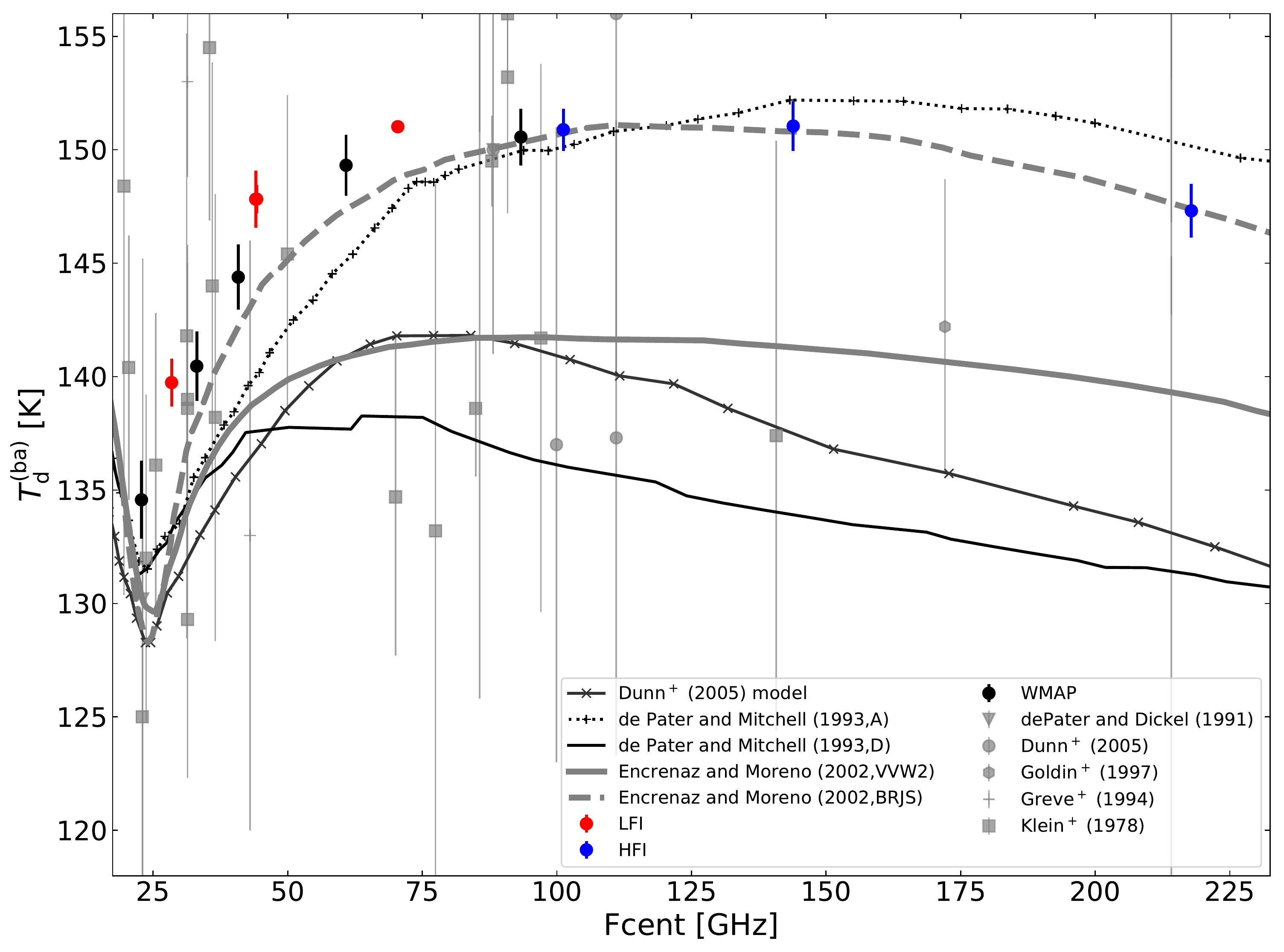}\\
        \caption{ \label{fig:saturn_tdisk_vz_models}
                Spectral energy distribution of Saturn disc from \Planck/LFI (red), \Planck/HFI (blue) and WMAP (black).
The model labelled dePater and Mitchel (1993,A) does not include clouds,
                while dePater and Mitchel (1993,D) includes clouds. 
                For other details about the other models, see the text.
        }
\end{figure*}

Figure~\ref{fig:saturn_tdisk_vz_models}
compares $\TbdiskBA$ for \Planck/LFI with WMAP 
 \citep{WMAP:PLANETS:2011,WMAP:PLANETS:2013}
and \Planck/HFI
 \citep{planck.intermediate.52.planet.flux.densities}.
There is a good agreement between the three datasets, 
even though our data seem to prefer a slightly warmer disc than those of WMAP.
Older data taken from literature
\citep{Klein:etal:1978,dePater:Dickel:1991,Greve:etal:1994,Goldin:etal:1997,Dunn:etal:2005}
\footnote{
Data presented in
\citet{Schloerb:etal:1979a}, 
\citet{Epstein:etal:1980}, 
\citet{Cunningham:etal:1981} and 
\citet{Grossman:etal:1989}
are outside our frequency range.
}
are very sparse in frequency coverage and exhibit wider error bars. 
In the frequency interval
50--150\,GHz, measurements from the literature span the range
135--160\,K, but  most of the measurements are in the lower side of the interval,
while measurements from WMAP and \Planck{} favour the upper side.
The reason could be in the absolute calibration of those old observations, as an absolute calibration error of the order
of ten percent is often quoted in these works, and observations are not usually calibrated 
against the same sources. This is the opposite of 
WMAP and \Planck{}, which share the same calibration.

Most of the models proposed in literature underestimate the 
combined WMAP and \Planck{} data; some of those models are presented in 
Fig.~\ref{fig:saturn_tdisk_vz_models}.
In all these models, the atmosphere is assumed to have 
abundances of 
NH$_3$, H$_2$O, H$_2$S, CH$_4$ enhanced with respect to the Sun by a factor of 3, 5, 11 and 5, respectively.
The first model presented in the figure \citep[][labelled ``de Pater and Mitchel (1993,A)'']{dePater:Mitchell:1993} 
does not include any contribution from cloud absorption, 
and this model underestimates the observed brightness below 90\,GHz; above this frequency, the first model matches our data, but overestimates the majority of the older measurements. The inclusion of 
clouds with NH$_3$ ice, H$_2$O liquid, and ice, NH$_4$SH ice
leads to the model labelled as ``de Pater and Mitchel (1993,D)'', which underestimates our measurements.
Similar behaviour appears with the models in
\citet{vanderTak:etal:1999}, with abundances of
NH$_3$, H$_2$O, H$_2$S, CH$_4$ enhanced by a factor of 1.9, 4, 11, and 4 with respect to solar values (not shown, for brevity), and in \citet{Dunn:etal:2005}, 
which is an improved version of the nominal model in \citet{dePater:Mitchell:1993}.
\citet{Encrenaz:Moreno:2002} 
 proposed two models with two different profiles for the extinction of the NH 1.28\,cm line:
 BRJS \citep{BenReuven:1966,Joiner:Steffes:1991} 
 and 
 VVW2, a Van Vleck-Weisskopf profile  
 \citep{dePater:Massie:1985,Lellouch:Destombes:1985,Moreno:1998}.
The authors favour VVW2, since it fits the data in literature better, 
 with the caveat that its model line profile heavily underestimates 
 \Planck{} and WMAP data, while the BRJS fits them much better.
We note that neither 
  \citet{dePater:Mitchell:1993} nor \citet{Dunn:etal:2005} include the  
  PH$_3$ absorption band at 263\,GHz, which is instead present in 
  the last two models.

In
 \citet{planck.intermediate.52.planet.flux.densities},
a model 
 named `ESA2 model'
 was used to compare \Planck{}/HFI results with WMAP and earlier \Planck{}/LFI results.
The predicted brightness temperature 
 is very similar to that predicted by the model in
 \citet{Encrenaz:Moreno:2002}, with 
 VVW2 profile.
The work also provides uncertainty limits; in particular, the upper limit is very similar
 to the Encrenaz \& Moreno model with 
 BRJS profile; this upper limit
 fits both WMAP and \Planck{} data.
Unfortunately no references or details are given about this model,
 and for this reason it is not presented in this work.
 
\begin{figure*}
\begin{center}
\includegraphics[width=0.75\textwidth]{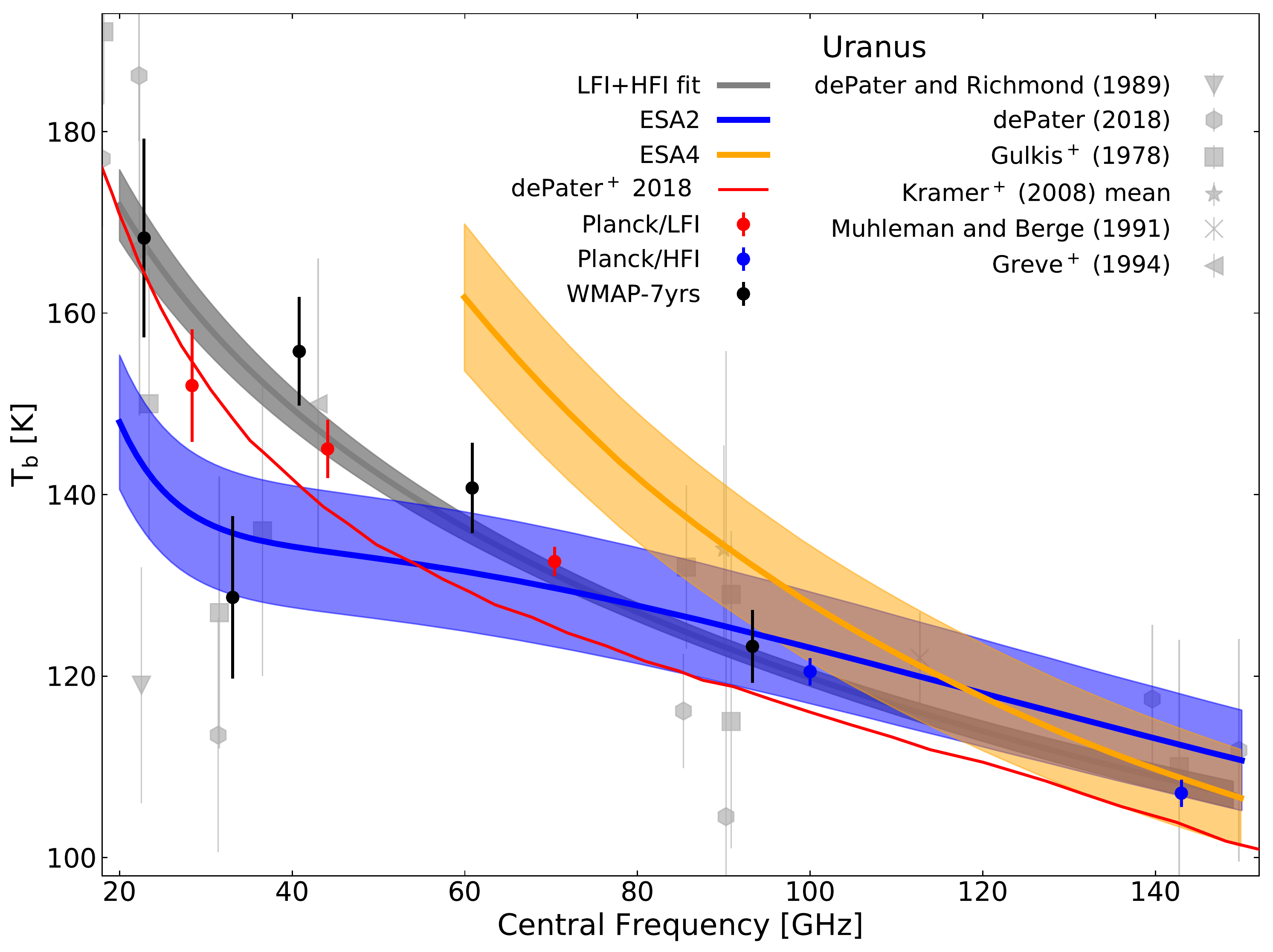}
\includegraphics[width=0.75\textwidth]{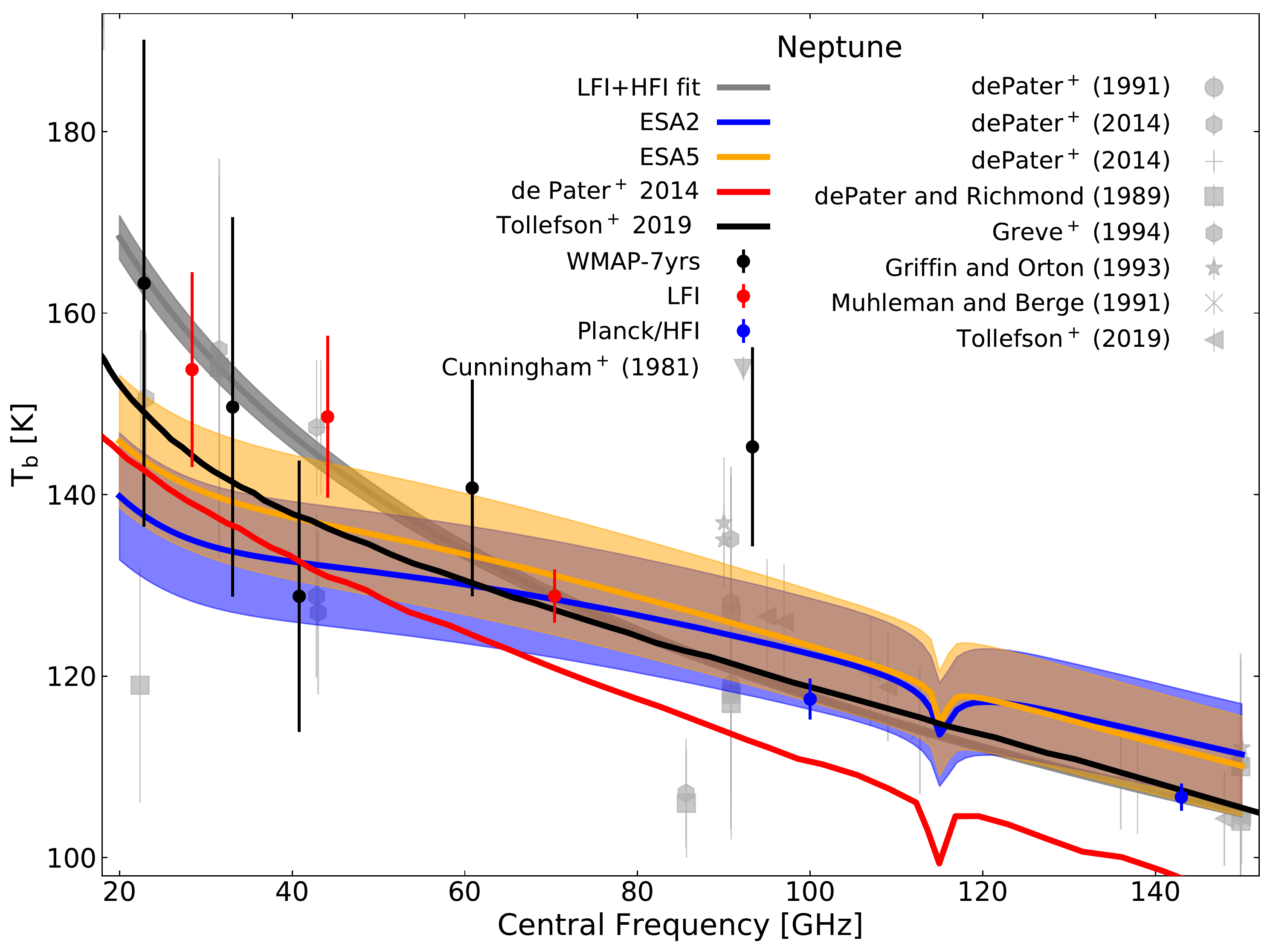}
\caption{ 
\label{fig:uranus:neptune:vz:models}
(Top) $\TbBA$ for Uranus and (bottom) Neptune, compared with representative models.
}
\end{center}
\end{figure*}

\subsection{Uranus and Neptune}\label{sec:Uranus:Neptune}

\begin{table*}\centering
  \footnotesize
  \caption{\label{tab:uranus:observing:conditions}
    Observing conditions of Uranus per transit.
  } 
 \begin{tabular}{rlrrrrrrrrrr}
   \hline\hline
   Transit&Epoch&PJD\_Start&PJD\_End&Nsmp&EcLon&EcLat&GlxLat&$\Rsun$&$\Delta$&$\AngDiam$&$\PlanetAspectAngle$\\
& & & & &[deg]&[deg]&[deg]&[AU]&[AU]&[arcsec]&[deg]\\
\midrule
1& 2009-12-09&     205.26&   214.03& 10727& 352.6& $-0.8$&  $-60.1$& 20.10& 20.01& 3.52&  5.21 \\
2& 2010-07-02&     410.43&   418.96& 10510&   0.5& $-0.8$&  $-60.9$& 20.09& 19.93& 3.54& 13.38 \\
3& 2010-12-14&     575.50&   584.31& 10914& 356.5& $-0.8$&  $-60.8$& 20.09& 20.02& 3.52&  9.32 \\
4& 2011-07-07&     780.66&   788.84& 10081&   4.4& $-0.7$&  $-60.6$& 20.08& 19.90& 3.54& 17.46 \\
5& 2011-12-25&     952.75&   958.41&  6487&   0.6& $-0.7$&  $-60.9$& 20.08& 20.11& 3.51& 13.48 \\
6& 2012-07-09&    1150.09&  1155.72&  6683&   8.4& $-0.7$&  $-59.8$& 20.07& 19.90& 3.54& 21.53 \\
7& 2012-12-27&    1321.50&  1327.14&  6762&   4.5& $-0.7$&  $-60.6$& 20.06& 20.08& 3.51& 17.55 \\
8& 2013-07-13&    1518.76&  1524.31&  6588&  12.3& $-0.7$&  $-58.6$& 20.05& 19.89& 3.54& 25.58 \\
\bottomrule
\end{tabular}
\end{table*}

 \begin{table*}
  \centering
  \footnotesize
  \caption{\label{tab:neptune:observing:conditions}
    Observing conditions of Neptune per transit.
  }
  \begin{tabular}{rlrrrrrrrrrr}
    \hline\hline
    Transit& Epoch& PJD\_Start& PJD\_End& Nsmp& EcLon& EcLat& GlxLat& $\Rsun$& $\Delta$& $\AngDiam$& $\PlanetAspectAngle$\\
           &  &  &  &  & [deg]& [deg]& [deg]& [AU]& [AU]& [arcsec]& [deg]\\
    \midrule
    1& 2009-11-03&  171.04&   177.90&  8439& 323.5& $-0.4$& $-44.5$& 30.03& 29.83& 2.29& $-28.80$\\
    2& 2010-05-19&  366.56&   374.99& 10239& 328.5& $-0.5$& $-48.0$& 30.02& 30.04& 2.27& $-28.43$\\
    3& 2010-11-06&  538.33&   545.34&  8705& 325.7& $-0.5$& $-46.1$& 30.02& 29.81& 2.29& $-28.61$\\
    4& 2011-05-22&  734.34&   742.96& 10462& 330.7& $-0.5$& $-49.6$& 30.01& 30.03& 2.27& $-28.14$\\
    5& 2011-11-19&  917.18&   922.35&  6192& 328.0& $-0.6$& $-47.8$& 30.00& 29.98& 2.28& $-28.38$\\
    6& 2012-06-05& 1115.51&  1122.59&  8597& 333.0& $-0.6$& $-51.1$& 30.00& 29.79& 2.29& $-27.79$\\
    7& 2012-11-23& 1286.19&  1293.91&  9436& 330.2& $-0.6$& $-49.3$& 29.99& 30.02& 2.27& $-28.10$\\
    8& 2013-06-08& 1483.26&  1490.15&  8343& 335.2& $-0.7$& $-52.5$& 29.99& 29.78& 2.29& $-27.41$\\
    \bottomrule
  \end{tabular}
\end{table*}

\begin{table}
  \centering
  \footnotesize
  \caption{\label{tab:uranus:channel:average}
    Channel-averaged results for Uranus.}
  \begin{tabular}{cr@{$\pm$}rr@{$\pm$}rr@{$\pm$}rr@{$\pm$}r}
    \toprule
    $\Fcent$& 
    \multicolumn{2}{c}{$\BrBAplanet$} &
    \multicolumn{2}{c}{$\TbRJ$} & \multicolumn{2}{c}{$\TbMono$} & \multicolumn{2}{c}{$\TbBA$} \\
    {[GHz]}&
    \multicolumn{2}{c}{[MJy/sr]} &
    \multicolumn{2}{c}{[K]} & \multicolumn{2}{c}{[K]} & \multicolumn{2}{c}{[K]} \\
    \midrule
28.4 &  3781.1 &       155.8 &  152.3 &        6.3 &    153.0 &          6.3 &  152.0 &        6.2 \\
44.1 &  8648.3 &       193.3 &  144.5 &        3.3 &    145.5 &          3.3 &  145.0 &        3.2 \\
70.4 & 20032.7 &       265.6 &  131.7 &        1.6 &    133.4 &          1.6 &  132.6 &        1.6 \\
    \bottomrule
  \end{tabular}
\end{table}

 \begin{table}
  \centering
  \footnotesize
  \caption{\label{tab:neptune:channel:average}
    Channel-averaged results for Neptune.}
  \begin{tabular}{cr@{$\pm$}rr@{$\pm$}rr@{$\pm$}rr@{$\pm$}r}
    \toprule
    \multicolumn{1}{c}{$\Fcent$} &
    \multicolumn{2}{c}{$\BrBAplanet$} & 
    \multicolumn{2}{c}{$\TbRJ$} & \multicolumn{2}{c}{$\TbMono$} & \multicolumn{2}{c}{$\TbBA$} \\
    \multicolumn{1}{c}{[GHz]}&
    \multicolumn{2}{c}{[MJy/sr]}&
    \multicolumn{2}{c}{[K]} & \multicolumn{2}{c}{[K]} & \multicolumn{2}{c}{[K]} \\
28.4 &        3827.2 &             267.1 &  154.1 &       10.8 &    154.8 &         10.2 &  153.8 &       10.8 \\
44.1 &        8852.7 &             538.0 &  148.0 &        9.0 &    149.1 &          9.0 &  148.6 &        8.9 \\
70.4 &       19445.7 &             458.5 &  127.9 &        2.9 &    129.6 &          2.9 &  128.2 &        2.9 \\
    \midrule
    \bottomrule
  \end{tabular}
\end{table}

Tables \ref{tab:uranus:observing:conditions} and \ref{tab:neptune:observing:conditions} describe the observing conditions for Uranus and Neptune.
There are eight transits in which both planets are observed; all the cases transits occurred far from the Galactic plane.
For both Uranus and Neptune, the signal is very weak, especially at 30\,GHz and 44\,GHz, and they are therefore difficult to detect.
For Uranus, $\DeltaTantPstar$ is in the range 
$6\times10^{-5} \div 5\times10^{-4}$~$\Kcmb$ at 30\,GHz,
$3\times10^{-4} \div 9\times10^{-4}$~$\Kcmb$ at 44\,GHz,
and
$7\times10^{-4} \div 3\times10^{-3}$~$\Kcmb$ at 70\,GHz,
While for Neptune 
$\DeltaTantPstar$ is in the range 
$2\times10^{-5} \div 3\times10^{-4}$~$\Kcmb$ at 30\,GHz,
$4\times10^{-5} \div 5\times10^{-4}$~$\Kcmb$ at 44\,GHz,
and
$6\times10^{-5} \div 1.3\times10^{-3}$~$\Kcmb$ at 70\,GHz.
In some cases, the result of the fit is $\DeltaTantPstar \le 0$, which means that the planet is not detected:
for Uranus,
this occurs in transits 2 and 3 for horns 25 and 26, and in transit 3 for horns 24, 27, and 28;
for Neptune, this occurs in transit 3 for radiometers 25M and 28M. We decided not to include these data in our analysis.

Figure~\ref{fig:synoptic_of_tb_ba_transit_channel_averages}
shows that the scatter in the channel averages for each transit are consistent with the error bars, except for $\TbBA$ at 30\,GHz in transit 2, and at 70\,GHz in transit 3. We removed these two data points before computing the channel averaged results presented in Table~\ref{tab:uranus:channel:average} (Uranus) and in Table~\ref{tab:neptune:channel:average} (Neptune).
It is known that Uranus has a significant time variability in microwave over a timescale
of decades, mainly connected to the change in the 
$\PlanetAspectAngle$ of the observation
\citep{Klein:Hofstadter:2006,Kramer:etal:2008}. 
A relative variation of $0.5\%/\mathrm{year}$ or $0.1\%$ at 90\,GHz for one degree of variation of $\PlanetAspectAngle$ was reported by \citet{Kramer:etal:2008}.
Assuming that the same numbers are valid 
at 70\,GHz and a time span between our observations of about 3.6 years, corresponding to a span 
of about $20.4^{\circ}$ in $\PlanetAspectAngle$,
we expect to detect a change of the order of 2--2.6\,\%.
As shown in 
Fig.~\ref{fig:synoptic_of_tb_ba_transit_channel_averages},
the scatter in our data is larger than this effect; therefore, we decided not to consider it.

Figure~\ref{fig:uranus:neptune:vz:models} 
compares \Planck/LFI\ Uranus and Neptune $\TbBA$ measurements with 
\Planck/HFI\ measurements at 100\,GHz and 143\,GHz from 
\citet{planck.intermediate.52.planet.flux.densities}, and WMAP seven-year measurements converted to 
$\TbBA$ .
In addition we show a selection of 
measurements published in the past literature.
The data for Uranus are taken from 
\citet{Gulkis:etal:1978,dePater:Richmond:1989}, \citet{Muhleman:Berge:1991}, \citet{Greve:etal:1994} and \citet{dePater:2018} 
\footnote{Data from 
        \protect\citet{Cunningham:etal:1981}, \citet{Griffin:Orton:1993} and  \citet{Klein:Hofstadter:2006}
        are not shown here since they are outside the frequency range of interest.};
for \citet{Kramer:etal:2008}; we only show the average of the data.
For Neptune data are taken from 
\citet{Cunningham:etal:1981}, \citet{dePater:Richmond:1989}, \citet{dePater:etal:1991}, \citet{Muhleman:Berge:1991}, \citet{Griffin:Orton:1993}, \citet{Greve:etal:1994,dePater:etal:2014} and \citet{Tollefson:etal:2019}.
Our measurements for both Uranus and Neptune are in general agreement with WMAP, \Planck/HFI, and the past literature.

In contrast to Jupiter and Saturn, where the 1.3\,cm $\Ammonia$ inversion line and 
$\Phosphine$ transitions 
significantly affect the spectrum, it has been observed that 
Uranus and Neptune spectra in the 20--150\,GHz 
smoothly decrease with frequency
\citep{dePater:etal:1991,Encrenaz:Moreno:2002}.
A simple law can be used to transfer observations from one frequency to the other,
similar to the fourth-order polynomial in $\log_{10} \nu$ provided by \citet{Griffin:Orton:1993};
in the range of frequencies of interest for this work, this 
can be reduced to the simpler form 
\begin{equation}\label{eq:uranus:neptuns:scaling}
\Tb = A\log_{10}\left(\frac{\nu}{100 \; \mathrm{GHz}}\right) + B,
\end{equation}
with $A$ and $B$ free parameters.
For Uranus
$A=-74.5 \pm 31.6 \,\mathrm{K}$, $B=118.9 \pm 0.9,\mathrm{K}$,
and for Neptune
$A=-72.3 \pm 9.9 \,\mathrm{K}$, $B=117.8 \pm 0.2 \,\mathrm{K}$.
The corresponding fit is shown as a grey band in the figures. 
For Uranus, the discrepancy between the 70\,GHz datum and WMAP is $\sim2\,\mathrm{K}$, a bit more than $1\,\%$.
On the contrary, for Neptune the WMAP V and W bands overestimate both our number at 70\,GHz and the \Planck/HFI estimate
at 100\,GHz.  

At odds with the expected smooth $\Tb$ variation with frequency,
the WMAP team noted that $\Tb$ in their
Ka band drops with respect to their K and Q bands \citep{WMAP:PLANETS:2011};
the significance of this drop is reinforced by comparison with 
other observations in the literature.
Our results does not disconfirm 
this finding,
as
our measure at 44\,GHz agree with WMAP Q band, while the 30\,GHz datum
is between the WMAP K and Ka bands.
The combination of all of the observations suggests the presence of a drop in the thermal 
emission of Uranus, which is centred at about 30\,GHz, of about 4--5\,GHz and a depth of 20--50\,K.
However, the uncertainty in the magnitude of the drop is very large; 
data in literature have wide error bars and use different calibrations. Therefore, we think that more data are needed to validate the existence of this spectral feature.

Models of the microwave emission of Uranus and Neptune are available in 
\citet{Griffin:Orton:1993}, \citet{Kramer:Moreno:Greve:2008}, \citet{Griffin:etal:2013} and \citet{Bendo:etal:2013}.
In this work, we consider the models in \citet{dePater:2018} for Uranus and in
\citet{dePater:etal:2014} and \citet{Tollefson:etal:2019} for Neptune, together with the ESA models used for the calibration of {\tt Herschel}; the latter has a quoted $5\,\%$ uncertainty
\citep{Moreno:1998,Herschel:Calibration}
\footnote{\url{https://www.cosmos.esa.int/web/herschel/calibrator-models}.}.
They are included in our figure because they are important for the inter-calibration between 
{\tt Herschel} and \Planck{}
\citep{Bertincourt:etal:2016,Muller:etal:2016,planck.intermediate.52.planet.flux.densities}.

The model of \citet{dePater:2018} for Uranus assumes 
abundances of $\water$, $\SulfidricAcid$, $\Methane$ 
enhanced of a factor 10
with respect to solar abundances of O, S, and C. Ammonia is kept at solar N abundance and
is captured in $\mathrm{NH}_4\mathrm{SH}$ clouds; therefore, it is depleted above the corresponding atmospheric layer. According to this model, at our frequencies
opacity is mainly due to $\SulfidricAcid$\ absorption and 
collisionally-induced absorptions from $\HidrogenTwo$.
The figure shows that the model essentially matches our scaling law 
and only slightly underestimates our 44 GHz and 70\,GHz data; it does not predict any 
 drop around 30\,GHz.

The ESA2 model for Uranus is an updated version of the model in
\citet{Moreno:1998} used for the calibration of \Herschel{}.
This model was used to validate \Planck/HFI data and for comparison with WMAP
\citep{planck.intermediate.52.planet.flux.densities}.
Our 70\,GHz measure is in close agreement with this model, and even the 30\,GHz datum agrees 
with this model.
It is interesting to note that the model predicts a decrease of signal  
around 30\,GHz, but it fails to follow the pattern of the spectrum below 45\,GHz. 
Unfortunately, few details are provided for this model, 
so it is not possible to push the analysis further.
After the release of the ESA2 model, the \Herschel{} collaboration proposed another
model of the spectrum of Uranus named ESA4, which includes
observations from Spitzer \citep{Orton:etal:2014} and extends down to 60\,GHz. It is evident that the model significantly overestimates
the brightness below 100\,GHz.

For Neptune, we considered the model in \citet{dePater:etal:2014}. The model featured abundances of $\SulfidricAcid$, $\water,$\ and $\Methane$\ enhanced by a factor of 30 with respect to solar abundances of S, O, and C, and a wet lapse rate. The model matches our scaling law but underestimated $\Tb$ of about 10\%.
We included in our analysis the model in \citet{Tollefson:etal:2019}\footnote{The model labelled {`30$\times$S dry'} in their Fig.~3.}. It features an abundance of $\SulfidricAcid$, $\Methane$, and $\water$ that is 30 times the proto-solar abundance. The model fits the data for \Planck/LFI 70\,GHz and \Planck/HFI 100\,GHz and 144\,GHz very well, but it underestimates the measurements at 30\,GHz and 44\,GHz. 
The ESA2 and ESA5 models are also shown in the figure as blue and orange lines, respectively. ESA5 was used to validate the \Planck/HFI data, but it has been not used for the final calibration of \Herschel\footnote{According to the {\tt Readme} file in the \Herschel{} models repository.} As already noted in \citet{planck.intermediate.52.planet.flux.densities}, the model slightly overestimated the brightness temperature in the range 70--200\,GHz and underestimated our 30\,GHz and 44\,GHz data. It is interesting to note the good agreement between ESA2 and our measurement at 70\,GHz. For Neptune, both ESA2 and ESA5 models marginally fit our results above 70\,GHz, and ESA5 looks slightly better than ESA2. 

\section{Conclusions}
\label{sec:conclusions}

We analysed the data in the \Planck\ 2018 public data release to characterize the emission of the planets Jupiter, Saturn, Uranus, and Neptune in the frequency range 30--70\,GHz, using all the data acquired by the LFI instrument during its four-year lifetime (August 2009--October 2013). 
In each transit, a planet was observed by \Planck/LFI for a few hours rather than days or weeks as in the case of WMAP. Within each transit, the cumulative integration time was about few seconds per planet, transit, and radiometer.
The LFI observed Jupiter seven times and the other planets eight times. 
In the past, just part of those transits were fully analysed in a self-consistent manner. On the contrary, we treated all the observations in a fully homogeneous manner. Moreover, we used our improved knowledge of the beam and bandpasses to refine our earlier analysis.

In the case of Jupiter and Saturn, the sensitivity of \Planck/LFI allowed us to reduce the impact of instrumental white noise to a small amount, and the dispersion of our measurements within each frequency channel after geometrical corrections shows a residual variability that is larger than the noise. Calibration uncertainties on individual radiometers could be a source of such variability: in particular, we cannot exclude that part of the radiometer-by-radiometer variability we observed in our sample of Jupiter's observations could be connected to uncertainties in the model of the bandpass of each radiometer. This could introduce small differences between the calculated radiometers central frequencies, bandwidths, or higher order bandpass moments and the real ones.
In principle, by comparing measures of a bright source, such as Jupiter, with a well-calibrated model or set of measures from another instrument, it would be possible to derive a correction for this effect. But if blindly applied, this method forces every other possible residual systematic in this correction. For this reason and because of a lack of a sufficiently accurate model of the emissivity of Jupiter, including non-thermal emission, we did not attempt to derive this kind of correction in this work. We guess that improved calibration methods, such as those described in \citet{BeyondPLANCK:I}, will improve this result.

Despite the difference in the time span of observations, our results are directly comparable 
to WMAP observations \citep{WMAP:PLANETS:2011,WMAP:PLANETS:2013}, which were
obtained in a similar range of frequencies. In particular, we confirm the good agreement between the \Planck/LFI and WMAP estimates of the SED of Jupiter. Our results improve the frequency coverage in the range 20--90\,GHz. A comparison with existing models below 70\,GHz allowed us to estimate Jupiter's synchrotron contribution in the 30\,GHz channel in the range 0.9--2.4\,Jy.
As \Planck/LFI and \Planck/HFI cover separate ranges of frequencies, to compare these values we had to rely on far-infrared emissivity models for giant planets. The result of our comparison shows a good agreement between the measurements of the two instruments.

Our estimates for Saturn's disc SED agree with WMAP at 30\,GHz, but our results favour a slightly warmer disc at 44\,GHz and 70\,GHz. 
With the present knowledge of the instrument it is not possible to assess whether the difference is due to some systematic in either Planck/LFI or WMAP, or if it is connected to the fact that WMAP observations were centred at negative planetocentric latitudes, while Planck/LFI observations were centred at positive latitudes. 
Given the large error bars of older measurements in literature, we can only say that our measures agree with most of the older measurements.
We compared our Saturn's measurements with known models published in the literature. All but two significantly underestimated the SED in the frequency range considered here.
About rings, we may note the excellent agreement of \Planck/LFI with both WMAP and existing estimates for frequencies below 100\,GHz.
Data below 100\,GHz show some discrepancy with \Planck/HFI, but
the existing data does not allow us to assess the significance of this mismatch.

Measures for Uranus and Neptune have very low S/N. For some transit and/or radiometer, confusion noise prevented a proper detection: consequently, error bars are larger than for Jupiter and Saturn, although they agree with those in literature.
In particular, our results are in agreement with WMAP and \Planck/HFI at 100\,GHz and 143\,GHz.
We compared a selection of existing models for the microwave emissivity of Uranus and Neptune with our data plus WMAP and \Planck/LFI. These comparisons show a good agreement in the 100--143\,GHz range, but significant discrepancies below 100\,GHz. 
In particular, the Uranus model presented in \citet{dePater:2018} and the Neptune model found in \citet{Tollefson:etal:2019} show a better agreement with our data.
For observers willing to use these planets as calibrators, we advise that a simple power law is very good at modelling the dependence on $\Tb$ for Uranus and Neptune in the frequency range 20--143\,GHz within the current error bars.

In earlier generations of CMB experiments, giant planets have been considered good beam calibrators and have been used as calibration sources between different experiments, 
thanks to their high S/N 
\citep{WMAP:PLANETS:2011,WMAP:PLANETS:2013,planck.2013.05.LFI.calibration,planck.2015.05.LFI.calibration}. 
Planetary observations will likely maintain the same importance in future missions, such as the planned LiteBIRD mission \citep{LiteBIRD:2019}.
The increased demand for accurate and sensitive CMB measurements will necessarily require more accurate models for the analysis of planetary emission in the microwave range. Outer planets, in particular Jupiter and Saturn, have complex spectra and no simple scaling law will work, especially when combining data from detectors with different bandpasses. In this case, people should use reliable models of planetary emissivities (including both thermal and non-thermal components); however, current models have uncertainties that are larger than the measurement errors. 
An observing campaign with ground-based instruments, coupled with progresses in modelling, could solve this problem.

\begin{acknowledgements}
This work is based on the 2018 Release of \Planck/LFI data.
The writers acknowledges the support of: \Planck{} Collaboration, ESA, ASI, CNR, Universit\`a di Milano and INAF (Italy); CNES and CNRS/INSU-IN2P3-INP (France);  NASA and DoE (USA); STFC and UKSA (UK); CSIC, MINECO, JA, and RES (Spain); Tekes, AoF, and CSC (Finland); DLR and MPG (Germany); CSA (Canada); DTU Space (Denmark); SER/SSO (Switzerland); RCN (Norway); SFI (Ireland); FCT/MCTES (Portugal); and ERC and PRACE (EU). A description of the Planck Collaboration and a list of its members, indicating which technical or scientific activities they have been involved in, can be found at \url{http://www.cosmos.esa.int/web/planck/}.
This work have been partially supported by INAF/Trieste Astronomical Observatory through ``Ricerca di Base 2019'' F.U. 1.05.01.01.
M.~M. acknowledges Maria Teresa Capria for useful suggestions about models of planetary atmospheres.
The authors acknowledge the staff of the Library of the INAF/Trieste Astronomical Observatory: Laura Abrami and Chiara Doz, for their kind help in recovering the needed bibliography during the lockdown imposed in Italy by the COVID 19 pandemic.
The authors thank the two anonymous referees for their suggestions, which helped to considerably improve the paper.
This research made use of 
{\tt Astropy},\footnote{http://www.astropy.org} a community-developed core {\tt Python} package for astronomy \citep{astropy:2013,astropy:2018};
{\tt EMCEE}\footnote{https://github.com/dfm/emcee} \citep{emcee} {\tt HEALPix}\footnote{https://healpix.jpl.nasa.gov} \citep{Healpix}; 
{\tt HEALPy}\footnote{https://github.com/healpy/healpy} \citep{Healpy}; 
{\tt IPython}\footnote{https://ipython.org} \citep{PER-GRA:2007};
{\tt MatPlotLib}\footnote{https://matplotlib.org} \citep{matplotlib};
{\tt NumPy}\footnote{https://numpy.org} \citep{2020SciPy-NMeth}; 
{\tt Pandas}\footnote{https://pandas.pydata.org} \citep{mckinney-proc-scipy-2010}.
{\tt PyQuarantine}\footnote{https://www.ict.inaf.it/gitlab/michele.maris/pyquarantine.git} 
{\tt SciPy}\footnote{https://www.scipy.org} \citep{2020SciPy-NMeth}; 
{\tt WebPlotDigitizer}\footnote{https://automeris.io/WebPlotDigitize} \citep{mckinney-proc-scipy-2010}.
\end{acknowledgements}
 
\appendix

\section{Technical aspects of the data analysis procedure}
\label{sec:technicalInformation}

In this appendix, we provide more information about some of the most technical issues we have tackled to produce estimates for brightness temperatures.

\subsection{Selection of samples and ROI}\label{sec:selection:of:samples}

We used the {\tt Horizons} web service\footnote{\url{https://ssd.jpl.nasa.gov/?ephemerides}} to compute the apparent position of the planet for each sample in the time-ordered data acquired by the \Planck/LFI radiometers. Using these positions, we selected those samples within the stability period of each pointing period according to the following criteria: they are not flagged as bad and their pointing direction in the sky is within $5\deg$ from the planet (the ROI). This radius limits the amount of data to process to a reasonable amount, and enables full coverage of the angular size of the main beam; moreover, it is large enough to estimate the contribution of the background. The $5\deg$ angular size separates the intermediate beam region and the far side-lobe region, for which the \Planck{} collaboration provided \GRASP{} beam maps \citep{planck.2015.04.LFI.beams}.

We divided the $5\deg$ radius ROI into three concentric regions: the planet ROI is the ring with
radius $\ROIplanet = 1.3\,\fwhm$ of the beam used to estimate $\DeltaTantP$; the avoidance ROI is the annulus between $\ROIplanet$
and  $\ROIavoid= 2\,\fwhm$; and finally, the background ROI is everything within $\ROIavoid$ and $\ROI$.
Typical values for $\ROIplanet$ are about $\approx0.7\deg$, $\approx0.5\deg$,  $\approx0.3\deg$ at 30, 44, and 70\,GHz respectively,
while for $\ROIavoid\approx1.1\deg$, $\approx0.8\deg$, $\approx0.4\deg$. The number of samples in the planet ROI is in the range $10^3\div 10^4$; the number of samples in the background ROI
is in the range $10^5\div 10^6$. Owing to changes in the scanning strategy during the mission, the density of samples in the ROI largely changed among different transits. 
As an example, Fig.~\ref{fig:Tant:classification}
shows the classification and masking of data in the first transit for radiometer LFI27-0 (30\,GHz).

\subsection{Background modelling}
\label{sec:background:modelling}

\begin{table}
  \centering
  \footnotesize
  \caption{\label{sec:typical:tant}
    Range of variability$^a$ for $\DeltaTantP$, in $\mKcmb$.
  }
  \begin{tabular}{lr@{.}l@{ -- }r@{.}lr@{.}l@{ -- }r@{.}lr@{.}l@{ -- }r@{.}l}
    \hline
    Planet& \multicolumn{4}{c}{30\,GHz}& \multicolumn{4}{c}{44\,GHz}& \multicolumn{4}{c}{70\,GHz}\\
    \hline
    Jupiter& 38&5& 42&7& 54&2& 99&1& 307&7& 368&5\\
    Saturn& 6&1& 7&0& 8&2& 1&6& 45&5& 55&6\\
    Uranus& 0&06& 0&5& 0&3& 0&9& 0&7& 2&7\\
    Neptune& 0&02& 0&3& 0&04& 0&5& 0&06& 1&3\\
    \hline
  \end{tabular}
  \tablenoteskip\\
  \tablenote{a}{Results represents the distribution over the whole set of transits and radiometers for each channel.}
\end{table}
 
In the \citet{planck.2013.05.LFI.calibration} and \citet{planck.2015.05.LFI.calibration}, 
the background was modelled as a constant derived from the median of the background ROI.
The constant included contributions from diffused foregrounds, CMB, point sources, and zero-point differences among different radiometers. 
However, after having masked point sources, the typical RMS of the diffuse background ($\DeltaTantP$) in the background ROI is $\rmsBackground\approx10^{-4}\,\Kcmb$
Compared to $\DeltaTantPstar$ for the planets observed by \Planck/LFI (Table~\ref{sec:typical:tant}), it is evident that this fluctuation is equivalent to 
$\rmsBackground/\DeltaTantPstar \approx (0.6 \cdots 5) \times 10^{-3}$ of the Jupiter signal, which is negligible.
For weaker planets, background fluctuations are more relevant:
for Saturn $\rmsBackground/\DeltaTantPstar \approx (0.4\div 3) \times 10^{-2}$, 
for Uranus $\rmsBackground/\DeltaTantPstar \approx 0.1\div 0.6$,
and
for Neptune $\rmsBackground/\DeltaTantPstar \approx 0.2\div 1.4$.
Proper background removal is mandatory for all the planets but Jupiter.

To remove the background, we used the \Planck{} 2018 sky maps to build a timeline $b_{\mathrm{k},t}$ for each transit and each radiometer. These timelines were computed using bilinear interpolation on the sphere at each pointing direction $\Pointingt$. 
As for planets, we considered smearing as well (see Sect.~\ref{sec:smearing}).
Since sky maps refer to the central frequency of the channel, the simulated timelines $b_{\mathrm{k},t}$ do not account for 
differences in bandpasses among different radiometers.
To fix this, we introduced a scaling parameter 
$\BackgroundScaling$
and a zero point $\BackgroundZero$ in the fit, so that 
$b_t = \BackgroundScaling b_{\mathrm{k},t} + \BackgroundZero$.
We determined the parameters $\BackgroundScaling$ and $\BackgroundZero$ for each radiometer by fitting the background model against the samples in the background ROI.
Typical $\BackgroundScaling$ varies from 0.5 to 1.3, while $\BackgroundZero$ varies within $\pm 0.2\, \mKcmb$.

Fig.~\ref{fig:Tant:classification} in the top right frame shows a background map derived from $b_t$. In the bottom left frame, we show the histogram of the background model using green dots, and we overlap a Gaussian distribution with the same mean and RMS (red line). The long tail in the right wing of the distribution is due to the bright source on the bottom left corner of the map. We used a simple $\sigma$-clipping, whose threshold is shown as a dashed blue line in the plot, to mask that region (bottom right part of the figure).

\subsection{Bandpasses and beam patterns}
\label{sect:bandpass:and:beam:pattern}

Equation~\eqref{eq:chisq:naive} shows that proper modelling of the beam shape is critical.
The results presented in this paper are based on the official band-averaged beam model, 
computed using \GRASP{}.
We derived a band-averaged map of the beam out of a set of gridded monochromatic maps, which were weighted according to the product between the the SED of the incoming radiation and the bandpass of the radiometer.
For planets, we used a $\nu^2$ SED, as it represents the SED of a planet emitting in the RJ regime.
To estimate $\gammabapt$ (Sect.~\ref{sec:estimationPlanetSignals}), we converted the instantaneous position of the planet in the $(u_t,v_t)$ coordinates using Eq.~\eqref{eq:pointing2uv}, and we recovered the beam response using bilinear interpolation.
There is a strong connection between the reduction of Jupiter observations and the estimation of the beam model because the former requires the latter, but the latter is usually validated through the former. Unfortunately, carrying on the two analysis tasks at the same time is prohibitive, owing to the computational time required by \GRASP{} to estimate beam maps. Therefore, in our analysis we had to assume the correctness of the \GRASP\  beam models produced by the \Planck/LFI collaboration.

\subsection{Smearing}\label{sec:smearing}

The signal acquired by \Planck/LFI radiometers was integrated over a discrete sampling time, $\tsmp$, which depended on the frequency of the detector (30, 44, or 70\,GHz) and was in the range 0.01--0.03\,s. In that time, the planet moved across the beam and causes smearing. 
Smearing smoothed the signal and it must be properly taken in account in data analysis because it reduced the value of $\DeltaTantPstar$ (about $1.4\%$ at 30\,GHz, $1.1\%$ at 44\,GHz, and 
$1.2\%$ at 70\,GHz).
The amount of smearing was constant, as $\tsmp$ for each radiometer was tuned to allow the beam to move by $\approx \fwhm/3$. 

To model the smearing effect, a common approach is to create a beam map by stacking and averaging a number of repetitions of the simulated \GRASP\ beam map, shifted along the direction of scan by a fixed amount. 
However, this approach does not account for the fact that the spin rate and the effective boresight angles can change during the mission. 
Therefore, we used a different strategy to deal with smearing. We over-sampled the modelled beam pattern along the path of the apparent motion of the planet in the beam reference frame and averaged the result.

To compute the planetary smearing for the $i$-th sample taken at time $t_i$, we took a triad of consecutive positions of the planet 
in the beam reference frame $(u,v)$ at times $t_{i-1}$, $t_i$ and $t_{i+1}$.
Given that the $u$ and $v$ directions are orthogonal, the motion is described by the equations
\begin{eqnarray}
u(l) &\approx& A_{u,i} l^2 + B_{u,i} l + C_{u,i},\\
v(l) &\approx& A_{v,i} l^2 + B_{v,i} l + C_{v,i},
\end{eqnarray}
where $l=(t-t_i)/\tsmp$, so that $l=-1, \, 0, \, +1$ for samples $i-1, \, i,\, i+1$, respectively. We derived the coefficients $A_{u,i}$, $B_{u,i}$, and  $C_{u,i}$ from the positions $u_{i-1}$, $u_i$, and $u_{i+1}$ using least-squares minimization. The result is
\begin{eqnarray}
A_{u,i}&=&\frac{u_{i+1}+u_{i-1}}{2} - u_{i},\\
B_{u,i}&=&\frac{u_{i+1}-u_{i-1} }{2},\\
C_{u,i}&=&u_{i},
\end{eqnarray}
and identical expressions can be derived for $A_{v,i}$, $B_{v,i}$, $C_{v,i}$ replacing $u$ with $v$.
We implemented over-sampling through an evaluation of the beam response over a number of positions $\Nsmear$ calculated for 
$-\frac{1}{2} \le l \le +\frac{1}{2}$ and including the background.
Our tests showed that $\Nsmear=11$ is sufficient.
We applied a similar procedure for the background calculation too, as mentioned in Sect.~\ref{sec:background:modelling}.

\subsection{Geometric corrections }
\label{sec:geometric:correction}

Geometric corrections have to be introduced to correct for different conditions of observations, in particular differences in planet-observer distances and planet aspect angles\footnote{Usually, analysis of planets assumes that a planet has a well-defined radius (i.e., the planet is a solid object). 
For a gas giant this is not true, since limb darkening and brightness temperature distribution across layers makes the radius a function of 
$\nu$, so that different instruments with different bandpass see different $\OmegaPlanet$. 
Analysis of this problem is postponed to another paper.
}.
The WMAP collaboration reduced all the observations to a fiducial distance
before computing $\Tb$ \citep{WMAP:PLANETS:2011,WMAP:PLANETS:2013}. 
While this step is not needed to recover $\Tb$, as those effects can be directly accounted in the fit, it is convenient for the discussion to add this step.
A geometrical correction factor is defined as 
\begin{equation}\label{eq:geometricalCorrectionFactor}
\geometricalCorrectionFactor=
\frac{\OmegaPlanetFiducial}{\OmegaPlanet} \frac{1}{1 + 
\effAsp
},
\end{equation}
where $\OmegaPlanet$ is the planet solid angle at the epoch of observation and $\OmegaPlanetFiducial$ is the planet solid angle at an arbitrary fiducial planet-planck distance,
in our case the distance of the first transit.
Planets are oblate spheroids, so that 
the solid angle depends on 
the latitude of the observer as seen from the Planet
$\PlanetAspectAngle$ (the sub-\Planck\ point).
Consequently, a fiducial $\OmegaPlanet$ may refer to an observer looking at the pole or at the equator.
The difference between the two conventions is $6.9\%$ for Jupiter $\Tb$,
$10.9\%$ for Saturn (only the disc),
$2.3\%$ for Uranus, and $1.7\%$ for Neptune.
We follow the convention in \citet{WMAP:PLANETS:2011}, \citet{WMAP:PLANETS:2013} and \citet{planck.intermediate.52.planet.flux.densities},
and we refer to observation at the equator\footnote{In \protect\citet{planck.2013.05.LFI.calibration}, the fiducial $\OmegaPlanet$ was taken at the pole.}.
In this way,
\begin{equation}
\OmegaPlanet = \frac{\pi \PlanetRadiusEquatorial \PlanetRadiusPolar} { \PlanetDistance^2},
\end{equation}
where $\PlanetDistance$ is the distance of the planet from \Planck\ at the epoch of observation,
and $\PlanetRadiusEquatorial$, $\PlanetRadiusPolar$ the equatorial and the polar radius of the planet.
In our observations, $\OmegaPlanet$ are in the ranges 
$(2.7 \div 3.1) \times 10^{-8}\,\sterad$ for Jupiter,
$(4.8 \div 5.4) \times 10^{-9}\,\sterad$ for Saturn,
$(2.2 \div 2.3) \times 10^{-10}\,\sterad$ for Uranus, and 
$(9.4 \div 9.6) \times 10^{-11}\,\sterad$ for Neptune.

The term $\effAsp$ 
accounts for the fact that 
the planet is not always seen with the same aspect angle, that is, the same ``sub-\Planck\ latitude'' $\PlanetAspectAngle$:
\begin{equation}
1 + 
\effAsp
= 
   \frac{
         \sqrt{
         (\PlanetRadiusPolar \cos \PlanetAspectAngle)^2 + (\PlanetRadiusEquatorial \sin \PlanetAspectAngle)^2 
         }
        }{
        \PlanetRadiusPolar
        }.
\end{equation}
The correction is tiny, as 
$\effAsp$
is  
$2.1\times10^{-6} \div 3\times10^{-4}$ 
for Jupiter,
$1.9\times10^{-4} \div 4.4\times10^{-3}$ 
for Uranus,
and 
$3.7\times10^{-3} \div 4.1\times10^{-3}$ 
for Neptune.
For Saturn, the disc would require a correction of the order $1.6\times10^{-4} \div 1.8\times10^{-2}$. However, having to account for the rings, we applied this correction together with that required for the rings (see Sect.~\ref{sec:Saturn}).

\subsection{Aperture correction}
\label{sec:aperture:correction}

Our fitting code assumes that every signal outside the $\ROIavoid$ is background. However, the beam extends outside $\ROIavoid$, so that the spilled signal is removed as background.
The aperture correction is defined as 
\begin{equation}
1 + \effAper = 
\frac{ 
        \int_{2\pi} \mathrm{d}\varphi \int_{0}^{\pi} \mathrm{d}\theta \sin\theta\,
        \BeamBAp(\Pointing)
}{ 
        \int_{2\pi} \mathrm{d}\varphi \int_{0}^{\ROIavoid} \mathrm{d}\theta \sin\theta\,
        \BeamBAp(\Pointing)
}.
\end{equation}
Typically, $\effAper \sim 10^{-3}$; they are listed in Table~\ref{tab:photo:parameters}.

\subsection{Beam model efficiency}\label{sec:beam:efficiency}
The integral over $4\pi$ of an ideal beam model must be normalized to some reference value, which is usually either 1 or $4\pi$. 
However, real models computed with \GRASP{} suffer numerical errors giving a slightly smaller results than the reference value. 
These numerical errors have many origins; the most important are the spatial resolution of the beam pattern, and the order of the approximation used by \GRASP{} to propagate the electromagnetic field through the telescope. 
The effect has a magnitude of some $10^{-3}$ and its value for each radiometer is reported in Table~\ref{tab:photo:parameters}, column $\effEtaBeam$. The quantity $\effEtaBeam$ is defined as 
\begin{equation}\label{eq:beam:efficiecy}
        \effEtaBeam = 1-\int_{4\pi} d\Omega \BeamGRASP(\Pointing)/4\pi.
\end{equation}
As this effect resembles a power loss in the beam, we dubbed $\effEtaBeam$ 
as {\em beam model efficiency}.

There are two possible corrections to this effect.
Firstly, we could assume that the \GRASP{} model is the correct beam model scaled by 
$1-\effEtaBeam$. In this case, $\OmegaBeamba$ is not affected by the beam model efficiency,
but the measured $\DeltaTantP$ is scaled up by a systematic factor $1/(1-\effEtaBeam)$.
This was the assumption used in
\citet{planck.2015.05.LFI.calibration}.
The problem with this approach is that the systematic effect leading to $ \effEtaBeam \ne 0$
is supposed to be equally distributed between the main beam and the side lobes. However, a \GRASP{} model is built to reproduce 
beam maps obtained from bright point sources, which are primarily sensitive to the shape of the main
beam. Therefore, the cause for $\effEtaBeam \ne 0$ is likely in the side-lobe pattern, which cannot be easily constrained
by observations.
In this case 
$\gammabat$ is unaffected by the problem, but the value of $\OmegaBeamba$ that has been derived from the model is erroneously scaled down by $1-\effEtaBeam$.

The truth is likely somewhere in the middle, and the
measured brightness must be corrected by an unknown factor 
$1+x$, with $-\effEtaBeam\le x \le +\effEtaBeam$.
As we could not tell what is the proper correction to apply, we avoided this step and left it as a source of uncertainty
with a flat distribution in the range 
$[-\effEtaBeam, \effEtaBeam]$. 
Appendix~\ref{appendix:averaged:brightness} provides more details of how 
this is accounted for in Monte Carlo simulations and bootstrap analyses.

Because it is a systematic effect, this error should be quoted separately as an unknown
scaling factor applied to 
$\BrBAplanet$, $\TbRJ$, $\TbMono$, or $\TbBA$. 
However, if we are estimating a total 
uncertainty budget, it is possible to consider this error as a random value with variance $\effEtaBeam^2/3$, which can be added to the noise variance.
This is clear in the analysis of 
$\BrBAplanet$, $\TbRJ$, $\TbMono$ or $\TbBA$ derived
from Monte Carlo simulations and bootstrap methods. A more conservative approach would be to drop the $1/3$ factor and simply add
$\effEtaBeam^2$ to the variance.

\subsection{Side lobes}
\label{sec:sidelobes}

It is customary to include only the main beam in the value of $\OmegaBeam$.
Correction for the power spilled in the side lobes is usually accounted with a term $1 + \effSL$ that corrects for the side-lobe efficiency \citep{planck.2013.05.LFI.calibration,planck.2015.05.LFI.calibration}. In this work, we do not follow this convention as we integrated $\OmegaBeamba$ over the whole $4\pi$ sphere; therefore, no $\effSL$ correction is needed.

\subsection{Blocking}\label{sec:blocking}

Blocking can be considered as a negative contribution to brightness, as in Eq.~\eqref{eq:TantToTbr}, or as a correction on the brightness temperature. The former quantity is listed in Table~\ref{tab:photo:parameters},
 in Table~\ref{tab:appendix:blocking} we provide the corresponding antenna temperatures.

 \begin{table}
   \centering
   \footnotesize
   \caption{\label{tab:appendix:blocking}
     Blocking corrections as antenna temperatures.
   }
   \begin{tabular}{lccc}
     \hline\hline
     & \multicolumn{3}{c}{$\DeltaTantBlk\, [\Kcmb]$} \\
     Planet  & 30 GHz                  & 44 GHz                   & 70 GHz\\
     \hline
     Jupiter & $6.0\times10^{-4}$ & $8.0\times10^{-4}$ & $3.0\times10^{-3}$ \\
     Saturn  & $1.1\times10^{-4}$ & $1.4\times10^{-4}$ & $4.9\times10^{-4}$ \\
Uranus  & $4.8\times10^{-6}$ & $6.2\times10^{-6}$ & $2.2\times10^{-5}$ \\
Neptune & $2.0\times10^{-6}$   & $2.6\times10^{-6}$ & $9.2\times10^{-6}$ \\
     \hline
   \end{tabular}
 \end{table}

\subsection{$\dBdTcmb$}\label{sec:dbdt_cmb}

The $\dBdTcmb$ factor is used to convert the antenna temperature $\DeltaTantP$ into a brightness, like in Eq.~\eqref{eq:DeltaTantPstar}.
The assumption that the CMB spectrum follows the RJ law, ($\dBdTcmb\propto\nu^2$), leads to overestimating the brightness of $\approx 12\%$ 
at 70\,GHz, $5\%$ at 44\,GHz, and $2.2\%$ at 30\,GHz.
Replacing $\dBdTcmbba$ with $\dBdTcmb$ has the effect of slightly underestimating the brightness
of $4.5\times10{-3} \cdots 6.9\times10{-3}$ at 30 GHz,
$2.1\times10{-3} \cdots 3.2\times10{-3}$ at 44 GHz, 
and
$1.7\times10{-3} \cdots 3.6\times10{-3}$ at 70 GHz.

\subsection{Band averaged $\Bnu$}\label{sec:band:averaged:bnu}

To solve for $\TbBA$ from Eq.~\eqref{eq:tb:ba}, we exploit the fact that 
$\BnuBA(\Tb)$ is a nearly linear increasing function of $\Tb$.
For each radiometer, we tabulate the quantity
\begin{equation}
\BnuBA(\Tb) = \frac{1}{\BandWidth} \int_{0}^{+\infty} \mathrm{d}\nu\, \tau(\nu) \Bnu(\nu,\Tb)\end{equation}
for $10\,\mathrm{K} \le \Tb \le 500\,\mathrm{K}$ in steps of 1\,K.
The tabulated function is then inverted by interpolating $\Tb$ as a function of measured $\BrBAplanet$, using the right side of Eq.~\eqref{eq:tb:ba} as input.

It is interesting to compare the difference in $\Tb$ accounting for band averaging versus simple analytical inversion of $\BrBAplanet=\Bnu(\Fcent,\Tb)$. For this, we can define a further correction factor $1 + \effTbBa = \TbBA / \invBnu(\Fcent,\BrBAplanet)$.
In all the cases, $\effTbBa < 0$: this means that neglecting band averaging causes an overestimation of $\Tb$. 
The quantity $\effTbBa$ varies between $3.9\times10^{-3}$ and $8.2\times10^{-3}$, depending on the radiometer.
Differences in brightness temperatures are between $-1.5\,\mathrm{K}$ and $-0.5\,\mathrm{K}$, depending on the radiometer and the planet.
Different planets and/or transits alter $\effTbBa$ by less than $10^{-5}$. 

\subsection{Averaged values}\label{appendix:averaged:brightness}

There are various ways to compute averaged values from our list of  measurements $\BrBAplanet$, $\TbRJ$, $\TbMono$ and $\TbBA$ for a planet from each transit and radiometer.
The simplest method is to compose a subset of measurements
specifying a list of transits and radiometers belonging to a given frequency channel 
and then to derive the weighted average of 
$\BrBAplanet$, $\TbRJ$, $\TbMono$, and $\TbBA$.
This is the approach used in \citet{planck.2015.05.LFI.calibration},
where channel averages were computed and then averaged across the transits.
The final uncertainty $\sigma_{\bar{x}}$ of $\bar{x}$ can be derived analytically using error propagation; however, if the distribution of $x$ is not Gaussian, then unreasonably small $\sigma_{\bar{x}}$ are obtained.
A better approach is to follow a least-squares minimization, fitting $\bar{x}$ to the list of $x_i$ in the subset 
\begin{equation}\label{eq:generic:weighted:average}
\chi^2(\bar{x})  = \Sigma_i 
                                \frac
                                {\left( \bar{x} - x_i\right)^2}
                                {\sigma_i^2},
\end{equation}
where $x$ is either $\BrBAplanet$, $\TbRJ$, $\TbMono$, or $\TbBA$, and 
$\sigma_i$ is the uncertainty.
As is well known, the minimization of $\chi^2$ gives the weighted average formula,
but the use of numerical minimization codes, such as the \texttt{curve\_fit} function in the \texttt{SciPy} package \citep{2020SciPy-NMeth}, 
can estimate 
$\sigma_{\bar{x}}$ from the covariance matrix of errors of fitted parameters,
leading to a more prudent estimate of the uncertainty.

Alternatively, a bootstrap of fitting residuals $r_i=(x_i-\bar{x})/\sigma_i$ can be used to resample the input $x_i$, and to derive a distribution of possible values 
of $\bar{x}$ from which $\sigma_{\bar{x}}$ can be obtained. In this case, we used the bootstrap algorithm provided by the \texttt{scikit-learn} package \citep{scikit-learn}. 
Finally by defining a likelihood for $\bar{x}$: $\log P(\bar{x}|{x_i} \propto -\chi^2(\bar{x})/2 $ and a prior for $\bar{x}$, a posterior probability for $\bar{x}$ can be formed and maximized.
Uncertainties can be estimated with Monte Carlo simulations.
 
In this paper we follow all of those approaches to define the uncertainties of our estimated averaged values, employing \texttt{emcee} \citep{emcee} for the estimation of uncertainties using Monte Carlo simulations. In general the results of the three methods are consistent each other, 
but in some cases the bootstrap approach provided larger uncertainties. 
We chose to take the largest uncertainty provided
by the three methods for each averaged quantity.

Since $\TbMono$ and $\TbBA$ are not additive quantities like $\BrBAplanet$ and $\TbRJ$, we estimated $\TbBA$ through the minimization of the quantity $\chi^2(\TbBA)$, defined as 
\begin{equation}\label{eq:averaging:tbba}
\chi^2(\TbBA)=
   \sum_{i}
   \frac{ 
   \left(
   \BnuBAi(\TbBA) 
   -
   \BrBAplaneti
   \right)^2
   }{
\sigma_i^2 
   },
\end{equation}
where $\sigma_i$ is the uncertainty on the measured $\BrBAplaneti$, and 
$\BnuBAi(T)$ is the black-body emissivity averaged over the bandpass for the radiometer that acquired the $i$-th sample.
The application of the bootstrap method requires to sample the residuals 
$r_i=(\BnuBAi(\TbBA) - \BrBAplaneti)/\sigma_i$. The application of the MCMC method requires to define the likelihood
\begin{equation}\label{eq:log:ml:brightness}
\log P({\BrBA} | \TbBA) =
   -\frac{1}{2} 
   \sum_i\frac{
           \left(
                   \BnuBAi(\TbBA) 
                   -
                   \BrBAplaneti
           \right)^2
   }{
           \sigma_i^2 
   }
   -
   \sum_i\frac{1}{2} 
   \log 2\pi
   \sigma_i^2,
\end{equation}
as well as the prior
\begin{equation}\label{eq:log:TbBA:prior}
\log P(\TbBA) = 
\left\{
\begin{array}{rl}
  -\infty, & \mathrm{if} \; \TbBA\le0 \\
    0, & \mathrm{if} \; \TbBA>0
\end{array}
\right.,
\end{equation}
which constrains $\TbBA>0$.
Similar formulas can be derived for $\TbMono$ by replacing $\BnuBAi(T)$ with 
$\Bnu(\Fcenti,T)$, where $\Fcenti$ is the central frequency of sample $i$.
 
To investigate the effect of the uncertainty in the correction for the beam numerical efficiency 
described in Sect.~\ref{sec:beam:efficiency}, we redefined the bootstrapped simulated brightness for sample $i$ as 
\begin{equation}
  B^{\mathrm{(ba)}}_{\mathrm{bstp},i} =  (1-z_{i} \effEtaBeami) (\BnuBAi(\TbBA) + \rho_{i}),
\end{equation}
where $\{\rho_i\}$ is a list of residuals sampled from the distribution of $\{r_i\}$,
$\{z_{i}\}$ is a list of random numbers taken from the uniform distribution $[-1,1]$.
Similarly, to investigate this effect using a Monte Carlo simulation
we modified the likelihood in Eq.~\eqref{eq:log:ml:brightness}
by replacing $\BnuBAi(\TbBA)$ with $\BnuBAi(\TbBA)/(1-z_{i} \effEtaBeami)$, where $\{z_i\}$ is a set of parameters with a flat distribution 
\begin{equation}\label{eq:log:csi:prior}
\log P(z_i) = \Sigma_i
\left\{
\begin{array}{rl}
    0, & \mathrm{if} \; -1 \le z_i \le 1; \\
  -\infty, & \mathrm{otherwise}
\end{array}
\right.,
\end{equation}
which multiplies the prior distribution for $\TbBA$.

Since we are dealing with averaged quantities, we had to define a reference frequency.
For each subset of measurements we took the weighted average of the $\Fcent$ of each measure, using the $\sigma_i$ as weights. 
As the relative errors for $\BrBAplanet$, $\TbRJ$, $\TbMono$, and $\TbBA$ are similar, 
the resulting averaged $\Fcent$ is nearly independent on 
the choice of the quantity to be averaged.

\subsection{Conversion of $\TbRJWMAP$ to $\Tb$}\label{appendix:WMAP:Tbrj:Tb}

The WMAP collaboration provided planets brightness temperatures in form of $\TbRJ$ and without any correction for blocking \citep[ex.]{WMAP:PLANETS:2011}.
To properly compare WMAP results to models, $\TbRJ$ must be converted either 
to $\TbMono$ or $\TbBA$.
In addition we must take in account the different value of the dipole amplitude used by \Planck{} and WMAP, as this leads to a mismatch in the absolute calibration level.
The \Planck{} team used the value $\DipolePlanck = 3364 \pm 2\,\muK$ 
\citep{planck.2013.05.LFI.calibration,planck.2015.05.LFI.calibration,planck.2018.lfi.processing},
while the WMAP  team used $\DipoleWMAP = 3355 \pm 8\,\muK$
\citep{wmap.maps.2009}. Therefore, we scaled the WMAP estimates of $\TbRJ$ by a factor of 
$\wmapTOplanck=1.002831 \pm 0.00246$. 

For $\TbMono$ we solved for 
\begin{equation}\label{eq:appendix:wmap:tbrj:to:tbc}
\Bnu(\Fcent, \TbMono) = \left(\TbRJ\wmapTOplanck + \DeltaTantBlk\right) \BrjOne(\Fcent),
\end{equation}
where $\DeltaTantBlk$ is provided by \citep{Page:etal:2003} 
for the bands K, Ka, Q, V, and W 
assuming values 2.2, 2, 1.9, 1.5 and 1.1\,K respectively.
We note that both $\BrjOne(\nu)$ and $\Bnu(\nu,T)$ depend on frequency;
we evaluated them at the central frequencies of each band as defined in Table~3 of 
\citet{WMAP:PLANETS:2011}.

For $\TbBA$, we follow what we explained in Sect.~\ref{sec:band:averaged:bnu} and solve for
\begin{multline}\label{eq:appendix:wmap:tbrj:to:tbba}
 \frac{1}{\BandWidth}\int_{0}^{+\infty} d\nu \; \BandPass(\nu) \Bnu(\nu,\TbBA)
    = \BrjOne(\Fcent) \left(\TbRJ\wmapTOplanck + \right. \\ \left. + \DeltaTantBlk \right),
\end{multline}
where the bandpass $\BandPass(\nu)$ for each band is taken from 
the Lambda website\footnote{\url{https://lambda.gsfc.nasa.gov/product/map/dr5/bandpass_info.cfm}}.

We note that other authors such as \citet{Gibson:dePater:2005} and \citet{Karim:etal:2018}
convert $\TbRJ$ to $\TbMono$ applying 
an additive correction 
\def\DeltaTrjTb{\Delta T_{\mathrm{rj}\rightarrow\mathrm{b}}} defined as follows:
\begin{equation}\label{eq:appendix:wmap:tbrj:to:tbc:mono}
\TbMono \approx \TbRJ+ \DeltaTantBlk + \DeltaTrjTb,
\end{equation}
where $\DeltaTrjTb$ is equal to 0.54, 0.79, 0.98, 1.46 and 2.23\,K for the bands from K to W; this is because of the assumption that 
$\invBnu(\nu,T) - T$ shows only a slight dependence on $T$ for $\nu < 100\,\mathrm{GHz}$ and $T>100\,\mathrm{K}$.
In this work however we prefer to apply Eq.~\eqref{appendix:WMAP:Tbrj:Tb}.

Finally, we want to underline the fact that the concise descriptions usually reported in the literature for this conversion leave some ambiguity in reproducing the published results. An example is Table~2 from \citet{Gibson:dePater:2005}. The authors quoted \citet{Page:etal:2003} and reported the value $\TbRJ=146.6\,\mathrm{K}$, but after the application of several corrections they end with 
a new estimate $\TbRJ=147.8\,\mathrm{K}$ that is converted in their final $T_{\mathrm{b}}^{\mathrm{new}}=148.4\,\mathrm{K}$ by adding 0.79\,K.
However, the 0.79\,K correction is the difference 
$T_{\mathrm{b},\mathrm{c},(\mathrm{Ka})}(148.4 \, \mathrm{K}) - 148.4\,\mathrm{K}$ 
that is derived from Eq.~\eqref{eq:appendix:wmap:tbrj:to:tbc}, and not the difference 
$T_{\mathrm{b},\mathrm{c},(\mathrm{Ka})}^{(\mathrm{ba})}(148.4 \, \mathrm{K}) - 148.4 \, \mathrm{K}$ 
from Eq.~\eqref{eq:appendix:wmap:tbrj:to:tbba}, which is 0.23\,K.
The authors state that they converted $\TbRJ$ to $\Tb$ through the integration of a black-body ideal brightness over the WMAP bandpass.
The difference is negligible when compared to the final uncertainties, so that the
conclusions in \citet{Gibson:dePater:2005} and \citet{Karim:etal:2018} (as well as other papers
that apply the same procedure) are not affected at all. 
But without the possibility to reconstruct the exact conversion procedure followed by other authors, it is difficult to judge whether small differences between our results and their results are significant or not.

\subsection{$\TbBA$ and $\TbRJ$ relations}\label{appendix:TbBA:and:TbRJ:relations}

\begin{figure}
  \centering
  \includegraphics[width=0.98\columnwidth]{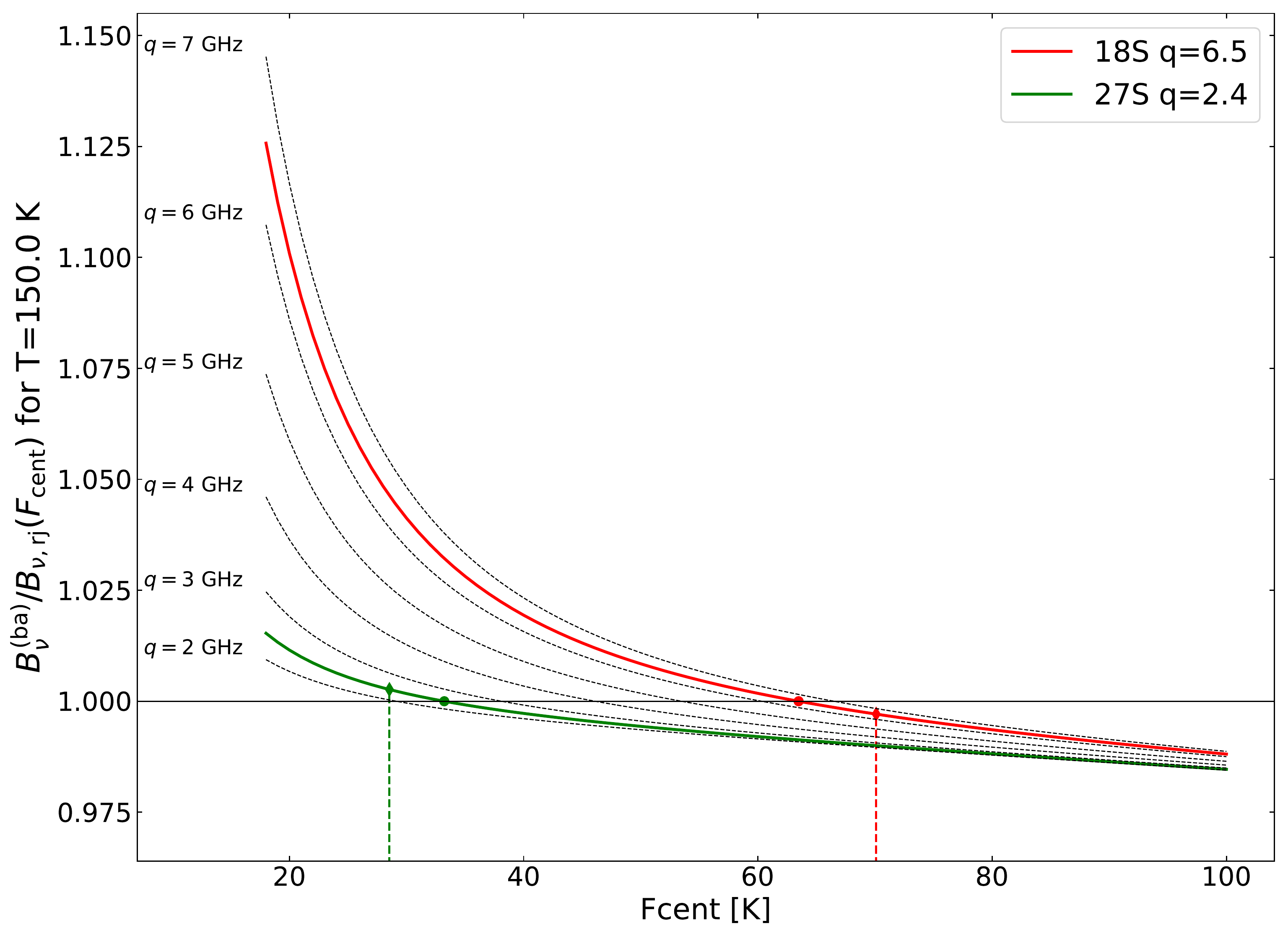}
  \caption{
    \label{fig:bnuba_brj:ratio}
Plot of the ratio $\BnuBA/\Brj$ computed
  for $T=150$\,K as a function of $\Fcent$ for the bandpasses of the
  radiometers 18S (thick red line) and 27S (thick green line),
  offseted in frequency, and for a set of top-hat bandpasses of
  different width (black thin lines).
The vertical dashed lines indicates the central frequencies of the
radiometers.
The bandwidth of the bandpasses are expressed as function of the
effective bandwidth $q$ defined in
Eq.~(\protect\ref{eq:effective:bandwidth:q}). The $q$-values for the
top-hat bandpasses are printed at the left of each curve, while
the \Planck/LFI bandpass is 6.5\,GHz for 18S and 2.4\,GHz for 27S.}
\end{figure}

It might sound surprising that $\TbRJ < \TbBA$ for 70\,GHz and 44\,GHz
data, while $\TbRJ > \TbBA$ for 30\,GHz data. However, this is
expected and is mainly a consequence of the interplay of bandwidth and
Planck's law.

Let us assume that we are observing a source with brightness $\cal{S}$.
Then, $\TbRJ$, $\TbMono$, and $\TbBA$ are solutions of the equations
$\cal{S} = \Brj(\TbRJ,\Fcent)$ and $\cal{S} = \BnuBA(\TbBA)$. Given
some temperature $T$, if $\Brj(T,\Fcent)>\BnuBA(T)$ then
$\TbRJ(\cal{S})<\TbBA(\cal{S})$, and vice versa.
We model bandpasses as top-hat functions with $\tau(\nu) = 1$ in the
frequency range $\Fcent-\delta/2\le\nu\le\Fcent+\delta/2$ and zero
otherwise.
However, to better model the nuances of the \Planck/LFI bandpasses,
instead of using $\Delta$ to characterize the bandwidth, we use the
parameter $q$, defined as follows:
\begin{equation}\label{eq:effective:bandwidth:q}
 q=\sqrt{\frac{\int \mathrm{d}\nu\;\tau(\nu)\;(\nu-\Fcent)^2}{\int \mathrm{d}\nu\;\tau(\nu)}},
\end{equation}
which reduces to $q=\delta/\sqrt{12}$ for a top-hat bandpass.
With those conventions, whenever the RJ approximation and Planck's law
agree, it follows that
\begin{equation}\label{eq:bandaverage:bnu}
 \BnuBA(T)
 \approx 
 \frac{2\kboltzman  T \Fcent^2 }{c^2} 
 \left(
      1+
\left(\frac{q}{\Fcent}\right)^2
      \right). 
\end{equation}

\begin{table}
  \caption{\label{tab:bandpass:critical:fcent}
    Effective bandwidth $q$ and critical central frequency $\FcentCrit$.
  }
  \centering
  \begin{tabular}{cccrccc}
    \toprule
    & $q$& \multicolumn{1}{c}{$\FcentCrit$}$^a$ 
    &&  
    & $q$& \multicolumn{1}{c}{$\FcentCrit$} 
    \\
    & [GHz]& \multicolumn{1}{c}{[GHz]} 
    &&  
    & [GHz]& \multicolumn{1}{c}{[GHz]} 
    \\
    \midrule
    18M&  6.01&  60.4&& 24M&  2.54&  34.2\\
    18S&  6.48&  63.5&& 24S&  2.76&  36.0\\
    19M&  4.94&  53.0&& 25M&  2.53&  34.0\\
    19S&  5.15&  54.5&& 25S&  2.43&  33.2\\
    20M&  5.18&  54.7&& 26M&  2.88&  37.1\\
    20S&  5.42&  56.4&& 26S&  2.31&  32.0\\
    21M&  5.74&  58.6&& 27M&  2.01&  29.2\\
    21S&  5.59&  57.6&& 27S&  2.44&  33.2\\
    22M&  5.46&  56.6&& 28M&  2.21&  31.2\\
    22S&  5.76&  58.7&& 28S&  2.47&  33.5\\
    23M&  4.52&  50.0&&    &      &      \\
    23S&  4.98&  53.4&&    &      &      \\
    \bottomrule
  \end{tabular}
  \tablenoteskip\\
  \tablenote{a}{$\FcentCrit$ is computed numerically at the reference
    value $T=150$\,K.}
\end{table}

\noindent
Given that the bandpass is symmetrical and receives more contribution
from the high-frequency side, the band-averaged brighntess is larger
than the monocromatic RJ. Consequently, in this approximation
$\TbBA<\TbRJ$.
However as $\Fcent$ increases, the RJ approximation overestimates the
true black-body brightness. Because of this, above some central
frequency $\FcentCrit$ we must have $\BnuBA<\Brj$ and $\TbBA>\TbRJ$.
The derivation of an approximated expression for $\FcentCrit$ is based
on the factorization of Planck's law as the product of the RJ law and
a dumping factor $x/(e^x-1)$ with $x=h\nu/\kboltzman T$.
Given that we are in the limits of small $x$ 
with a much smaller $\delta$, we may assume that over the bandpass
$x/(e^x-1)\approx x_c/(e^{x_c}-1)\approx 1/(1+x_c/2)$, where
$x_c=h\Fcent/\kboltzman T$.
Therefore, the band-averaged brightness has again the form of the
right side of Eq.~\eqref{eq:bandaverage:bnu}, but scaled by the factor
$x/(e^x - 1)$.
The critical frequency is such that
$\BnuBA(T,\FcentCrit)/\Brj(\FcentCrit)=1$, and therefore
\begin{equation}\label{eq:fcent:crit}
  \FcentCrit \approx 
  \sqrt[3]{\frac{2\kboltzman T}{h}} q^{2/3};
\end{equation}
if $T$ is expressed in K and $q$ in GHz, then
$\FcentCrit \approx 3.47\,T^{1/3}q^{2/3}\,\mathrm{GHz}$.
A representative case for \Planck/LFI observations of Jupiter is
$T=150$\,K and $q$ in the range 1\,GHz--6.5 GHz, resulting in
$\FcentCrit$ in the range 18\,GHz--64\,GHz.
A more accurate calculation can be easily obtained numerically, and this
is shown in Fig.\,\ref{fig:bnuba_brj:ratio}, where the ratio
$\BnuBA/\Brj(\Fcent)$ is calculated for the representative case
$T=150$\,K. As expected from our calculation, since the dumping factor
decreases with $\Fcent$, the ratio of the band-averaged
brightness to the RJ brightness decreases too; the main parameter
describing the curve is $q$. For radiometer 27S, $q\approx2.4$\,GHz:
in fact, the corresponding line fits nicely between the top-hat
bandpasses with $q=2$\,GHz and $q=3$\,GHz. We find a similar behaviour
for 18S, where $q\approx6.5$\,GHz.

For the \Planck/LFI radiometers, $\FcentCrit$ always falls within the
range 29.1\,GHz--66.5\,GHz. In particular for the 27S, we derive numerically $\FcentCrit=29.2$\,GHz, while the analytical approximation gives $30$\,GHz. For the 18S, we derive 63.5\,GHz numerically and 64\,GHz analytically.

The explanation of the inverted behaviour of $\TbBA$ and $\TbRJ$ at
30\,GHz and at 70\,GHz is now clear.
For 27S, $\Fcent=28.5$\,GHz, just below its critical value, so for
this channel $\BnuBA>\Brj(\Fcent)$ and $\TbBA<\TbRJ$.
On the contrary, for 18S $\Fcent=70.1$\,GHz, slightly above its
critical value: this results in $\BnuBA<\Brj(\Fcent)$ and
$\TbBA>\TbRJ$.
Last but not least, the 44\,GHz radiometers have $q$ comparable to
those of the 30\,GHz, since $\FcentCrit\approx30$\,GHz, smaller than
their central frequencies: therefore, they behave as the 70\,GHz
radiometers.

Table~\ref{tab:bandpass:critical:fcent} provides the estimates for $q$
and $\FcentCrit$ for all the \Planck/LFI radiometers. We computed the
values for $\FcentCrit$ by numerical integration at the reference
temperature of 150\,K. These values can be scaled to different
temperatures in the range 125\,K--175\,K by using the $\sqrt[3]{T}$
dependence of Eq.~\eqref{eq:fcent:crit} within a two percent accuracy.

\bibliographystyle{aa}
\bibliography{lfi_planets_paper}

\end{document}